\documentclass[longauth]{aa}  
\usepackage{graphicx}
\usepackage{txfonts}
\bibpunct{(}{)}{;}{a}{}{,} 
\usepackage[normal]{threeparttable}
\usepackage[nottoc, notlof, notlot]{tocbibind}   

\usepackage{amsmath} 
\usepackage{float}
\usepackage{mathabx}
\usepackage{tabularx}
\usepackage{siunitx}
\usepackage{colortbl}
\usepackage{soul}
\usepackage[caption = false]{subfig}
\usepackage[inkscapeformat=png]{svg}
\usepackage{hyperref}
\hypersetup{colorlinks,linkcolor={blue},citecolor={blue},urlcolor={red}} 
\usepackage{multirow}
\usepackage{orcidlink}
\newcommand{\vlsr}{\mbox{$V_{\rm LSR}$\;}}

\newcommand{\kms}{\mbox{km~s$^{-1}$\,}}

\usepackage{bm}

\newcommand{\col}[1]{\multicolumn{1}{c}{#1}}
\newcolumntype{N}{S[table-format=2.1]}

\begin{document} 

   \title{Challenges in probing turbulent and magnetic support in cores: the W43-MM1 protocluster case study}

   \author{M. Valeille-Manet\inst{1, 2}\orcidlink{0009-0005-5343-1888} \and
          F. Louvet \inst{1}\orcidlink{0000-0003-3814-4424} \and 
          F. Motte \inst{1}\orcidlink{0000-0003-1649-8002} \and 
          A. M. Stutz \inst{3}\orcidlink{0000-0003-2300-8200} \and
          C. Arce-Tord \inst{4}\orcidlink{0000-0002-0176-4331} \and 
          P. C. Cortés \inst{5,6}\orcidlink{0000-0002-3583-780X} \and
          M. Fernandez-Lopez \inst{7}\orcidlink{0000-0001-5811-0454} \and 
          N. A. Sandoval-Garrido \inst{3}\orcidlink{0000-0001-9600-2796} \and 
          P. Sanhueza \inst{8}\orcidlink{0000-0002-7125-7685} \and 
          R. H. Álvarez-Gutiérrez \inst{9}\orcidlink{0000-0002-9386-8612}\and
          S. Chevalier \inst{1} \and 
          A. Ginsburg \inst{10}\orcidlink{0000-0001-6431-9633} \and
          A. Koley \inst{3}\orcidlink{0000-0003-2713-0211} \and
          P. Saha \inst{11,12}\orcidlink{0000-0002-0028-1354} \and
          S. Savorgnano \inst{13, 14}\orcidlink{0009-0008-6617-7972} \and 
          R. Veyry \inst{1}}
   \institute{Univ. Grenoble Alpes, CNRS, IPAG, 38000 Grenoble, France             
    \and Laboratoire d’astrophysique de Bordeaux, Univ. Bordeaux, CNRS, B18N, all\'ee Geoffroy Saint-Hilaire, 33615 Pessac, France
    \and Departamento de Astronom\'{i}a, Universidad de Concepci\'{o}n,Casilla 160-C, Concepci\'{o}n, Chile
    \and Instituto de Radioastronomía, Pontificia Universidad Católica del Perú, Lima, Perú
    \and Joint ALMA Observatory, Alonso de Córdova 3107, Vitacura, Santiago, Chile
    \and National Radio Astronomy Observatory, 520 Edgemont Road, Charlottesville, VA 22903, USA
    \and Instituto Argentino de Radioastronomía (CCT-La Plata, CONICET; CICPBA), C.C. No. 5, 1894, Villa Elisa, Buenos Aires, Argentina
    \and Department of Astronomy, School of Science, The University of Tokyo, 7-3-1, Hongo, Bunkyo-ku,
Tokyo 113-0033, Japan 
    \and Scottish Universities Physics Alliance (SUPA), School of Physics and Astronomy, University of
    St. Andrews, North Haugh, St. Andrews KY16 9SS, UK
    \and Department of Astronomy, University of Florida, PO Box 112055, FL, USA
    \and Academia Sinica Institute of Astronomy and Astrophysics, No.1, Sec. 4., Roosevelt Road, Taipei 10617, Taiwan
    \and National Astronomical Observatory of Japan, National Institutes of Natural Sciences, 2-21-1 Osawa, Mitaka, Tokyo 181-8588, Japan
    \and LPSC – IN2P3 – CNRS, 53 Av. des Martyrs, 38000 Grenoble, France
    \and Institut Néel – INP – CNRS, 13 Av. des Martyrs, 38000 Grenoble, France}

   \date{Received ...; accepted ...}

 
  \abstract
    {Estimating the level of non-thermal support in cores is both challenging but crucial for constraining the earliest stages of star formation and improving our understanding of the underlying physical processes.}
    {We aim to quantify the kinetic and magnetic support operating within the cores of the high-mass protocluster W43-MM1, and to test the simple assumptions behind the approximated virial theorem, which is used to interpret observations.}
    {We used ALMA 12\,m molecular line observations of DCN\,(3–2), $\mathrm{^{13}CS}$\,(5–4), and $\mathrm{CH_3CN}\,(5_3$–$4_3)$ to estimate the kinetic support. The plane-of-sky magnetic field strength ($B_\mathrm{POS}$) was derived from ALMA 12\,m dust-polarization observations using the Davis–Chandrasekhar–Fermi (DCF) method. $B_\mathrm{POS}$ estimates are obtained in areas with a diameter of $\sim12\,500$\,au ($\sim2.3^{\prime\prime}$, the three-beam scale of polarimetric observations), which ensures statistically meaningful measurements for the DCF analysis. These values are then extrapolated to core scales ($\sim2\,500$\,au) using the density–field strength relation.}
    {We derive kinetic support estimates for 45 cores (21 prestellar and 24 protostellar), of which 21 also have magnetic field support estimates. The velocity dispersions range from 0.34 to 4.48\,$\mathrm{km\,s^{-1}}$. The three-beam scale $B_\mathrm{POS}$ values span 1.1–22.5 mG, which extrapolate to 1.1–49.3 mG at the core-scale. Using the virial theorem, we find 16 out of the subset of 24 cores with turbulence alone ($\sim70\,\%$) to be stable against gravitational collapse. From the subset of 21 cores with both estimates, 10 cores ($\sim 50\,\%$) appear to be magnetically supported against gravity when considering magnetic fields alone. Combining both estimates, 18 cores ($\sim 85\,\%$) are found with $\alpha_{\mathrm{vir,B}} > 1$. These numbers are unexpectedly high, especially for protostellar cores that are expected to be undergoing collapse.} 
    {We conclude that the contamination of the measured linewidths by organized motions (of order 1-3\,$\mathrm{km\,s^{-1}}$ and consistent with previous observational studies), together with the omission of surface terms in the observational virial theorem, prevents us from accurately measuring the non-thermal support in cores. This highlights that any simplified framework of a virial analysis can introduce significant biases when assessing the physical support mechanisms within cores.}

    \titlerunning{Challenges in probing turbulent and magnetic support in cores}
    \authorrunning{M. Valeille-Manet et al.}
    \maketitle

\section{Introduction}
The interplay of gravity, turbulence, magnetic fields, thermal pressure, and dynamical events from cloud formation or stellar feedback is complex and regulates dense gas concentration across scales. 
Observationally, some high-mass star-forming regions are known to be highly dynamical, undergoing global infall and shocks (e.g., \citealp{Schneider2010, Peretto2013, Louvet2016, Avison2021}). Moreover, they are often illuminated by massive stellar clusters with strong UV radiation fields, and associated with  strong magnetic fields (e.g., \citealp{Saha2024, Cortes2024, Sanhueza2025}). Unlike in low-mass star formation, hierarchical gas inflows often observed in high-mass ($M>8\,M_\odot$) star-forming regions (e.g., \citealp{Alvarez-Gutierrez2024, Sandoval-Garrido2025}) are thought to interact strongly with the magnetic-field topology and strength across scales—from parsec-scale clouds to 0.1 pc clumps and down to $\sim$1000 au cores (e.g., \citealp{Li2014, Padoan2020, Rosen2020}). These large-scale dynamics, as proposed in the clump-fed view \citep{Peretto2013, Motte2018}, strongly suggest that turbulence and magnetic fields, although driven on larger scales, cascade down to the core scale where they are expected to provide support against 
gravitational collapse \citep[e.g.,][]{McKee&Ostriker2007, Padoan2011}. 

At the core-scale (few thousand au), the magnetic field is known to increase with density (\citealp{Crutcher2012}) and to influence gas structure during gravitational collapse \citep{LeGouellec2024}. Ideal and non-ideal MHD models show that magnetic fields tend to reduce the fragmentation level in the formation of stars, allowing the formation of stars with higher masses within a core mass reservoir (e.g., \citealp{Commercon2011, Matsushita2017, Mignon-Risse2021a, Commercon2022}). Magnetic fields are also expected to play a crucial role in the accretion–ejection process at protostellar scales (a few hundred au). The magneto-centrifugal mechanism \citep{Blandford&Payne1982, Ferreira2006}, is proposed to remove angular momentum (via ejections along magnetic field lines) and facilitate accretion onto the protostar. Consequently, the magnetic field and the turbulence are expected to be key regulators in the dynamics of structures that lead to the birth of high-mass stars, potentially from the scale of clumps to that of cores, and  within them.

While turbulent and magnetic support in the context of star formation have been investigated for several decades (e.g., \citealp{Larson1981, Myers&Goodman1988, Bertoldi&McKee1992, Ballesteros-Paredes2006, Kauffman2013, Pillai2015, Liu2020, Li2022, Wang2024}), only a limited number of studies have quantified the relative importance of combined turbulent and magnetic support on an individual core basis \citep{Liu2020, Sanhueza2021, Cortes2021, Palau2021, Cortes2024, Zapata2024, Saha2024, Sanhueza2025, Hwang2026}.
This scarcity arises from several observational and methodological challenges. In particular, disentangling micro-turbulence which provides effective support against gravitational collapse, from organized motions (infall, rotation, streamers/converging flows) that broaden linewidths without contributing to support is non-trivial. Such analyzes further require high-angular-resolution spectral line and polarization observations. In a purely collapsing core, infall motions can indeed artificially increase the observed velocity dispersion and thus be misinterpreted as turbulent support \citep{Ballesteros-Paredes2018}. Altogether, these effects significantly complicate a reliable assessment of non-thermal support within cores.

In the high-mass star-forming regime, the W43-MM1 massive protocluster is an excellent target to investigate the influence of non-thermal support in cores. It is the youngest and most massive protocluster of the Galactic mini-starbust W43, defined as such because of its very high star formation activity in the Milky Way ($\mathrm{SFR} \sim 6000\, M_{\odot}\,\mathrm{Myr}^{-1}$, \citealp{Louvet2014}). Located at a distance of 5.5 kpc, W43-MM1 has an estimated mass of $1.7 \times 10^4\, M_{\odot}$, and an estimated size of 7\,$\mathrm{pc}^2$ (see \citealp{Nguyen2013, Dell'Ova2023}), emphasizing its strong potential for high-mass star formation. It also has high polarization fractions, with an average value of 5\,\% \citep{Cortes2016}.
At the scale of clumps, that is at $\sim0.1$\,pc, \cite{Cortes2016} previously made a study on the magnetic field supports in the protocluster W43-MM1 using a common velocity dispersion for every object, where they found that the magnetic field was dominated by gravity. 
This protocluster has also been extensively studied within the ALMA-IMF project. \cite{Motte2018} and \cite{Nony2023} showed that its core mass function (CMF) is top-heavy. \cite{Nony2020} further classified the cores into prestellar and protostellar objects based on their association with outflows, and \cite{Motte2025} provided refined temperature and mass estimates for each core and protostellar luminosities. Finally, \cite{Koley2025} found a parsec scale velocity gradient of $\sim 3$\,km/s/pc using the $\mathrm{C^{18}O}$\,(2--1) line.

Here we propose to study the support inside individual cores, at the scale of a few thousands au. We take the advantage of molecular lines from the ALMA-IMF project (see \citealp{Motte2022, Cunningham2023}) to measure the turbulence, using linewidths as a proxy, and polarization data \citep{Arce-Tord2020} to estimate the magnetic field strength in the plane of the sky (POS).
After presenting the dataset used in this work in Sect.\,\ref{sec:dataset}, we describe the methods and results of the turbulence analysis in Sect.\,\ref{sec:Turbulence}, followed by the methods and results used to derive magnetic field strength estimates in Sect.\,\ref{sec:Mag_field}. We then discuss the dynamical equilibrium of the cores in Sect.\,\ref{sec:discussion} before concluding.


\section{Dataset} \label{sec:dataset}

\subsection{Core catalog} \label{subsec:Core_catalog}
We make use of the core catalog provided in \cite{Motte2018} and further extended in \cite{Nony2023} with a mass completness of 1.6\,$M_\odot$. The cores were extracted using the algorithm \textit{getsf} \citep{Men'shchikov2021}, applied on the denoised 1.3\,mm continuum map with a resolution of $0.51^{\prime\prime}\times0.36^{\prime\prime}$. The nature of cores was defined in \cite{Nony2020} leading to the detection of 27 protostellar cores using protostellar outflows by studying the CO\,(2--1) and SiO\,(5--4) lines and the hot cores identified by \cite{Brouillet2022} and \cite{Bonfand2024}. For this work, we consider core \#14\footnote{This core is labeled as core \#30 in \cite{Valeille-Manet2025}} as prestellar, following \cite{Valeille-Manet2025}, due to its insufficient evidence of association with an outflow. We note that this change of classification does not affect any statistical result in the following. \cite{Valeille-Manet2025} also identified four high-mass prestellar cores ($M>16,M_\odot$) in W43-MM1, namely cores \#6, \#17, \#134, and \#136\footnote{These correspond to cores \#6, \#18, \#21, and \#32 in \cite{Valeille-Manet2025}, respectively.}, which show no evidence of outflow emission in CO or SiO. We note, however,
that core \#136 no longer meets the 16\,$M_\odot$ threshold used in the analysis by \cite{Valeille-Manet2025} due to differences in beam size in between the two catalogs. The core temperatures, masses and protostellar luminosities have been estimated by \cite{Motte2025} and span the 0.3-115\,$M_\odot$ and 0.14-12.21$\times 10^3\,L_\odot$ ranges.

\subsection{ALMA-IMF molecular lines} \label{subsec:linedata}
The line data used for this paper are 12\,m only line datacubes from the Large Program ALMA-IMF\footnote{\url{https://www.almaimf.com}: ALMA transforms our view of the origin of stellar masses} (\#2017.1.01355.L, PIs: Motte, Ginsburg, Louvet, Sanhueza). We used here the DCN\,(3--2) and $\mathrm{^{13}CS}$\,(5--4) lines (previously observed in the projects \#2013.1.01365.S and \#2015.1.01273.S, and then reprocessed in the ALMA-IMF pipeline) which are two dense gas tracers with high critical densities, to estimate the turbulence inside the cores of W43-MM1. These tracers were already used in previous studies to estimate the velocity dispersion inside cores (e.g., \citealp{Nony2018, Cunningham2023}). We note that we did not identify any superior tracers that more reliably peak on and trace the cores. \cite{Mininni2025} also showed that DCN is one of the molecular lines that best matches the dust emission toward cores. For two hot cores (core \#1 and core \#4 of the catalog, defined as hot cores in \citealp{Brouillet2022} and \citealp{Bonfand2024}) for which the velocity dispersion could not be estimated with these two tracers, due to highly complex line profiles, we used the hot core tracer $\mathrm{CH_3CN\,(5_3}$--$4_3)$. The central frequency, spectral resolution, beam size and critical density for each line are provided in Table\,\ref{tab:lines_spec}. The data reduction of the line cubes is described in detail in \cite{Cunningham2023}. The \vlsr of each core is derived using the fit of each line (see Sect.\,\ref{sec:Turbulence}).

\begin{table*}[htbp!] 
    \centering
    \begin{threeparttable}[c]
    \caption{Spectral lines used along with their spectral resolution, beam size and critical density.}
    \label{tab:lines_spec}
    \begin{tabular}{ccccccc}
    \hline \noalign {\smallskip}
    Line & Frequency & \multicolumn{2}{c}{Resolution} & \multicolumn{2}{c}{Beam size} & Critical density\\
        &  $\left[\mathrm{GHz}\right]$ & $\left[\mathrm{kHz}\right]$ & $\left[\kms\right]$ & $\left[^{\prime\prime} \times ^{\prime\prime}\right]$ & $\left[\mathrm{au}\right]$& $\left[\mathrm{cm^{-3}}\right]$\\
    \hline \noalign {\smallskip}
    $\mathrm{CH_3CN\,(5_3}$--$4_3)$ & 91.971 & 488 & 1.6 & $0.59\times 0.37$ & 2590 & $10^5-10^6\,\mathrm{cm^{-3}}$\\
    DCN\,(3-2) & 217.238 & 244 & 0.34 & $0.65 \times 0.47$ & 3030 & $10^7\,\mathrm{cm^{-3}}$\\
    $\mathrm{^{13}CS}$\,(5--4) & 231.221 & 244 & 0.32 & $0.59 \times 0.43$ & 2750 & $10^6\,\mathrm{cm^{-3}}$\\
    \hline
    \end{tabular}
    \end{threeparttable}
\end{table*}

\subsection{Polarization maps}
\begin{figure*}[htbp!]
    \centering
    \begin{minipage}[c]{0.74\textwidth}
        \includegraphics[width=\linewidth]{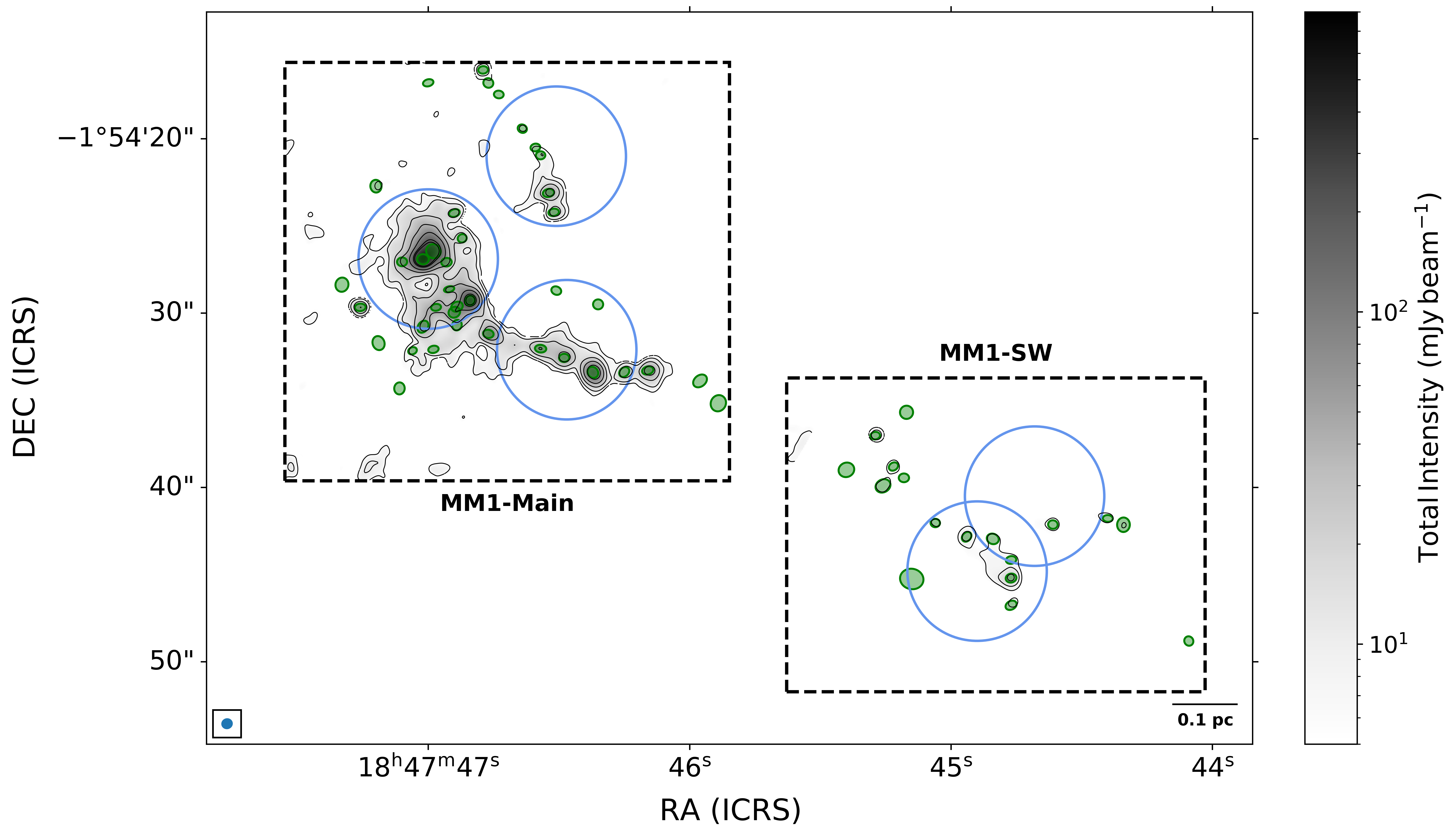}
    \end{minipage}%
    \hfill
    \begin{minipage}[c]{0.25\textwidth}
        \caption{Continuum emission at 1.3\,mm of the MM1-Main and MM1-SW mosaics from project \#2015.1.01020.S. Contours are 3, 5, 10, 20, 40, 80, 100 in units of $\sigma$, with $\sigma=2.1\,\mathrm{mJy\,beam^{-1}}$. Blue circles represent one third of primary beam of the five pointings used for the mosaics (see \citealp{Arce-Tord2020}). Green ellipses display the cores of \cite{Nony2023}. The blue ellipse in the bottom left represents the $0.66^{\prime\prime}\times0.63^{\prime\prime}$ beam.}
        \label{fig:two_mosaics}
    \end{minipage}
\end{figure*}

To map the magnetic field orientation in the plane of the sky toward W43, we used the 12\,m ALMA 1.3\,mm continuum observations in full polarization from project \#2015.1.01020.S, whose data were first presented in \cite{Arce-Tord2020}. We combined three and two pointings toward MM1-Main and MM1-SW to clean and image the different stokes parameters in mosaic mode, using the \textit{tclean} procedure from CASA 5.1.2 \citep{McMullin2007}. Imaging details are the same as in \cite{Arce-Tord2020}. 
The centers of the pointings towards MM1-Main and MM1-SW, respectively, are separated by less than $11.2^{\prime\prime}$, respecting a Nyquist stokes $I$ sampling because the primary beam is about $23^{\prime\prime}$. The two mosaics are presented in Fig.\,\ref{fig:two_mosaics}. The noise levels of Stokes $I$, $Q$ and $U$, and the beam size are presented in Table\,\ref{tab:sigma}. The noise levels for individual pointings are reported in \cite{Arce-Tord2020}.

From these observations, we have the Stokes $U$, $Q$ and $I$ maps from which we can compute the linear polarization map as $P = \sqrt{Q^2 + U^2}$. However, towards low signal-to-noise ratio (SNR) regions ($\leq 5\sigma_{\mathrm{p}}$, where $\sigma_{\mathrm{p}}$ is the rms noise level of the linearly polarized emission), the linear polarization is subject to a systematic overestimation arising from the non-linear combination of noisy Stokes $Q$ and $U$ parameters \citep{Vaillancourt2006}. To correct for this bias in low SNR regions, we applied the pixel-by-pixel method proposed by \cite{Hull&Plambeck2015}. For high SNR polarized emission ($>5\,\sigma_\mathrm{P}$), we simply use $P = \sqrt{Q^2 + U^2 - \sigma_{\mathrm{P}}^2}$. The noise level of the linear polarization map of each mosaic is presented in Table\,\ref{tab:sigma}, which is used to define the $3\,\sigma_\mathrm{P}$ threshold to plot the polarization vectors.

\begin{figure*}
    \centering
    \includegraphics[width=0.49\textwidth]{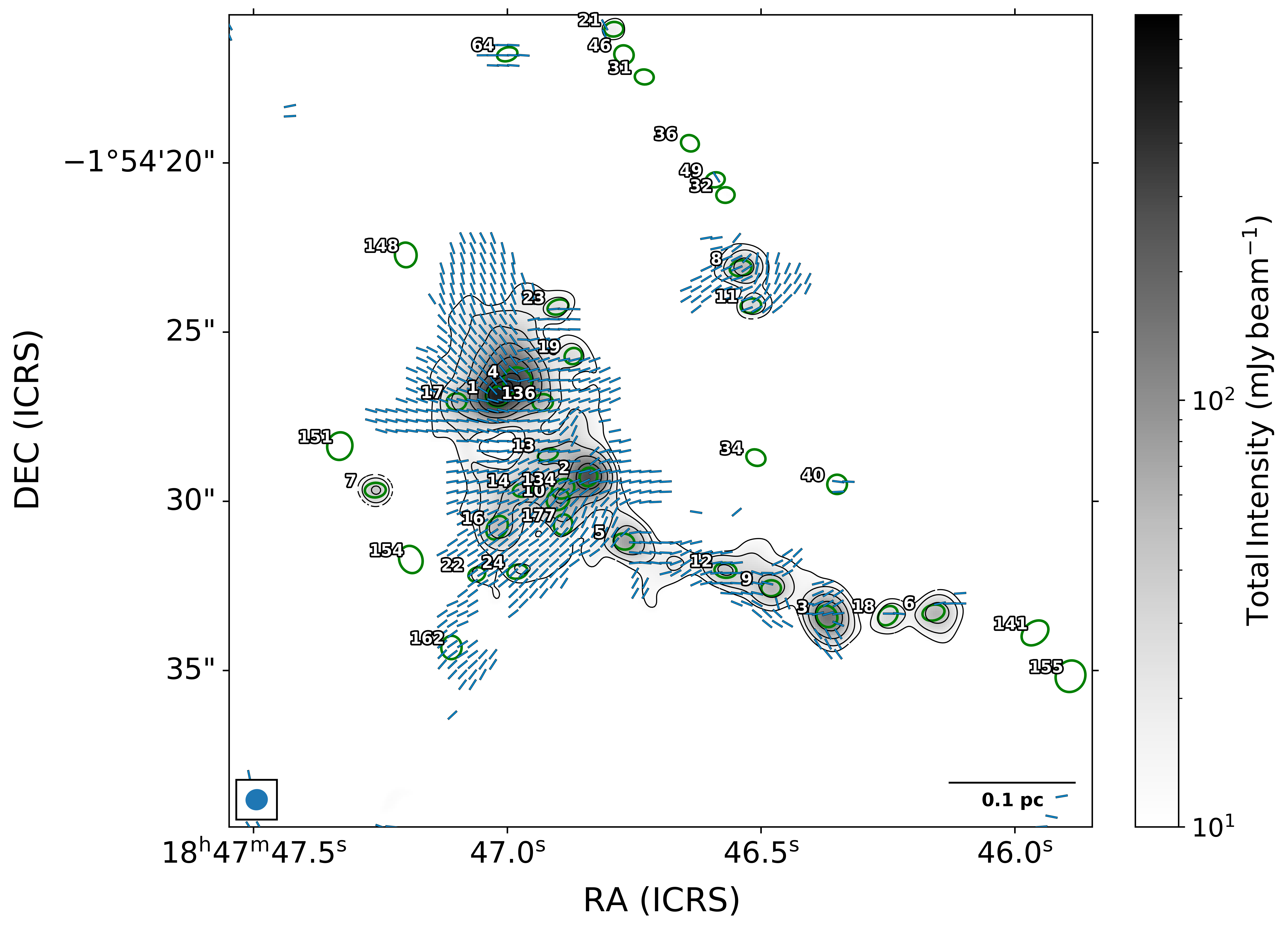}
    \includegraphics[width=0.49\textwidth]{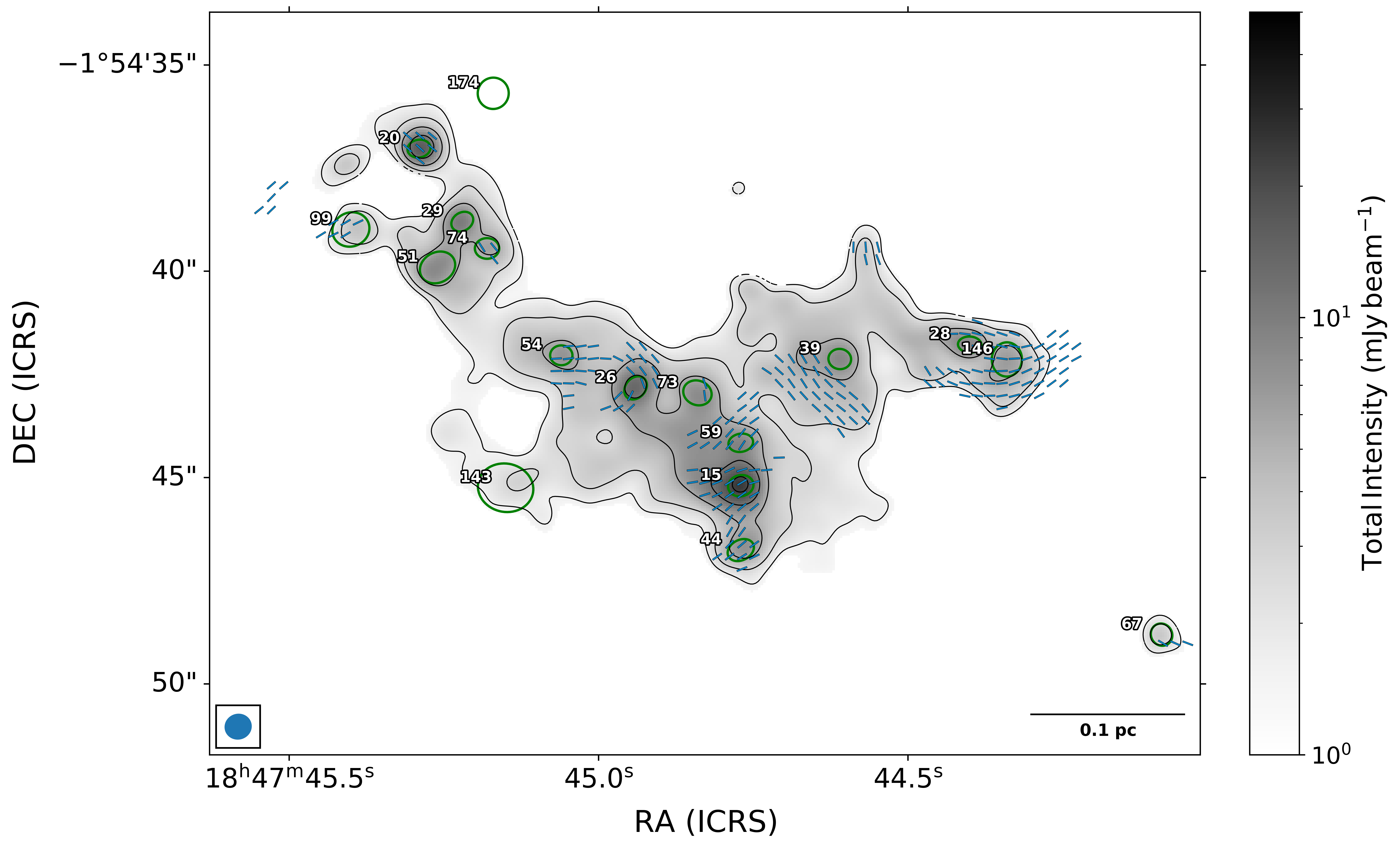}
    \caption{Zoom in the MM1-Main and MM1-SW mosaics. Continuum emission at 1.3\,mm is shown in grayscale. Contours are 3, 5, 10, 20, 40, 80, 100 in units of $\sigma$ (values are given in Table\,\ref{tab:sigma}). The blue semi-vectors show the magnetic field orientation (dust polarization semi-vectors rotated by $90^{\circ}$) where the polarized intensity exceeds a noise level of $3\,\sigma_{\mathrm{p}}$. Green ellipses display the cores of the catalog of \cite{Nony2020} with their corresponding number. The blue ellipses in the bottom left represent the continuum beam.}
    \label{fig:mosaic_Bfield}  

\end{figure*}

The polarization angle maps are derived following $\vcenter{\hbox{$\chi$}} = 0.5\,\mathrm{arctan}\left(U/Q\right)$. The “radiative alignment torque” mechanism (RAT) predicts that an external radiation field causes the dust grain to spin-up and that the short axis of these dust grains would tend to be aligned with magnetic field lines \citep{Lazarian2007, Lazarian&Hoang2007}. As a result, aligned elongated oblate grains emit polarized thermal radiation, with the polarization oriented perpendicular to the magnetic field. By rotating the observed polarization angle by $90^{\circ}$, we can thus infer the morphology of the magnetic field projected onto the POS. We present in Fig.\,\ref{fig:mosaic_Bfield} the magnetic field orientation measured over the MM1-Main and MM1-SW mosaics.

\begin{table*}[htbp!] 
    \centering
    \begin{threeparttable}
    \caption{Central coordinates, noise values and beam size for each mosaic.}
    \label{tab:sigma}
    \begin{tabular}{ccccccccc}
    \hline \noalign {\smallskip}
    Mosaic & RA & DEC & $\sigma_\mathrm{I}$ & $\sigma_\mathrm{Q}$ & $\sigma_\mathrm{U}$ & $\sigma_\mathrm{P}$ & \multicolumn{2}{c}{Beam size}\\
     & $\left[\mathrm{ICRS}\right]$ & $\left[\mathrm{ICRS}\right]$ & \multicolumn{4}{c}{$\left[\mathrm{mJy\,beam^{-1}}\right]$} & $\left[^{\prime\prime} \times ^{\prime\prime}\right]$ & $\left[\mathrm{au}\right]$\\
    \hline \noalign {\smallskip}
    W43-Main & 18:47:47.00 & -1:54:26.90 & 3.86 & 0.11 & 0.12 & 0.11 & \multirow{2}{*}{0.66$\times$0.63} & \multirow{2}{*}{3500}\\
    W43-SW   & 18:47:44.68 & -1:54:40.50 & 0.53 & 0.04 & 0.04 & 0.04 \\
    \hline
    \end{tabular}
    \end{threeparttable}

    \vspace{0.5em}
    \noindent
    \begin{minipage}{\textwidth}
    \footnotesize
    \textbf{Notes.} Noise values of the Stokes $I$, $Q$ and $U$ mosaics along with those of the linear polarization maps ($\sigma_\mathrm{P}$).  The noise values of individual pointings are reported in \cite{Arce-Tord2020}.
    \end{minipage}
\end{table*}

\section{Turbulence strength estimation}\label{sec:Turbulence}

\subsection{Method} \label{subsec:turb_method}
To investigate the turbulent support at the core-scale, we used the lines presented in Sect.\,\ref{subsec:linedata} and measured the velocity dispersion of the line using a simple Gaussian fit. 
The velocity dispersion of the gas corresponds to the sum of the non-thermal (which include turbulence) and the thermal motions:

\begin{equation}
    \label{eq:kin_Vdisp}
    \sigma_{\mathrm{v}} = \sqrt{\sigma_{\mathrm{v,NT}}^2 + \sigma_{\mathrm{th}}^2},
\end{equation}
with $\sigma_{\mathrm{th}} = \sqrt{k_{\mathrm{B}}T / \mu m_{\mathrm{H}}}$, where $\mu$ is the mean molecular weight of the line specie used to trace the velocity dispersion, $k_{\mathrm{B}}$ is the Boltzmann constant and $T$ is the gas temperature, which we assume coincides with the dust temperature (full thermalization). The turbulence, within non-thermal motions, and thermal supports combined act against gravitational collapse.

We used the same On–Off method as described in \cite{Valeille-Manet2025} (see their Sect.\,3.1), which allows to isolate the emission of the cores by subtracting the contribution of their local environment. The On spectrum is computed as the mean of the pixels within the elliptical area defining the source, while the Off spectrum corresponds to the mean signal in an annulus located between 2.5 and 3.5 times the \textit{getsf} catalog source size (we discuss the annulus parameters in Sect.\,\ref{subsec:turbulence_estimates}, see also \citealp{Valeille-Manet2025}). This method relies on the assumption that the vicinity of each core is representative of its local environment, such that subtracting the Off spectrum isolates the emission arising from the core itself from the surrounding environmental contribution. We also note that pixels belonging to neighboring cores that overlap with the annulus are masked from the Off spectrum to avoid contamination of 
the environmental contribution by emission from adjacent cores.


For most cores, we use the DCN\,(3--2) and $\mathrm{^{13}CS}$\,(5--4) lines, except for cores \#1 and \#4 (the two brightest hot cores of W43-MM1, \citealp{Brouillet2022, Bonfand2024}) for which we use the $\mathrm{CH_3CN\,(5_3}$--$4_3)$ line. The use of different tracers may introduce small systematic differences in linewidth values as they do not necessarily trace the same gas, but it is essential in order to make a statistical study of cores.
The fitting procedure follows this hierarchy:
\begin{itemize}
\item If the DCN\,(3--2) On–Off spectrum exhibits a clean, single-Gaussian profile, we use it for the velocity dispersion measurement.
\item If not, we use the $\mathrm{^{13}CS}$\,(5--4) On–Off spectrum, provided it meets the same criterion.
\item If no On–Off spectrum is suitable, we fit the On spectrum of the DCN\,(3--2) line.
\item Finally, if the DCN On spectrum cannot be reliably fitted, we use the $\mathrm{^{13}CS}$\,(5--4) On spectrum.
\end{itemize}

\subsection{Turbulence level in cores} \label{subsec:turbulence_estimates}
Out of the 56 cores identified in the two mosaics shown in Fig.\,\ref{fig:two_mosaics}, we were able to estimate the level of non-thermal motions for 45 of them (21 prestellar and 24 protostellar). We will refer this as the turbulence level but we put forward that, within the cores, we cannot disentangle turbulence from other types of gas motion. The remaining 11 cores do not show significant emission in the DCN\,(3--2) or $\mathrm{^{13}CS}$\,(5--4) lines, preventing any reliable measurement of their velocity dispersion. These 11 sources are preferentially isolated cores (prestellar and protostellar) of low masses (below $3\,M_\odot$ except for core \#21).
We also searched for emission in the $\mathrm{N_2D^+}$\,(3--2) and $\mathrm{HC_3N}$\,(24--23) transitions, other efficient tracers of dense gas available in the ALMA-IMF dataset, but without success.
Among the 45 cores with a turbulence estimate, the DCN\,(3--2) line is used for 30 sources, the $\mathrm{^{13}CS}$\,(5--4) line for 13, and the $\mathrm{CH_3CN\,(5_3}$--$4_3)$ line for cores \#1 and \#4.
The On–Off method could be applied to 27 cores, while 18 cores have measurements from their On spectra. The 18 cores with On spectra measurements are marked with a dagger in Table\,\ref{tab:Core_properties} and Table\,\ref{tab:alpha_vir}. A Kolmogorov–Smirnov (KS) test was performed to compare the On–Off and On samples, yielding a p-value greater than 0.05. This result indicates that both samples cannot be differentiated and are plausibly drawn from the same parent distribution (likely due to the limited sample size).  The values of the velocity dispersion for each core (i.e., the standard deviation of the Gaussian fit) are presented in Table\,\ref{tab:Core_properties}. We show in Fig.\,\ref{fig:fits_maps_core2} and in App.\,\ref{app:fits+mom_maps} the spectra and Gaussian fits for cores \#2 and \#1, \#3, \#13 and \#17 respectively (top left plots). We note that our Gaussian fits are not noticeably contaminated by protostellar outflows, as line wings do not affect the fitting procedure. This can be seen in Fig.\,\ref{fig:fits_maps_core2} and App.\,\ref{app:fits+mom_maps} for cores \#1, \#2, \#3, and \#13, which are protostellar cores known to drive outflows \citep{Nony2020}. 

We first examine the velocity dispersions obtained for each tracer individually. For the 30 cores with DCN detections, we find a mean velocity dispersion of 0.98\,$\mathrm{km\,s^{-1}}$. This is consistent with the mean linewidths of 1.14\,$\mathrm{km,s^{-1}}$ measured toward five cores by \citet{Cunningham2016} in the intermediate-mass star-forming regime, and 0.91\,$\mathrm{km,s^{-1}}$ measured toward four cores by \citet{Sakai2022} in the high-mass star-forming regime using the same transition. It is also slightly lower than the values reported by \citet{Cunningham2023} for the ALMA-IMF evolved (1.2\,$\mathrm{km\,s^{-1}}$), young, and intermediate protoclusters (1.5\,$\mathrm{km\,s^{-1}}$). \\
For the 13 cores with $^{13}$CS detections, the mean velocity dispersion is 1.28\,$\mathrm{km\,s^{-1}}$, consistent with the values reported by \citet{Nony2018} for one protostellar and one prestellar core in W43-MM1 using the same line. \\
The mean velocity dispersion derived from $\mathrm{CH_3CN}$ is 3.82\,$\mathrm{km\,s^{-1}}$. However, cores \#1 and \#4 are the two main hot core sources in W43-MM1 and exhibit particularly complex line profiles, which limits the comparison of their velocity dispersions with the rest of the sample. \\
The average higher velocity dispersions obtained with $^{13}$CS compared to DCN likely reflect the fact that the two lines trace different layers of gas within the cores. Since DCN traces denser gas and is detected in a larger number of cores, we consider it the more reliable tracer for this sample. We therefore apply a correction factor of 0.98/1.28 to the 
velocity dispersions derived from $^{13}$CS to place them on the same scale as the DCN measurements before combining the two sets of values.

We end up with velocity dispersions for the whole sample between 0.34 and 4.48\,$\mathrm{km\,s^{-1}}$ with a mean of 1.10\,$\mathrm{km\,s^{-1}}$. These values are slightly higher than the velocity dispersions measured in the cores of two massive protoclusters in the study of \cite{Saha2022} and  in the cores of the ASHES \citep{Li2023} survey, where typical values remain below 1\,$\mathrm{km\,s^{-1}}$. We note that different tracers were used in these studies, $\mathrm{H^{13}CO^+}$ in \cite{Saha2022}, and $\mathrm{N_2D^+}$ and $\mathrm{DCO^+}$ in \cite{Li2023}. However, W43-MM1 is a much more dynamically active region than those studied by \cite{Saha2022}. In addition, the ASHES survey \citep{Sanhueza2019, Morri2023} focused exclusively on Infrared Dark Clouds (IRDCs), which are typically less evolved and less dynamically active than W43-MM1 and the ALMA-IMF protoclusters.

We tested several annulus configurations in addition to the default 2.5--3.5 source size to assess the sensitivity of the derived On-Off linewidths to the annulus parameters. We considered four additional configurations (1.5--2.5, 2.0--4.0, 3.5--4.5, and 4.5--5.5 core sizes) and applied them to five cores with existing DCN On-Off measurements (\#2, \#7, \#11, \#14, and \#15). These cores span a mass range of ${\sim}2$--$50\,M_\odot$, velocity dispersions between 0.5 and 2.4\,$\mathrm{km\,s^{-1}}$, and are distributed across different regions of the W43-MM1 protocluster, making them a representative subsample of the full core population studied here. All configurations yield linewidths consistent with those obtained with the 2.5--3.5 annulus, with a maximum discrepancy of 0.17\,$\mathrm{km\,s^{-1}}$, which is smaller than the spectral resolution of the DCN cube. The choice of annulus size therefore has no significant impact on the derived linewidths, 
and we retained the 2.5--3.5 configuration for consistency with the approach adopted in \cite{Valeille-Manet2025}.

\section{Magnetic field strength estimation} \label{sec:Mag_field}
\subsection{Method} \label{subsec:mag_field_method}
We used the Davis-Chandrasekhar-Fermi method \citep{Davis1951, Chandrasekhar&Fermi1953} to obtain the magnetic field strength in the POS ($B_{\mathrm{POS}}$) from the dust polarized emission, following:

\begin{equation}
    \label{eq:Bpos}
    B_{\mathrm{POS}} = Q \sqrt{\mu_0 \, \rho} \, \frac{\sigma_{\mathrm{v, NT}}}{\sigma_{\theta}} \left(\mathrm{SI}\right) = Q \sqrt{4\pi \, \rho} \, \frac{\sigma_{\mathrm{v, NT}}}{\sigma_{\theta}} \left(\mathrm{cgs}\right),
\end{equation}
where $\rho$ is the density of the gas, $\sigma_{\mathrm{v, NT}}$ is the non-thermal velocity dispersion in the gas traced by the linewidth of a molecular line (see Sect.\,\ref{subsec:turb_method}), $\sigma_{\theta}$ is the dispersion in polarization position angle, and $Q$ is an empiric correction factor to account for overestimations of the magnetic field strength ($Q=0.5$, \citealp{Ostriker2001, Padoan2001, Heitsch2001, Liu2019}). \citet{Skalidis2021} showed that the factor $Q$ depends on the turbulent regime and varies with the Alfvénic Mach number. However, since estimating the Alfvénic Mach number itself requires an assumed value of $Q$, adopting a Mach-number-dependent $Q$ would introduce circularity into our analysis. We therefore opted to keep $Q=0.5$, consistent with the value adopted in much of the literature. \cite{Huang2025} also compared the classical DCF method with the more sophisticated variants of \cite{Heitsch2001} and \cite{Falceta-Goncalves2008} to estimate magnetic field strengths toward 26 protostars in Orion. All methods yielded consistent results, with differences smaller than 0.1\,mG for field strengths of a few mG ($<5\,\%$). For one protostar with sufficient statistics, they further compared their estimates with the ADF method of \citet{Houde2009}, finding a difference of 0.3\,mG ($\sim10\,\%)$. These results suggest that the classical DCF method with a correction factor $Q$, as adopted in this work, provides robust magnetic field strength estimates compared to more elaborate methods. 

As shown in \cite{LeGouellec2020}, the angle dispersion can be computed following:

\begin{equation}
    \label{eq:disp_angle}
    \sigma_{\theta} = \sqrt{\frac{1}{4n} \sum_{i=1}^n \left[ \mathrm{arctan}\left( \frac{U_i \, \langle Q \rangle - Q_i \, \langle U \rangle}{Q_i \, \langle Q \rangle + U_i \, \langle U \rangle}\right)\right]^2},
\end{equation}
where $n$ represents the number of pixels over which the dispersion is computed, and $U$ and $Q$ the respective Stokes. This classical DCF method is typically used in regions where $\sigma_\theta < 25^{\circ}$ \citep{Ostriker2001}.
 
To satisfy the Nyquist spatial sampling criterion, we consider four semi-vectors (representing the magnetic field orientation) per beam (two in each direction, see Fig.\,\ref{fig:mosaic_Bfield}).
To ensure a statistically meaningful estimate of the angular dispersion within each core, we computed it over an area that encompasses 10 beam-independent points. It assures a dispersion value within 15\,\% from the true dispersion \citep{Cortes2019}. This area corresponds to a diameter of 3.6 beams major axis centered on the core and will be referred to as the three-beam scale or three-beam area from now on.
At this scale ($\sim 2.3^{\prime\prime}$), the dispersion estimate typically includes about 45 semi-vectors, which corresponds to measuring the magnetic field strength on surroundings more diffuse gas.

To maximize the number of cores for which a magnetic field strength can be derived, we included all cores containing at least 20 semi-vectors.
We quantified the bias of using fewer semi-vectors on cores \#1 and \#14, both containing 45 semi-vectors above 3\,$\sigma_{\mathrm{P}}$ in their three-beam area, and allowing us to probe this effect in two different areas of the MM1-Main mosaic. We recalculated the dispersion after randomly removing between 1 and 25 semi-vectors (performing 1000 random draws for each case).
As shown in Appendix\,\ref{fig:app_sampling_effect}, the resulting angle dispersion varies by 10\,\% at most, demonstrating that the magnetic field strength estimate is robust against moderate incompleteness in the sampling.

Since the dispersion of polarization position angles is measured over a region corresponding to 3.6$\times$beam in diameter, the non-thermal velocity dispersion must be estimated over the same spatial scale.
To properly trace the diffuse gas surrounding each core, and not artificially inflate its non-thermal velocity dispersion, we excluded all pixels belonging to the central core itself as well as to any neighboring cores that overlap with the three-beam area. 
As described in Sect.\,\ref{subsec:turb_method}, we then extracted the spectra of each source within this area and attempted to fit them with a single-Gaussian profile.
This selection ensures that the velocity dispersion we measure reflects the non-thermal motions of the surrounding medium (three-beam scale) rather than those of the cores.

To compute the density used in Eq.\,\ref{eq:Bpos}, we measured the integrated 1.3\,mm flux within the previously defined area (i.e., the three-beam area with pixels from the central core and neighboring cores removed) using the bsens–denoised continuum map from \cite{Nony2023}. We used this map because it provides higher data quality than the Stokes $I$ map presented above and ensures consistency with the core fluxes, which were derived from the same dataset. We then converted it into a mass using Eq.\,\ref{eq:Mass}.
For this conversion, we adopt the dust temperature ($T_{\mathrm{dust}}$) derived by \cite{Dell'Ova2023} at a spatial scale of $2.5^{\prime\prime}$ and listed in \cite{Motte2025}:

\begin{equation}
    \label{eq:Mass}
    M_{\mathrm{3beam}} = \frac{S_{1.3 \,\mathrm{mm}}^{\mathrm{int}} \;d^2}{\kappa_{1.3 \,\mathrm{mm}}\; B_{1.3 \,\mathrm{mm}}\left(T_{\mathrm{dust}}\right)},
\end{equation}
where $S_{1.3 \,\mathrm{mm}}$ is the integrated flux at 1.3\,mm, $d=5.5$\,kpc is the distance of the W43-MM1 protocluster, $\kappa_{1.3 \,\mathrm{mm}}$ is the dust $+$ gas mass opacity set to $0.01 \, \mathrm{cm^2\, g^{-1}}$ \citep{Ossenkopf1994} assuming a gas-to-dust ratio of 100, and $B_{1.3 \,\mathrm{mm}}\left(T_{\mathrm{dust}}\right)$ is the Planck function.
Assuming spherical symmetry, we derived the volume corresponding to the three-beam area and thus estimated the average gas density, which in turn allows us to evaluate the magnetic field strength in the surrounding diffuse medium.

Finally, to relate this large-scale estimate to the magnetic field strength at the scale of the cores, we extrapolate these three-beam scale values. We perform this extrapolation down to the core scales (${\sim}2500$\,au) in order to compare the magnetic, kinetic, and gravitational energies at a common physical scale. For that, we compute each core density using their mass and size, and we apply the field–density relation, $B \, \propto\, n^{\beta}$ with $\beta = 0.53^{+0.09}_{-0.07}$ \citep{Whitworth2025}. We note that this observational relation relies on Zeeman measurements which are more reliable than ones with the DCF method.

\subsection{Velocity structures in the three-beam area}\label{subsec:turb_refinement}
We measured the velocity dispersion at the three-beam scale for all the cores presenting at least 20 polarization semi-vectors. In total we are able to recover the magnetic field strength for 21 cores. The $\mathrm{CH_3CN\,(5_3}$--$4_3)$ line is used for cores \#1 and \#4, $\mathrm{^{13}CS}$\,(5--4) for cores \#17 and \#22 (which have their velocity dispersion corrected by the same factor as in Sect.\,\ref{subsec:turbulence_estimates}) and DCN\,(3--2) for the remaining 17 cores. 

To ensure that our linewidth measurements truly trace turbulent motions at both the core and three-beam scales, we constructed moment-0 (integrated intensity) and moment-1 (velocity) maps around each core. The three-beam spectra with Gaussian fits and the moment maps for cores \#2, and \#1, \#3, \#13 and \#17 are shown in Fig.\,\ref{fig:fits_maps_core2} and in Appendix\,\ref{app:fits+mom_maps} (top right and bottom panels), respectively. In these figures, the ellipses representing the sizes of the core are overlaid on the moment-0 maps, while the masks used to extract the three-beam spectra (and thus the three-beam velocity dispersion) and densities are shown on the moment-1 map. As seen for cores \#13 and \#17, the tracer does not always peak on the core itself and may instead trace the surrounding material. Approximately 50\,\% of the cores display moment-0 maps with extended emission. This extended emission has also been observed for DCN in the study of \cite{Mininni2025}.

The moment-1 maps clearly reveal organized velocity structures and gradients across the three-beam masks, which indicate that large-scale motions add to micro-turbulence to widen the line width. If the gas kinematics at this scale were purely turbulent, we would expect the velocity field within the mask to appear spatially homogeneous, with no obvious gradients. This directly highlights a limitation of the DCF method, which implicitly assumes that the observed velocity dispersion is dominated by isotropic turbulent motions rather than ordered large-scale flows. This behavior is observed systematically across the sample, as illustrated in Fig.\,\ref{fig:fits_maps_core2}, and in Appendix\,\ref{app:fits+mom_maps}. Similar velocity structures have also been reported on a few core scales belonging to some ALMA-IMF protoclusters \citep{Alvarez-Gutierrez2024, Sandoval-Garrido2025}, where it is referred to as a “V-shape”.
At the core scale, the moment-1 maps appear comparatively homogeneous.
Given that the beam size is comparable to the core size, the observed velocity field mainly reflects averaged motions at this scale, while potential sub-beam velocity structures would remain unresolved.

To refine our estimate of the turbulent velocity dispersion at the three-beam scale, we therefore account for the contribution of these large-scale motions. We computed the standard deviation of the pixel velocities inside each three-beam mask from the moment-1 map, which quantifies the magnitude of these coherent gradients. This value is then subtracted in quadrature from the velocity dispersion derived from the Gaussian line fits (see Sect.\,\ref{subsec:turb_method}), to avoid as much as possible large-scale motions in this three-beam area. We also subtract the thermal velocity component (in quadrature) from the three-beam velocity dispersion, as the magnetic field strength estimate should only account for non-thermal turbulent motions. The resulting corrected velocity dispersion provides a refined estimate of the random, turbulent component of the gas motions, reducing the contribution from large-scale velocity gradients or ordered flows. These corrected velocity dispersions at the three-beam scale range from 0.63 to 3.12 \,$\mathrm{km\,s^{-1}}$ with a mean value of 1.70 \,$\mathrm{km\,s^{-1}}$ (see Table\,\ref{tab:Core_properties}). Values increased by 5\,\% on average would have been found without this refinement.

We also compared our three-beam scale velocity dispersion estimates with the method proposed by \citet{Polychronakis2025}, which consists in taking the mean of the second moment map within the three-beam mask of each core as a proxy for the velocity dispersion used in magnetic field estimates. Applying this method to the 17 cores with DCN measurements, we find that the resulting velocity dispersions are on average ${\sim}20$\,\% larger than those derived with our approach. In the following, we adopt the results from our method presented above.

\subsection{Magnetic field strength in cores}\label{subsec:mag_field_estimate}

Using the velocity dispersions derived above and presented in Table\,\ref{tab:Core_properties}, we estimated the magnetic field strength in the POS at the three-beam scale using Eq.\,\ref{eq:Bpos}. To obtain the dispersion of polarization angles required in this calculation, we apply Eq.\,\ref{eq:disp_angle}, which directly computes the angular dispersion from the Stokes $Q$ and $U$ parameters. This computation is performed over all pixels in the three-beam area where the polarized intensity exceeds $3\,\sigma_\mathrm{P}$. The values of $\sigma_\theta$ are presented in Table\,\ref{tab:Core_properties}. As explained in Sect.\,\ref{subsec:mag_field_method}, the DCF method is reliable for regions with $\sigma_\theta < 25^{\circ}$, which here is the case for all the cores except cores \#3, \#4, \#5 and \#9 that we decided to keep in the following analysis in order to preserve the statistical significance of the sample. The slightly higher angle dispersions observed in these latter cores could be related to gravitational distortions of the magnetic field morphology, which would in turn lead to an underestimation of the field strength. A similar trend was reported by \citet{Cortes2019} in other protoclusters of the W43 complex. We discuss again these four cores in Sect.\,\ref{subsec:mag_support_and_contribution}.

We investigated the impact of the assumed geometry on our results. Since the adopted geometry can significantly affect the density estimates used in the magnetic field strength computation, we assessed the influence of adopting an ellipsoidal geometry in place of the default spherical one. Specifically, we modeled each core as an oblate disk (e.g., \citealp{Myers2018}), taking the semi-major axis as the third dimension, and defined a three-beam ellipsoidal area with the same 
surface area as its spherical counterpart. We find that ellipsoidal geometries yield three-beam scale densities that are on average 10\,\% smaller than the spherical estimates, translating into $B_\mathrm{POS}$ values that are on average 5\,\% smaller. To account for this geometrical uncertainty, we adopt the mean value between the spherical and ellipsoidal estimates for the three-beam scale $B_\mathrm{POS}$, with its associated uncertainty.

We derive $B_\mathrm{POS}$ at the three-beam scale ranging from 1.1 to 22.5\,mG. These values are higher than those reported by \citet{Cortes2016} for W43-MM1 and other W43 protoclusters \citep{Cortes2019}. 
This difference likely arises from the differences in our approach. We use individual velocity dispersion measurements for each core, derived from spectral line fitting, whereas \cite{Cortes2016} and \cite{Cortes2019} adopted a single, region-averaged velocity dispersion (1.3\,$\mathrm{km\,s^{-1}}$ in W43-MM1 and $<1\,\mathrm{km\,s^{-1}}$ in the other protoclusters) for all cores within each protocluster. We have also smaller core-specific angle dispersions on average, and our cores are less dense due to the individual temperatures adopted from \cite{Motte2025}. Magnetic field strengths of the order of $\sim$ a few mG have also been observed in other high-mass star-forming regions (e.g., \citealp{Zapata2024, Cortes2024, Cortes2021, Sanhueza2021, Liu2020}).

We also compared our results with the method of \citet{Skalidis_Tassis2021}, which proposes to replace $Q/\sigma_\theta$ by $1/\sqrt{2\sigma_\theta}$ in the DCF formula to account for non-Alfv\'enic compressible modes. Applying this method to our sample of 21 cores, we obtain magnetic field strengths that are on average ${\sim}20$\,\% smaller than those derived with our approach. Interestingly, this difference is comparable to the 17\% underestimation that \cite{Skalidis_Tassis2021} report for their method based on comparisons with simulated data, although we note that this does not constitute an independent validation of either approach, since our DCF estimates are not a ground-truth reference. We decided to retain our results from the method presented above, as the differences in the two methods are small and do not change the conclusions of this paper presented in Sect.\,\ref{sec:discussion} and \ref{sec:conclusion}.

Using these three-beam scale $B_\mathrm{POS}$, together with the density of each core and its corresponding three-beam area (see Sect.\,\ref{subsec:mag_field_method}), we extrapolated  the $B_\mathrm{POS}$ values to the core-scale. We obtained resulting $B_\mathrm{{POS}}$ values from 1.1 to 49.3\,mG. We note that three cores (cores \#4, \#10, \#146) exhibit core-scale $B_\mathrm{{POS}}$ slightly smaller than their three-beam scale estimates due to higher density surroundings. Among the 21 cores, 14 have field strengths above 10\,mG, and nine of those exceed 20\,mG. These are the first extrapolated estimates of magnetic field strengths reaching such magnitudes, but this is also the first study that attempts to extrapolate the field strengths down to core-scales using a density–field scaling relation. This observational density–magnetic field extrapolation can be compared with theoretical expectations. Under the assumption of flux-freezing (i.e., ideal MHD; \citealp{Mestel1966}), 
the magnetic field strength scales with density as $B \,\propto\, n^{2/3}$. When ambipolar diffusion is taken into account (i.e., resistive effects in non-ideal MHD; \citealp{Mouschovias1999}), the scaling becomes shallower, typically $B \,\propto\, n^{0.5}$, which is close to the observational trend adopted here. In the flux-freezing case, our extrapolated $B_{\mathrm{POS}}$ values would range from 1 to 70\,mG, with five cores exceeding 40\,mG. However, as shown by \citet{Tritsis2015}, the relation $B \,\propto\, n^{2/3}$ is only valid in the case of spherical collapse, which is not expected when magnetic fields are dynamically important and resistive effects are present. In such conditions, the relation $B \,\propto\, n^{0.5}$ is theoretically preferred. We therefore adopted the observational scaling relation of \citet{Whitworth2025}, which is consistent with this expectation. With these results in hand, we can now perform a simplified virial analysis to investigate the dynamical equilibrium of the cores. 

\begin{figure*}[h]
\centering
\includegraphics[width=0.49\textwidth]{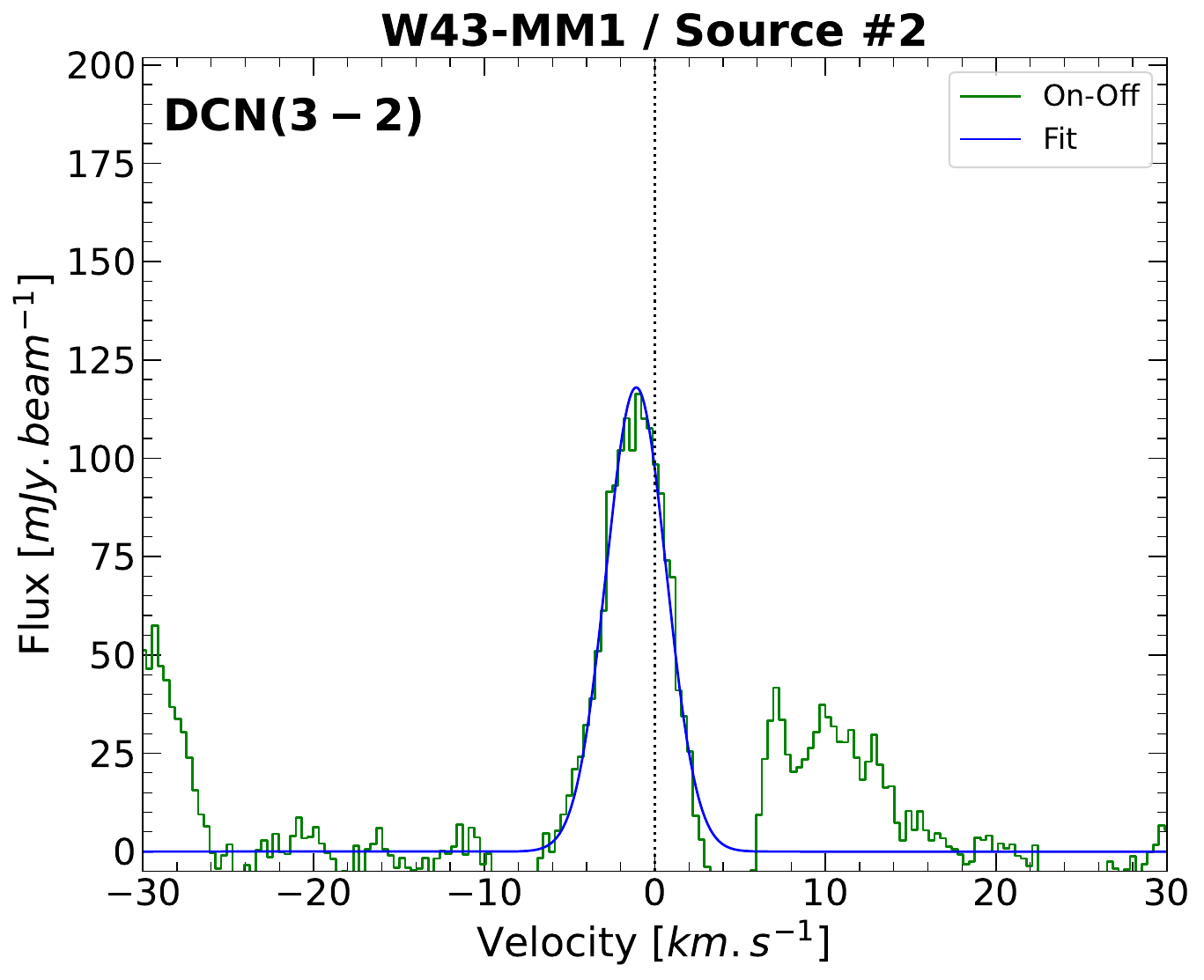}
\includegraphics[width=0.48\textwidth]{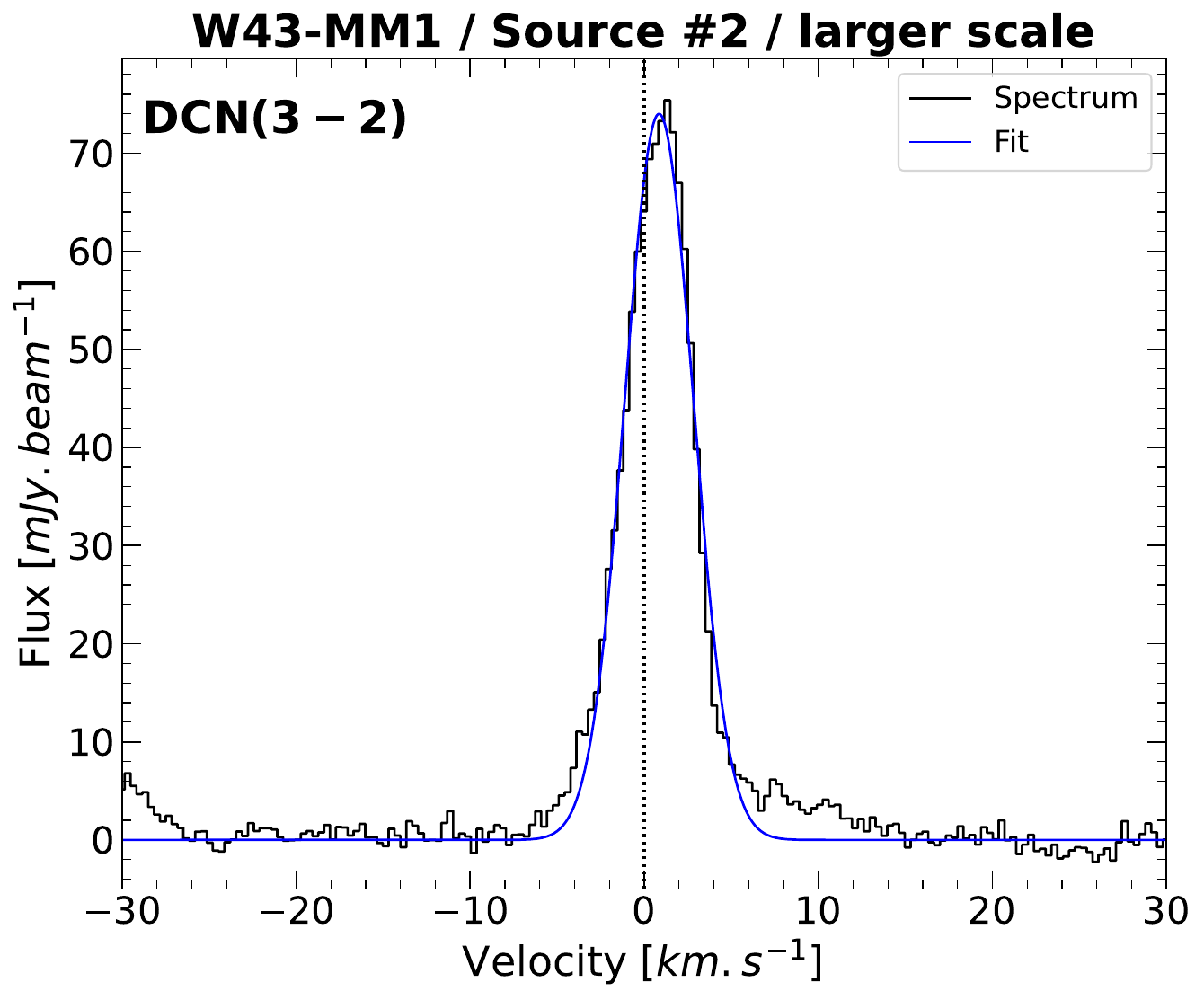}
\includegraphics[width=0.49\textwidth]{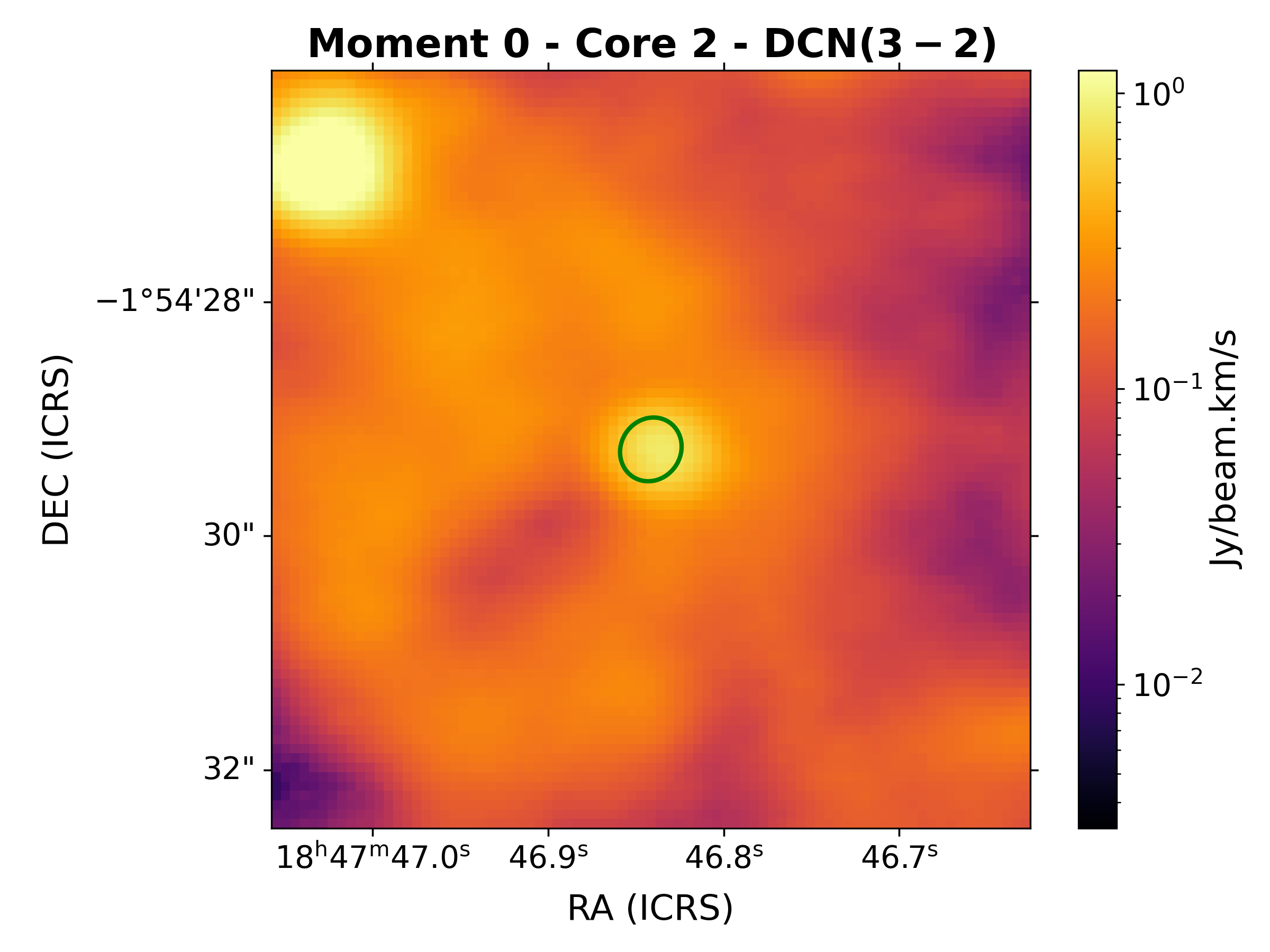}
\includegraphics[width=0.49\textwidth]{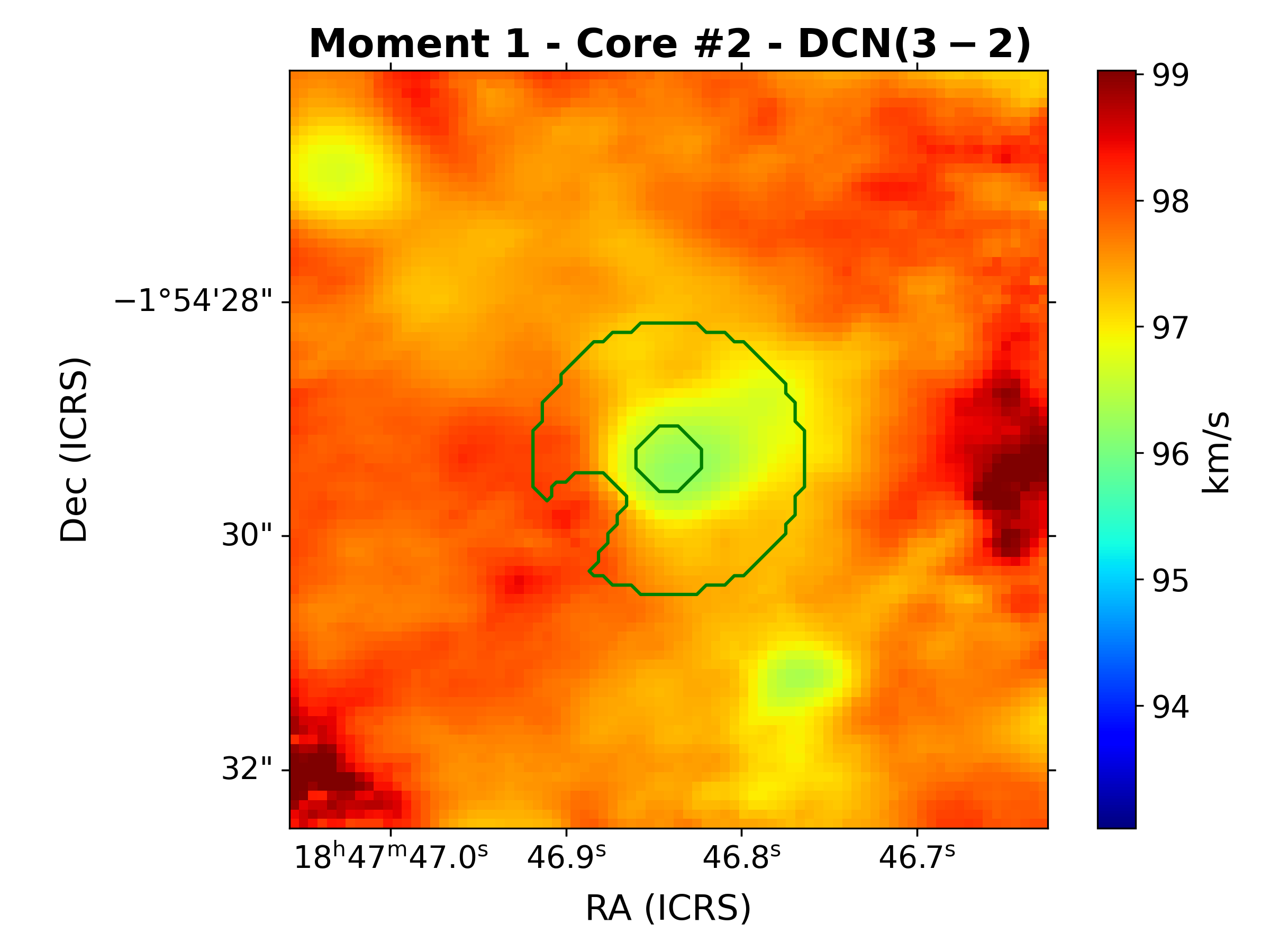}
\caption{Gaussian fits and moment maps for core \#2. Top row: Gaussian fits and spectra used to estimate the velocity dispersions at the core-scale (left) and at the three-beam scale (right). Bottom row: moment-0 map (left) of the tracer used to derive the core-scale turbulence, and moment-1 map (right) of the tracer used for the three-beam-scale turbulence. The green ellipse in the moment-0 map marks the continuum core, while the dark-green region in the moment-1 map indicates the mask used to extract the density and spectra at the three-beam scale.}
\label{fig:fits_maps_core2}  
\end{figure*}




\section{Dynamical equilibrium of the W43-MM1 cores}\label{sec:discussion}

\subsection{Thermal and turbulent supports} \label{subsec:thermal_turbulent_support}
As individual temperatures are available for all cores from \cite{Motte2025}, we can first estimate the thermal contribution to the velocity dispersions derived in Sect.\,\ref{subsec:turbulence_estimates}. 
The mean molecular weights are 28 for DCN, 45 for $^{13}$CS, and 41 for $\mathrm{CH_3CN}$. Mass averaged dust temperature of cores range from 17 to 69\,K, which correspond to thermal velocity dispersions of 0.06 to 0.14\,$\mathrm{km\,s^{-1}}$ depending on the tracer and the source. These values, of the order of $0.1\,\mathrm{km\,s^{-1}}$, are significantly lower than the observed line-of-sight dispersions reported in Table\,\ref{tab:Core_properties}. Thermal motions therefore contribute only marginally to the total linewidths, and are dynamically negligible. In the following, we thus refer to the kinetic support as the combination of thermal and turbulent components, with turbulence largely dominating.

Using the velocity dispersions and the magnetic field strengths derived in the sections above, we can apply the simplified virial theorem, which is limited to volume terms (e.g., \citealp{Dib2007}, Chevalier et al. in prep) because only these can be constrained by observations. We then compute the parameter $\alpha$ to evaluate whether non-thermal support is strong enough to counteract gravity. If $\alpha < 1$, the core is subvirial, meaning that gravity dominates and the core is prone to collapse. Conversely, if $\alpha > 1$, the core is supervirial, indicating that volume support provided by either turbulence or magnetic support is strong enough to prevent collapse.
If we only account for kinetic support, the virial parameter can be expressed as:
\begin{equation}
    \label{eq:alpha_kin}
    \alpha_{\mathrm{kin}} = \frac{M_{\mathrm{kin}}}{M_{\mathrm{core}}}=3\left(\frac{5-2 p}{3-p}\right) \frac{F W H M^{\mathrm{dec}}\,\sigma_\mathrm{v}^2}{G\, M_{\mathrm{core}}},
\end{equation}
where $p$ is the power-law index of the density profile, $FWHM^{\mathrm{dec}}$ is the deconvolved size of the core, $G$ is the gravitational constant, $\sigma_\mathrm{v}$ is the velocity dispersion of the core estimated with the Gaussian fit (which corresponds to $\sigma_\mathrm{v} = \sqrt{\sigma_\mathrm{v,NT}^2 + \sigma_\mathrm{th}^2}$), and $M_{\mathrm{core}}$ is the mass of the core. The mass of each core is computed using its flux measured by \cite{Nony2023} and its temperature from \cite{Motte2025}. The velocity dispersion for each core is taken from Table\,\ref{tab:Core_properties}. This methodology builds upon the approach adopted in \cite{Cortes2016} but uses, for the first time, individual values for the linewidth and the temperature of cores. We note that, for protostellar cores, the mass of the central protostar is not included in the total mass, as reliable measurements are not available. However, this omission is not expected to significantly affect the results presented below because W43-MM1 protostars correspond to young, Class 0-like protostars with most of their mass still in the protostellar envelope (e.g., \citealp{Motte2025}). The power-law index $p$ will be chosen equal to 0 (flat density profile) for prestellar cores and equal to 2 (peaked density profile) for protostellar cores for the rest of the study. If $p=2$ were adopted for prestellar cores and $p=0$ for protostellar cores, their virial parameter would be reduced by a factor $3/5$ and increased by a factor $5/3$ respectively.

\begin{figure*}[h!]
    \centering
    \includegraphics[width=0.49\textwidth]{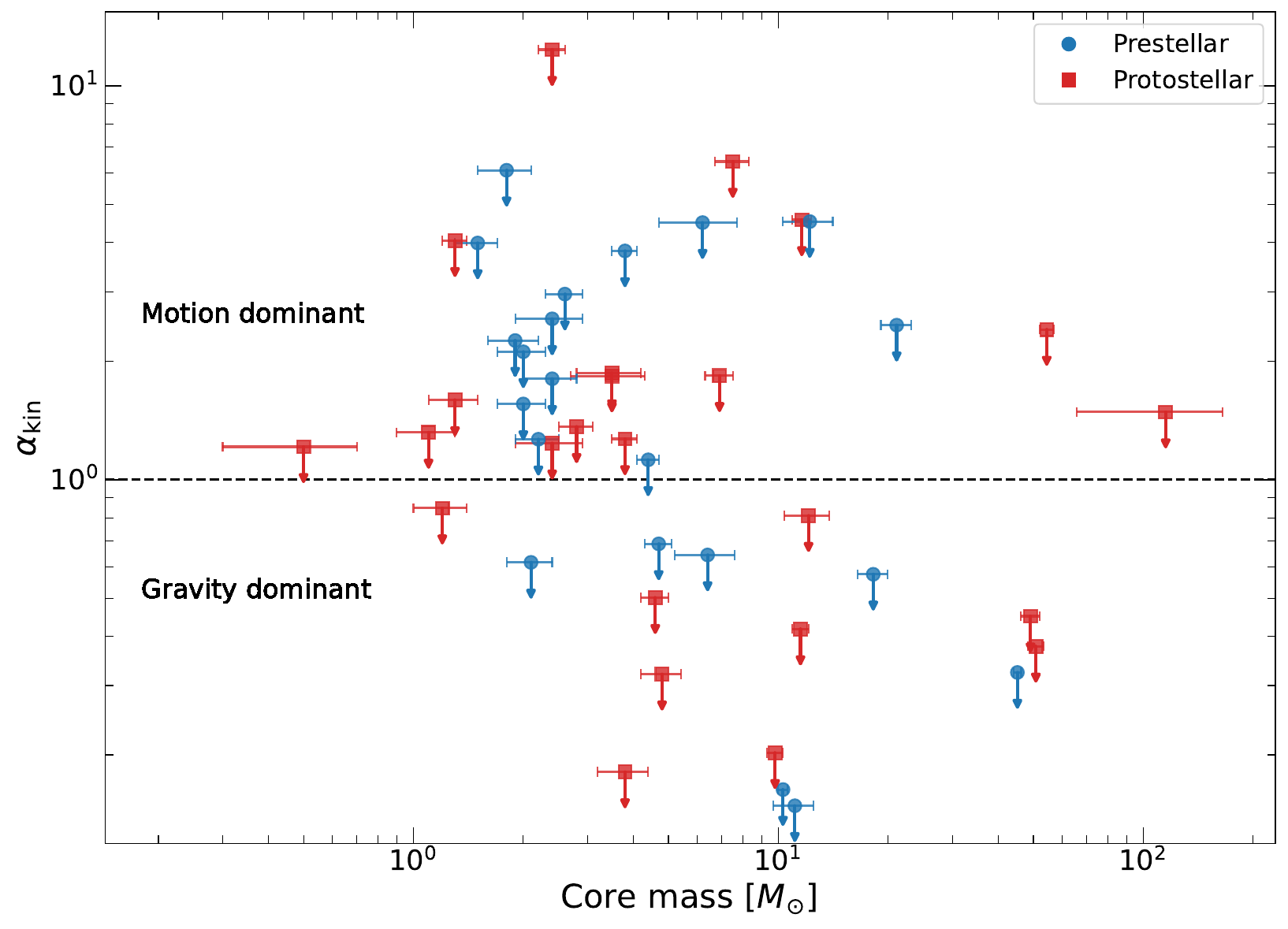}
    \includegraphics[width=0.49\textwidth]{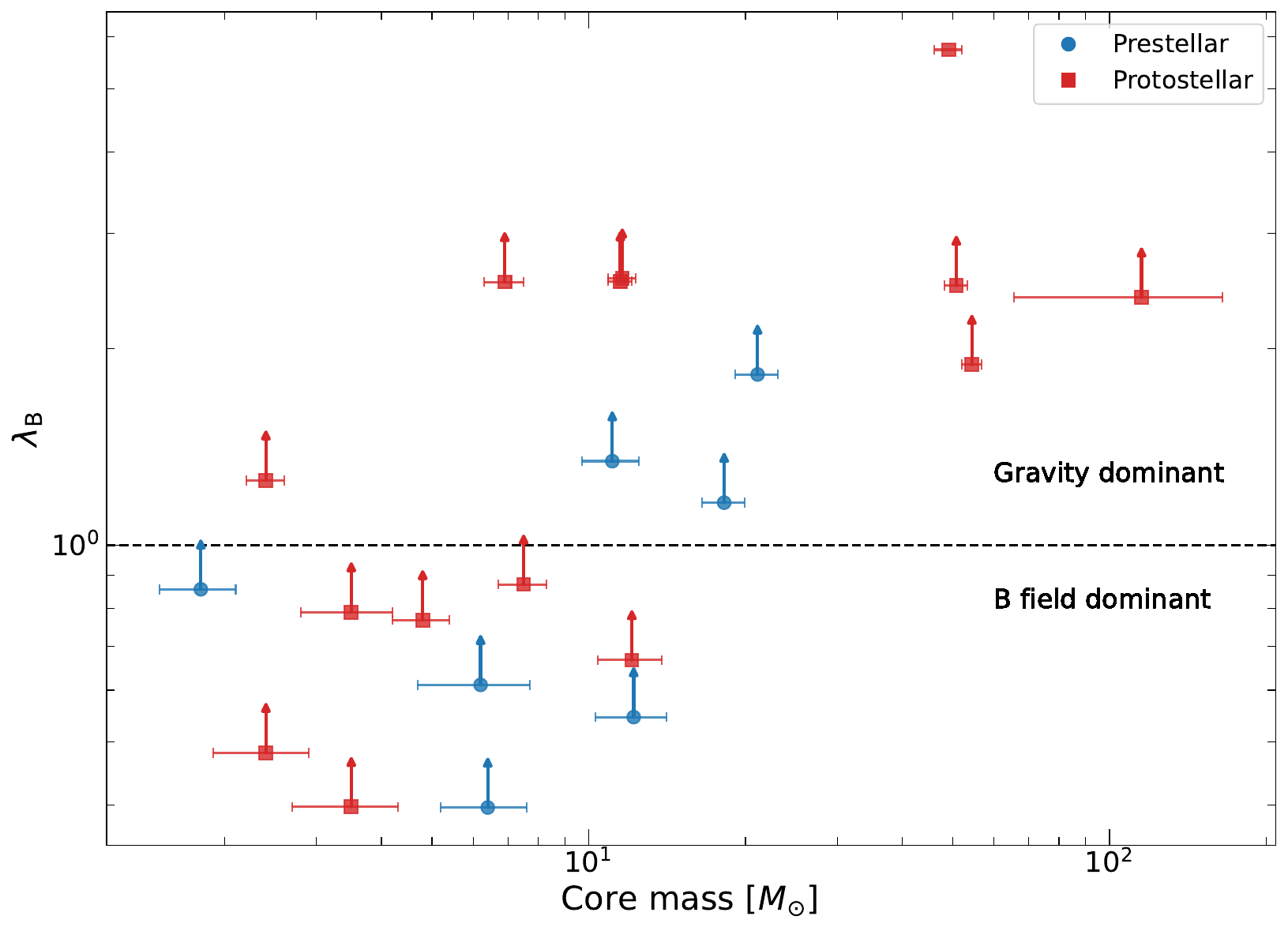}
    \caption{Parameter $\alpha_{\mathrm{kin}}$ (\textbf{left}) of the virial theorem (only taking kinetic motions into account) and mass-to-flux ratio $\lambda_{\mathrm{B}}$ (\textbf{right}) as a function of core masses. Prestellar cores are presented as blue dots and protostellar cores with red squares. The black dashed lines indicate $\alpha_{\mathrm{kin}} = 1$ and $\lambda_{\mathrm{B}} = 1$, corresponding to the virial and magnetic critical thresholds, respectively. The values are presented as upper limits for $\alpha_{\mathrm{kin}}$ and lower limits for $\lambda_{\mathrm{B}}$ as we likely overestimate the turbulence (see Sects.\,\ref{subsec:thermal_turbulent_support} and \ref{subsec:caveats}). A large fraction of the cores appears to be individually supported by either kinetic motions or magnetic fields.} 
    \label{fig:alpha_kin+ratio} 
\end{figure*}

Figure\,\ref{fig:alpha_kin+ratio} (left panel) presents the $\alpha_{\mathrm{kin}}$ values for each core, where mass errorbars are reported from \cite{Motte2025} and points are presented as upper limits (see following). The values are also listed in Table\,\ref{tab:Core_properties}. We find that 16 out of the 45 cores have $\alpha_{\mathrm{kin}} < 1$, while the remaining 29 cores exhibit values ranging from 1 to 12.  From visual inspection, no clear correlation is found between the level of kinetic support and the location of the cores within the region (whether the core is located in the massive central clump or more isolated), and no trend is observed with the mass of the cores. We note also that similar percentages of prestellar and protostellar cores are found across the two regimes: $1/3$ below and $2/3$ above $\alpha_{\mathrm{kin}}=1$. This result is somewhat unexpected, as protostellar cores are expected to display a collapsing envelope and outflows. Moreover, 15 cores show $\alpha_{\mathrm{kin}} > 2$, suggesting that they are gravitationally unbound when only turbulent motions are considered. 
This suggests that the velocity dispersions we measure likely include contributions from larger-scale motions, rather than purely microturbulent support within the cores. This effect is well illustrated in \cite{Hacar2016}, who showed that line velocity dispersions can be artificially broadened by multiple velocity components within the integrated area. Contamination of non-thermal motions by gravitationally driven flows (such as infall) has been observed at the clump scale in \cite{Traficante2018}. In that study, the authors found a positive correlation between the gravitational acceleration, used as a tracer of large-scale gravitational collapse, and the kinetic acceleration, which reflects both thermal and non-thermal motions. This correlation suggests that part of the non-thermal component may actually originate from gravitational collapse rather than from true turbulent support. At the core-scale, the recent study by \cite{Morii2025} similarly revealed a correlation between the infall velocity and the non-thermal velocity dispersion, suggesting that part of the observed velocity dispersions may in fact be driven by gravitational infall. For this reason, the points presented in Fig.\,\ref{fig:alpha_kin+ratio} (left panel) are upper limits, as our turbulence estimates are likely overestimated.

\subsection{Magnetic support and its relative contribution} \label{subsec:mag_support_and_contribution}
To quantify the magnetic field support alone, we can use the mass-to-magnetic flux ratio $\lambda_\mathrm{B}$ following \cite{McKee&Ostriker2007} and \cite{Pillai2015}:
\begin{equation}
    \label{eq:mass-to-flux-ratio}
    \lambda_{\mathrm{B}} = \frac{\left(M / \Phi_B\right)}{\left(M / \Phi_B\right)_{\mathrm{cr}}}=0.76\left(\frac{\left\langle N_{\mathrm{H}_2}\right\rangle}{10^{23} \mathrm{~cm}^{-2}}\right)\left(\frac{B_{\mathrm{POS}}}{1000 \,\mu \mathrm{G}}\right)^{-1},
\end{equation}
where $\left(M / \Phi_B\right)_{\mathrm{cr}} = 1/(2\pi\,G^{1/2})$ with $\left\langle N_{\mathrm{H}_2}\right\rangle$ the mean column density of each core, which can be approximated as $M_{\mathrm{core}}/(\pi\,R^2)$ assuming cylindrical geometry (for column density), with $R$ the radius of the core (assumed to be the deconvolved $FWHM$). It is important to note that this mass-to-flux ratio, which only accounts for the plane-of-sky magnetic field, represents a lower limit to the true mass-to-flux ratio that includes the full three-dimensional magnetic field. Values of $\lambda_\mathrm{B} < 1$ indicate that magnetic support dominates over gravity, while $\lambda_\mathrm{B} > 1$ corresponds to magnetically supercritical cores where gravity prevails. To quantify the relative contribution of magnetic field to kinetic support, we can also compute the magnetic mass following:

\begin{equation}
    M_\mathrm{\phi} = 3\left(\frac{5-2 p}{3-p}\right) \frac{F W H M^{\mathrm{dec}}}{G}\frac{1}{6}\sigma_\mathrm{alf}^2,
\end{equation}
with $\sigma_{\mathrm{a l f}}$ (in cgs) being the alfvenic velocity defined as $\sigma_{\mathrm{a l f}} = B/\sqrt{4\pi\rho}$ with $B$ being the magnetic field strength which we only measure here in the POS, and $\rho$ the mass volume density of the core.

Figure\,\ref{fig:alpha_kin+ratio} (right panel) displays the mass-to-flux ratios for the 21 cores with magnetic field strength estimates. We find that 10 out of the 21 cores (four prestellar and six protostellar) have mass-to-flux ratios below unity, while the remaining 11 cores (two prestellar and eight protostellar) are magnetically supercritical. These $\lambda_{\mathrm{B}}$ values are listed in Table\,\ref{tab:alpha_vir}. For the most massive cores, the magnetic field cannot counteract gravity whereas in lower-mass cores (and up to about 20\,$M_\odot$), a fraction appears magnetically subcritical, but this difference is independent of the evolution and stage of the core (i.e., prestellar or protostellar). However, as discussed in Sect.\,\ref{subsec:turb_refinement}, large-scale velocity gradients are present at the three-beam scale. This suggests that, in addition to overestimating the core-scale turbulence, we may also overestimate the turbulence at the three-beam scale, which would in turn lead to inflated estimates of $B_{\mathrm{POS}}$. We also note that, as pointed out by \citet{Tritsis2026}, mass-to-flux ratio estimates derived from $B_{\mathrm{POS}}$ are subject to significant limitations. Using simulations, 
they demonstrate that only ${\sim}20\,\%$ of the estimated $\lambda_B$ values fall within the true range of mass-to-flux ratios, and that this method is not a reliable diagnostic, notably because clouds appear more magnetically supported at later evolutionary stages, 
contrary to physical expectations. We therefore caution that our conclusions drawn from these mass-to-flux ratio estimates should be interpreted carefully, and that they are mostly useful for comparisons between cores.

\begin{table} [htbp!]
    \centering     
    \begin{threeparttable}[c]
    \caption{Virial parameters measured for the W43-MM1 cores.}
    
    \begin{tabular}{llcccc}
    \hline \noalign {\smallskip}
     \col{Core ID} &  \col{Class}  &  \col{$\alpha_{\mathrm{kin}}$}  & \col{$\lambda_{\mathrm{B}}$} & \col{$\alpha_{\mathrm{vir,B}}$} & \col{$M_{\mathrm{kin}}/M_{\mathrm{\phi}}$} \\ \noalign {\smallskip}
    \hline \noalign {\smallskip}

1 & Proto & 1.49 & 2.40 & 1.71 & 6.53 \\
2 & Proto & 0.38 & 2.50 & 0.59 & 1.80 \\
3 & Proto & 0.45 & 5.74 & 0.49 & 11.34 \\
4 & Proto & 2.41 & 1.89 & 2.77 & 6.59 \\
5 & Proto & 4.57 & 2.56 & 4.77 & 22.99 \\
6 & Pre & 0.32 & - & 0.32 & - \\
7 & Proto & 0.20 & - & 0.20 & - \\
$8^{\dagger}$ & Proto & 0.42 & 2.54 & 0.62 & 2.06 \\
9 & Proto & 1.84 & 2.53 & 2.04 & 9.01 \\
$10^{\dagger}$ & Proto & 6.43 & 0.87 & 8.15 & 3.72 \\
11 & Proto & 12.36 & 1.26 & 13.19 & 14.96 \\
12 & Proto & 0.32 & 0.77 & 2.53 & 0.14 \\
$13^{\dagger}$ & Proto & 1.87 & 0.79 & 3.96 & 0.89 \\
$14^{*}$ & Pre & 0.15 & 1.34 & 1.35 & 0.12 \\
15 & Proto & 1.27 & - & 1.27 & - \\
16 & Proto & 0.81 & 0.67 & 3.75 & 0.28 \\
17 & Pre & 0.58 & 1.16 & 2.19 & 0.36 \\
18 & Proto & 0.50 & - & 0.50 & - \\
$19^{\dagger}$ & Proto & 1.83 & 0.40 & 10.09 & 0.22 \\
20 & Pre & 0.16 & - & 0.16 & - \\
$22^{\dagger}$ & Proto & 1.24 & 0.48 & 6.87 & 0.22 \\
23 & Proto & 0.18 & - & 0.18 & - \\
24 & Pre & 0.64 & 0.40 & 14.46 & 0.05 \\
$28^{\dagger}$ & Pre & 1.80 & - & 1.80 & - \\
$29^{\dagger}$ & Proto & 4.05 & - & 4.05 & - \\
$32^{\dagger}$ & Pre & 0.69 & - & 0.69 & - \\
$34^{\dagger}$ & Pre & 3.81 & - & 3.81 & - \\
$36^{\dagger}$ & Proto & 1.32 & - & 1.32 & - \\
$40^{\dagger}$ & Pre & 1.12 & - & 1.12 & - \\
44 & Proto & 0.85 & - & 0.85 & - \\
46 & Pre & 2.96 & - & 2.96 & - \\
49 & Proto & 1.59 & - & 1.59 & - \\
$51^{\dagger}$ & Proto & 1.36 & - & 1.36 & - \\
54 & Pre & 0.62 & - & 0.62 & - \\
59 & Proto & 1.21 & - & 1.21 & - \\
$73^{\dagger}$ & Pre & 2.11 & - & 2.11 & - \\
$74^{\dagger}$ & Pre & 1.56 & - & 1.56 & - \\
$134^{\dagger}$ & Pre & 2.47 & 1.83 & 3.12 & 3.78 \\
$136^{\dagger}$ & Pre & 4.52 & 0.55 & 11.84 & 0.62 \\
$141^{\dagger}$ & Pre & 1.27 & - & 1.27 & - \\
146 & Pre & 6.11 & 0.86 & 9.07 & 2.06 \\
148 & Pre & 2.56 & - & 2.56 & - \\
151 & Pre & 3.99 & - & 3.99 & - \\
$154^{\dagger}$ & Pre & 2.26 & - & 2.26 & - \\
177 & Pre & 4.50 & 0.61 & 10.31 & 0.77 \\
    \hline
    \end{tabular}
    \label{tab:alpha_vir}
\begin{tablenotes}
\item[*] This core was classified as protostellar in \cite{Nony2020} and as prestellar in \cite{Valeille-Manet2025} (see text).
\item[$\dagger$] These cores have their core velocity dispersion measured with their On spectra. 

\end{tablenotes}    
\end{threeparttable}
\end{table}

\begin{figure}[h!]
    \centering
    \includegraphics[width=0.49\textwidth]{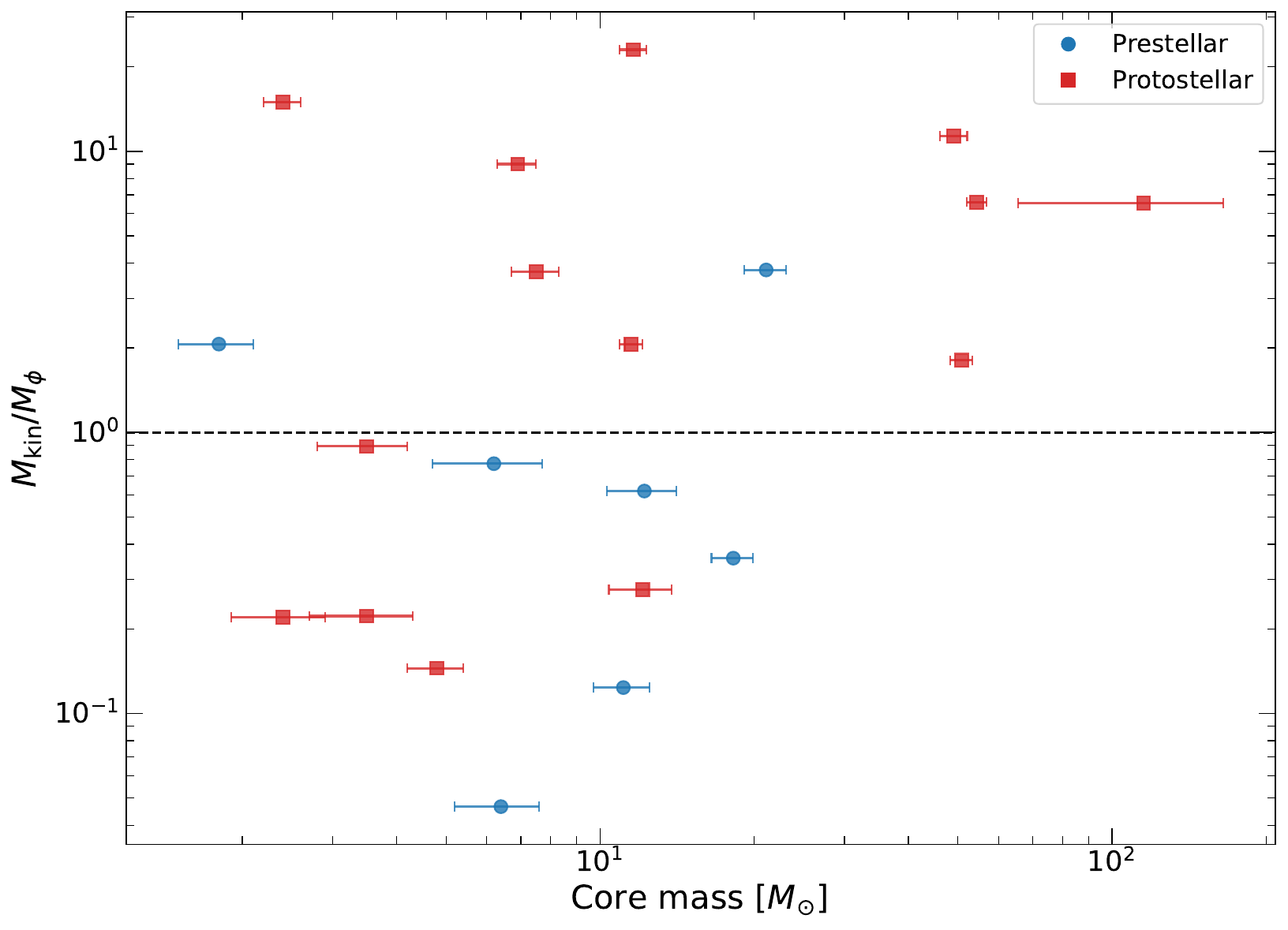}
    \caption{Ratio between the kinetic mass $M_{\mathrm{kin}}$ and the magnetic mass $M_{\mathrm{B}}$ as a function of core masses. Prestellar cores are presented as blue dots and protostellar cores with red squares. The black dashed line represents the equipartition threshold between kinetic and magnetic support. Kinetic motions seem to dominate in the most massive cores while no trend is observed below 20\,$M_\odot$.}
    \label{fig:Mkin/Mphi}
\end{figure}

Using the kinetic and magnetic masses derived from the virial theorem, we can compare the relative importance of turbulence and magnetic fields in providing support within the cores of our sample. Figure \ref{fig:Mkin/Mphi} shows the ratio of kinetic to magnetic mass for the same sample of 21 cores where both quantities could be estimated (values are also displayed in the last column of Table\,\ref{tab:alpha_vir}). Ratio values above unity indicate that kinetic support dominates over magnetic support, while the opposite is true for values below unity. Here again, we do not observe any clear trend with the evolutionary stage of the cores. However, for the most massive objects (five cores above $20\,M_\odot$, one prestellar and four protostellar), turbulence appears to provide a more significant support than magnetic fields. This trend is opposite to that reported by \citet{Cortes2019}. As mentioned above, this discrepancy likely arises from methodology as \citet{Cortes2019} adopted a single, uniform velocity dispersion for their entire core sample, whereas in our analysis, we adopt individual velocity dispersions for each core. This results in an opposite trend for the most massive cores, which in our case exhibit larger velocity dispersions. Turbulence dominating over magnetic fields has also been recently reported in the massive core MM2 of the G11.92 region \citep{Sanhueza2025}. For cores \#3, \#4, \#5, and \#9, which were discussed in Sect.\,\ref{subsec:mag_field_estimate} as exhibiting large dispersions in polarization position angles, we find that they also display high ratios of kinetic to magnetic mass. Therefore, their dominant support is clearly kinetic. Consequently, keeping these cores in the virial analysis, despite the less reliable magnetic field strength estimates, does not bias our results. 

\subsection{Comparison with low-mass studies and models}
On the turbulence side, our results strongly differ from those obtained in low-mass star-forming regions. We find velocity dispersions indicative of strong turbulence within the cores of a high-mass protocluster, which includes both low- and high-mass objects (from 0.3 to 115\,$M_\odot$). This behavior contrasts with the turbulence dissipation typically observed in low-mass cores of low-mass star-forming regions (see the review by \citealp{Pineda2023} and references therein). For instance, \citet{Pineda2010} showed in the Perseus molecular cloud a sharp transition from a turbulent medium to coherent, subsonic cores, at a scale of 0.04\,pc (resolved in our observations) consistent with turbulence dissipation. Our results therefore emphasize that massive star-forming environments operate in a qualitatively different dynamical regime mimicking a different turbulent regime from their low-mass counterparts.

On the magnetic field side, the values found in low-mass star-forming cores are several orders of magnitude lower than those derived in our sample. Using the DCF method in such regions, $B_{\mathrm{POS}}$ typically ranges from a few tens to a few hundreds of $\mu$G (see the review by \citealp{Pattle2023} and references therein, Sect. 5.1.2). These low-mass cores are generally close to magnetic criticality. Extrapolating our mean three–beam–scale $B_{\mathrm{POS}}$ value of 4.5\,mG to the typical density of low-mass cores in these studies ($10^5\,\mathrm{cm^{-3}}$) yields a magnetic field strength of about 1.1\,mG. This is between one and nearly two orders of magnitude higher than the values typically reported for such regions. Like for motions, this emphasizes the different magnetic regimes operating in low- and high-mass star-forming environments.

We can also compare our findings with recent models of high-mass star-forming cores. \citet{Mignon-Risse2021a} showed in their non-ideal MHD simulations that when turbulence is super-Alfvénic (that is, when turbulent motions dominate over magnetic support) the resulting stellar multiplicity increases. This trend is consistent with previous MHD studies by \citet{Commercon2011}, and later confirmed in the non-ideal MHD simulations of \citet{Matsushita2017} and \citet{Rosen2020}, where magnetic fields were found to suppress the fragmentation of massive cores and disks. In comparison, our analysis (Sect.\,\ref{subsec:mag_support_and_contribution}) indicates that magnetic fields dominate only in some cores below 20\,$M_\odot$, implying that the most massive cores in our sample should be more prone to fragmentation. 

\subsection{Virial equilibrium in the cores of W43-MM1}\label{subsec:virial_equilibrium}

\begin{figure}[h!]
    \centering
    \includegraphics[width=0.49\textwidth]{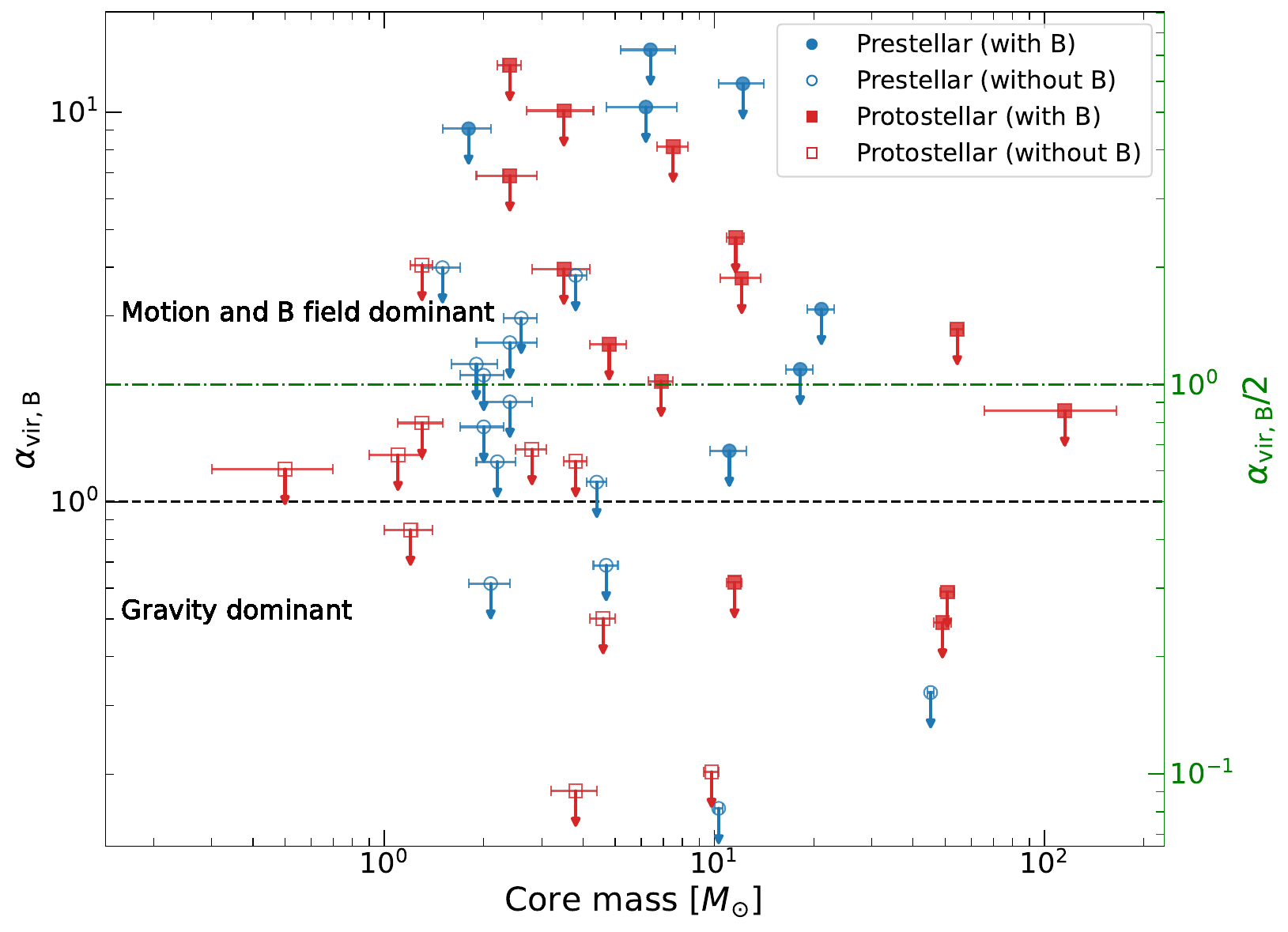}
    \caption{Total virial parameter $\alpha_{\mathrm{vir,B}}$ as a function of core masses. Prestellar cores are presented as blue dots and protostellar cores with red squares. Filled symbols correspond to cores for which a magnetic field strength could be estimated, while open symbols indicate cores where only kinetic (thermal $+$ non-thermal) motions are considered. The black dashed line displays the critical threshold $\alpha_{\mathrm{vir,B}} = 1$. The values of $\alpha_{\mathrm{vir,B}}$ are presented as upper limits as we probably overestimate the turbulence (see Sect.\,\ref{subsec:thermal_turbulent_support} and \ref{subsec:caveats}). The second y-axis in green shows the same parameter $\alpha_{\mathrm{vir,B}}$ divided by a factor of two. The green dotted-dashed line represents the same threshold in this case.} 
    \label{fig:alpha_tot}

\end{figure}

When taking into account both the magnetic field and the kinetic supports we can rewrite Eq.\,\ref{eq:alpha_kin} as (e.g., \citealp{Sanhueza2021}):
\begin{equation}
    \label{eq:kin_Vdisp}
    \alpha_{\mathrm{vir,B}} = 3\left(\frac{5-2 p}{3-p}\right) \frac{F W H M^{\mathrm{dec}}}{G\, M_{\mathrm{core}}}\left(\sigma_\mathrm{v}^2+\frac{1}{6} \sigma_{\mathrm{a l f}}^2\right).
\end{equation}
We apply this analysis to the subset of 21 cores for which both turbulence and $B_{\mathrm{POS}}$ estimates are available. For the remaining 24 cores, for which only turbulence measurements exist, we instead use Eq.\,\ref{eq:alpha_kin}. Figure\,\ref{fig:alpha_tot} displays the final virial parameter value of each core as a function of core masses.

We first examine in more detail the 24 cores for which only turbulent support could be estimated. These objects, shown with empty markers in Fig.\,\ref{fig:alpha_tot}, span a range of $\alpha_\mathrm{kin}$ values from 0.16 to 4 and have masses between 0.5 and 45\,$M_\odot$. They can be separated into two groups: cores for which turbulence alone provides sufficient support against gravity, and cores for which the absence of magnetic information prevents us from determining whether they are gravitationally unstable or could still be supported. Sixteen cores fall into the first category (six protostellar and ten prestellar), with seven cores having $\alpha_{\mathrm{kin}} > 2$, while no firm conclusion can be reached for the remaining eight cores (four prestellar and four protostellar). Among these eight ambiguous cases is core \#6, one of the high-mass prestellar cores of W43-MM1 discussed in Sect.\,\ref{subsec:Core_catalog} and in \cite{Nony2018} and \cite{Valeille-Manet2025}. With a value of $\alpha_\mathrm{kin}=0.55$, this core cannot be supported by turbulence alone, suggesting that additional support would be required to counteract gravitational collapse.

We then focus on the 21 cores with both turbulence and magnetic field estimates, shown with filled markers in Fig.\,\ref{fig:alpha_tot}. They span a range of $\alpha_\mathrm{vir,B}$ values from 0.5 to 14 and have masses between 1.8 and 115\,$M_\odot$. No clear trend is observed for the level of support with the core mass or the evolutionary stage (prestellar versus protostellar). Among these 21 cores, only three cores could be prone to collapse with an $\alpha_\mathrm{vir,B} <1$, protostellar cores \#2, \#3 and \#8. Their overall level of volume support seems in adequation with their protostellar status. However, the 18 remaining cores (seven prestellar and 11 protostellar) have an $\alpha_\mathrm{vir,B}>1$ which suggests an apparent stabilization when considering only the volume terms. Among the 18 supervirial cores, 16 have $\alpha_{\mathrm{vir,B}} > 2$ which suggests that they may be gravitationally unbound, unless additional constraints are considered (see Sect.\,\ref{subsec:caveats}). As shown in Appendix\,\ref{app:alpha_tot_appendix}, when adopting a single velocity dispersion for all cores to estimate the kinetic support at the core-scale (and likewise a single velocity dispersion to derive $B_{\mathrm{POS}}$ at the three-beam scale for all cores), we recover an anti-correlation between the total level of support and core mass, with the most massive cores being gravitationally dominated, consistent with \cite{Cortes2016}, \cite{Cortes2019}, and \cite{Li2023}. This underscores the importance of using core-specific velocity dispersions to obtain a more reliable virial assessment. \\
Among these 18 cores, there are three high-mass prestellar cores of the region, cores \#17, \#134 and \#136 (see Sect.\,\ref{subsec:Core_catalog}, \citealp{Valeille-Manet2025}). The total virial parameter $\alpha_{\mathrm{vir,B}}$ reaches values of about two and three for cores \#17 and \#134, respectively, while it exceeds ten for core \#136. A difference in the dominant support mechanism is also observed among these three cores. Cores \#17 and \#136 appear to be magnetically dominated, with $M_{\mathrm{kin}}/M_{\phi}$ ratios below unity (consistent with strong magnetic field strengths exceeding 30\,mG, see Table\,\ref{tab:alpha_vir}). In contrast, core \#134 is kinetically dominated, with $M_{\mathrm{kin}}/M_{\phi} = 3.8$ and a lower magnetic field strength of 12.1\,mG.
Overall, the limited statistics prevent us from firmly assessing the relative importance of turbulent and magnetic support in high-mass prestellar cores. Although the virial framework used here is subject to large uncertainties, our results do not contradict the interpretation of these objects as massive, relatively stable gas reservoirs, in line with expectations for their prestellar nature \citep{Valeille-Manet2025}.

\vspace{-0.3cm}
\subsection{Limitations of current observational estimates of the non-thermal supports}\label{subsec:caveats}

Like any other observational studies in high-mass protoclusters, it is likely that we overestimate the non-thermal supports within the cores, resulting in $\alpha_{\mathrm{vir,B}}$ values that are systematically too high. In particular, protostellar cores are associated with outflows and are thus expected to have a collapsing envelope. Consequently, they should naturally appear sub-virial in the virial analysis. 

We estimated by how much the measured non-thermal support may be contaminated by gravitational infall motions for protostellar cores with $\alpha_{\mathrm{vir,B}} > 1$ for which we have both turbulent and magnetic support estimates (11 cores in our sample). We computed the velocity that would need to be subtracted in order to bring them back to a state of gravitational equilibrium (i.e., $\alpha_{\mathrm{vir,B}} = 1$). We then compared these velocities with the infall motions or converging supersonic flows typically observed in high-mass star-forming regions. The resulting velocities range from 1.0 to 3.1\,$\mathrm{km\,s^{-1}}$, depending on the core.
The upper end of this range is consistent with the converging flows reported by \cite{Csengeri2011} in Cygnus X in clumps of 0.1\,pc, with velocities of about 2–3\,$\mathrm{km\,s^{-1}}$. This is also in good agreement with the velocity gradients of $\sim2$–3\,$\mathrm{km,s^{-1}}$ observed by \cite{Beuther2025_noreview} along the filaments of the G351 protocluster at $\sim0.2$\,pc scales. The lower values are comparable to the infall velocities of $\sim$0.7\,$\mathrm{km\,s^{-1}}$ observed by \cite{Peretto2013} at the parsec size in the IRDC SDC335. Similarly, \cite{Morii2025} recently measured infall velocities between 0.38 and 1.45\,$\mathrm{km\,s^{-1}}$ toward cores (thousands of au scale) in the high-mass star-forming region G333.541-00.082, which are of the same order of magnitude as our estimates. A purely collapsing core may indeed be misinterpreted as being supported, since infall motions can broaden the observed linewidths and thus be incorrectly attributed to turbulence. \cite{Ballesteros-Paredes2018} showed that the gas accelerates under gravity until the kinetic and gravitational energies become comparable, reaching the point where the core appears supported even though it is not in equilibrium.

We may therefore overestimate both the turbulent level and the magnetic field strengths in our analysis. We note that the relatively large velocity dispersions measured in our cores contrast with the conclusions of the review by \cite{Beuther2025} (see their Sect.\,2.4), in which turbulence in high-mass cores is generally found to be predominantly subsonic. 
However, most of these previous studies targeted IRDCs, which are likely less dynamically active, thus either less evolved or forming lower mass stars than the "mini-starbust" W43-MM1 investigated here. A future study will be required to investigate in detail the kinematics of W43-MM1, in particular to search for velocity structures such as the “V-shapes” reported in other ALMA-IMF protoclusters \citep{Alvarez-Gutierrez2024, Sandoval-Garrido2025}. Such an analysis is beyond the scope of the present paper.

In addition to the possible contamination of the measured linewidths by organized motions (infall, rotation, streamers/converging flows), our virial analysis includes only the volume terms of the virial theorem. The surface terms (such as the confining pressure exerted by the surrounding medium) are extremely difficult to be constrained observationally and are therefore not included here, even though they would further reduce the effective values of $\alpha_{\mathrm{vir,B}}$. In principle, both the thermal, kinetic and magnetic contributions to the virial theorem contain volume and surface components, but only the former are accessible with observations. Using 3D MHD simulations, \cite{Shadmehri2002}, \cite{Dib2007}, and Chevalier et al. (in prep.) showed that surface terms are between roughly half and of the same order of magnitude as the volume terms. Thus, by applying a virial analysis that neglects surface contributions, we inevitably miss an important part of the underlying physics at play within the studied cores. \\
Given the absence of surface terms and the potential contamination of our linewidths by bulk motions, our virial parameters are likely overestimated by at least a factor of two. On the second y-axis of Fig.\,\ref{fig:alpha_tot}, we show the resulting $\alpha_\mathrm{vir,B}$ values after reducing the non-thermal support (kinetic only, or kinetic plus magnetic) by a factor of two. Under this correction, the number of protostellar cores supported solely by turbulence with $\alpha_\mathrm{kin}<1$ increases from four to nine. Conversely, the number of protostellar cores with both support estimates and in a state of gravitational collapse increases from three to four. Nevertheless, even with a factor-of-two reduction, 11 protostellar cores still appear to have sufficient non-thermal support to remain stable against gravitational collapse. This highlights that we must be overestimating the supports by even a larger factor.  In addition, uncertainties beyond the interpretation of linewidths and the neglect of surface terms, such as assumptions on temperature, geometry, and density structure, can significantly affect virial analyses. Because virial estimates rely on multiple simplifying assumptions, they can result in a large dispersion of virial parameter values, as already suggested by, e.g., \cite{Kauffman2013}.

All these considerations demonstrate the intrinsic difficulty of accurately quantifying non-thermal supports in cores, even when relying on a state-of-the-art dataset such as the one used in this study. This work also suggests that non-thermal support is likely to be overestimated in other studies of cores that employ line fitting and a virial analysis limited to volume terms.

\section{Conclusion}\label{sec:conclusion}
In this work, we have presented an analysis of molecular line and dust polarization observations of the high-mass protocluster W43-MM1 to estimate the kinetic and magnetic support in its cores, ranging from 0.3 to 115\,$M_\odot$. Our main results are the following: 
\begin{itemize}
    \item Using DCN\,(3--2), $\mathrm{^{13}CS}$\,(5--4) and $\mathrm{CH_3CN\,(5_3}$--$4_3)$ lines, we obtained velocity dispersions at the core-scale for 45 cores ranging from 0.34 to 4.48\,$\mathrm{km\,s^{-1}}$ (Sect.\,\ref{subsec:turbulence_estimates}). Using the same tracers, we measured velocity dispersions at the three-beam scale to compute $B_{\mathrm{POS}}$ using the DCF method. We could estimate $B_{\mathrm{POS}}$ for 21 out of the 45 cores. At the three-beam scale the obtained $B_{\mathrm{POS}}$ range from 1.1 to 22.5\,mG. Extrapolating these values to the core-scale using the magnetic field-volume density relation based on Zeeman measurements, we obtained $B_{\mathrm{POS}}$ from 1.1 to 49.3\,mG (Sect.\,\ref{subsec:mag_field_estimate}). 

    \item Applying the virial theorem we estimated the kinetic support for 24 cores and the overall kinetic and magnetic support for 21 cores (see Figs.\,\ref{fig:alpha_kin+ratio} and \ref{fig:alpha_tot}). Within the framework of this simplified virial analysis, sixteen cores supported only by kinetic motions and eighteen cores supported by both kinetic and magnetic components appear to be sufficiently supported against gravitational collapse. This is unexpected for the protostellar cores within these groups, which would instead be expected to show reduced support consistent with a partly collapsing envelope. 
    
    \item For the 11 supported protostellar cores for which both turbulent and magnetic estimates are available, we computed that contaminant velocities from large-scale convergent flows and/or infall in the range of 1.0 to 3.1\,$\mathrm{km\,s^{-1}}$ could broaden our velocity dispersions and thus artificially increase the measured level of support inside the cores of W43-MM1 (see Sect.\,\ref{subsec:caveats}). Such infall velocities are plausible given the results from previous observational studies. We also put forward that omitting the surface terms on the virial theorem immediately bias the virial numbers towards higher values. 

    \item Knowing these uncertainties, we find that kinetic motions provide the dominant contribution to the support budget in the most massive cores, with no clear trend emerging toward lower-mass objects (see Fig.\,\ref{fig:Mkin/Mphi}). The high-mass prestellar cores in W43-MM1 appear to be sufficiently supported when both turbulent and magnetic contributions are considered; however, dominant support between turbulence and magnetic fields in the high-mass prestellar phase could not be firmly established with the present data.
\end{itemize}

Overall, our results underscore the intrinsic difficulty of reliably measuring individual support terms in cores embedded within a highly dynamical and massive environment such as the W43-MM1 protocluster. The possible contamination of our linewidths by motions associated with convergent flows, infall, and rotation together with the absence of surface terms in our analysis of non-thermal support, may contribute to this observational discrepancy in the virial theorem. These results also suggest that a similar overestimation of non-thermal support is likely present in other core studies that rely on the same methods and analysis. Nevertheless, this work represents a big step forward toward a complete virial analysis of statistical core samples at a few thousand au scales, in which both turbulence and magnetic field strengths are estimated on a core-by-core basis, and where possible observational biases are taken into account.

\begin{acknowledgements}
We thank the anonymous referee for his comments that helped to improve the quality of this work. This paper makes use of the following ALMA data: \#2013.1.01365.S, \#2015.1.01273.S, \#2017.1.01355.L, and \#2015.1.01020.S. ALMA is a partnership of ESO (representing its member states), NSF (USA) and NINS (Japan), together with NRC (Canada), MOST and ASIAA (Taiwan), and KASI (Republic of Korea), in cooperation with the Republic of Chile. The Joint ALMA Observatory is operated by ESO, AUI/NRAO and NAOJ.
This project has received funding from the European Research Council (ERC) via the ERC Synergy Grant \textsl{ECOGAL} (grant 855130) and from the French Agence Nationale de la Recherche (ANR) through the project \textsl{COSMHIC} (ANR-20-CE31-0009). AS gratefully acknowledges support by the Fondecyt Regular (project code 1220610), and ANID BASAL project FB210003. N.S.G. gratefully acknowledges support from ANID Beca Doctorado Nacional 21250244. PS was partially supported by a Grant-in-Aid for Scientific Research (KAKENHI Number JP23H01221) of JSPS. R.A.G. acknowledges support from the STFC (grant ST/Y002229/1). A.K. would also like to acknowledge Fondecyt postdoctoral fellowship (project id: 3250070, 2025). P.S. was partially supported by a Grant-in-Aid for Scientific Research (KAKENHI No JP24K17100) of the Japan Society for the Promotion of Science (JSPS).
\end{acknowledgements}

\bibliographystyle{aa}
\bibliography{biblio}

@ARTICLE{Whitworth2025,
       author = {{Whitworth}, D.~J. and {Srinivasan}, S. and {Pudritz}, R.~E. and {Mac Low}, M. -M. and {Eadie}, G. and {Palau}, A. and {Soler}, J.~D. and {Smith}, R.~J. and {Pattle}, K. and {Robinson}, H. and {Pillsworth}, R. and {Wadsley}, J. and {Brucy}, N. and {Lebreuilly}, U. and {Hennebelle}, P. and {Girichidis}, P. and {Gent}, F.~A. and {Marin}, J. and {S{\'a}nchez Valido}, L. and {Camacho}, V. and {Klessen}, R.~S. and {V{\'a}zquez-Semadeni}, E.},
        title = "{On the relation between magnetic field strength and gas density in the interstellar medium: a multiscale analysis}",
      journal = {\mnras},
         year = 2025,
        month = jul,
       volume = {540},
       number = {3},
        pages = {2762-2786},
          doi = {10.1093/mnras/staf901},
archivePrefix = {arXiv},
       eprint = {2407.18293},
 primaryClass = {astro-ph.GA},
       adsurl = {https://ui.adsabs.harvard.edu/abs/2025MNRAS.540.2762W}
}

@ARTICLE{Motte2025,
       author = {{Motte}, F. and {Pouteau}, Y. and {Nony}, T. and {Dell'Ova}, P. and {Gusdorf}, A. and {Brouillet}, N. and {Stutz}, A.~M. and {Bontemps}, S. and {Ginsburg}, A. and {Csengeri}, T. and {Men'shchikov}, A. and {Valeille-Manet}, M. and {Louvet}, F. and {Bonfand}, M. and {Galv{\'a}n-Madrid}, R. and {{\'A}lvarez-Guti{\'e}rrez}, R.~H. and {Armante}, M. and {Bronfman}, L. and {Chen}, H. -R.~V. and {Cunningham}, N. and {D{\'\i}az-Gonz{\'a}lez}, D. and {Didelon}, P. and {Fern{\'a}ndez-L{\'o}pez}, M. and {Herpin}, F. and {Kessler}, N. and {Koley}, A. and {Lefloch}, B. and {Le Nestour}, N. and {Liu}, H. -L. and {Moraux}, E. and {Nguyen Luong}, Q. and {Olguin}, F. and {Salinas}, J. and {Sandoval-Garrido}, N.~A. and {Sanhueza}, P. and {Veyry}, R. and {Yoo}, T.},
        title = "{ALMA-IMF: XVI. Mass-averaged temperature of cores and protostellar luminosities in the ALMA-IMF protoclusters}",
      journal = {\aap},
         year = 2025,
        month = feb,
       volume = {694},
          eid = {A24},
        pages = {A24},
          doi = {10.1051/0004-6361/202451931},
archivePrefix = {arXiv},
       eprint = {2412.02011},
 primaryClass = {astro-ph.GA},
       adsurl = {https://ui.adsabs.harvard.edu/abs/2025A&A...694A..24M}
}

@ARTICLE{LeGouellec2020,
       author = {{Le Gouellec}, V.~J.~M. and {Maury}, A.~J. and {Guillet}, V. and {Hull}, C.~L.~H. and {Girart}, J.~M. and {Verliat}, A. and {Mignon-Risse}, R. and {Valdivia}, V. and {Hennebelle}, P. and {Gonz{\'a}lez}, M. and {Louvet}, F.},
        title = "{A statistical analysis of dust polarization properties in ALMA observations of Class 0 protostellar cores}",
      journal = {\aap},
         year = 2020,
        month = dec,
       volume = {644},
          eid = {A11},
        pages = {A11},
          doi = {10.1051/0004-6361/202038404},
archivePrefix = {arXiv},
       eprint = {2009.07186},
 primaryClass = {astro-ph.GA},
       adsurl = {https://ui.adsabs.harvard.edu/abs/2020A&A...644A..11L}
}

@ARTICLE{Heitsch2001,
       author = {{Heitsch}, Fabian and {Zweibel}, Ellen G. and {Mac Low}, Mordecai-Mark and {Li}, Pakshing and {Norman}, Michael L.},
        title = "{Magnetic Field Diagnostics Based on Far-Infrared Polarimetry: Tests Using Numerical Simulations}",
      journal = {\apj},
         year = 2001,
        month = nov,
       volume = {561},
       number = {2},
        pages = {800-814},
          doi = {10.1086/323489},
archivePrefix = {arXiv},
       eprint = {astro-ph/0103286},
 primaryClass = {astro-ph},
       adsurl = {https://ui.adsabs.harvard.edu/abs/2001ApJ...561..800H}
}

@ARTICLE{Padoan2001,
       author = {{Padoan}, Paolo and {Goodman}, Alyssa and {Draine}, B.~T. and {Juvela}, Mika and {Nordlund}, {\r{A}}ke and {R{\"o}gnvaldsson}, {\"O}rn{\'o}lfur Einar},
        title = "{Theoretical Models of Polarized Dust Emission from Protostellar Cores}",
      journal = {\apj},
         year = 2001,
        month = oct,
       volume = {559},
       number = {2},
        pages = {1005-1018},
          doi = {10.1086/322504},
archivePrefix = {arXiv},
       eprint = {astro-ph/0104231},
 primaryClass = {astro-ph},
       adsurl = {https://ui.adsabs.harvard.edu/abs/2001ApJ...559.1005P}
}

@ARTICLE{Ostriker2001,
       author = {{Ostriker}, Eve C. and {Stone}, James M. and {Gammie}, Charles F.},
        title = "{Density, Velocity, and Magnetic Field Structure in Turbulent Molecular Cloud Models}",
      journal = {\apj},
         year = 2001,
        month = jan,
       volume = {546},
       number = {2},
        pages = {980-1005},
          doi = {10.1086/318290},
archivePrefix = {arXiv},
       eprint = {astro-ph/0008454},
 primaryClass = {astro-ph},
       adsurl = {https://ui.adsabs.harvard.edu/abs/2001ApJ...546..980O}
}

@ARTICLE{Chandrasekhar&Fermi1953,
       author = {{Chandrasekhar}, S. and {Fermi}, E.},
        title = "{Problems of Gravitational Stability in the Presence of a Magnetic Field.}",
      journal = {\apj},
         year = 1953,
        month = jul,
       volume = {118},
        pages = {116},
          doi = {10.1086/145732},
       adsurl = {https://ui.adsabs.harvard.edu/abs/1953ApJ...118..116C}
}

@ARTICLE{Davis1951,
       author = {{Davis}, Leverett},
        title = "{The Strength of Interstellar Magnetic Fields}",
      journal = {Physical Review},
         year = 1951,
        month = mar,
       volume = {81},
       number = {5},
        pages = {890-891},
          doi = {10.1103/PhysRev.81.890.2},
       adsurl = {https://ui.adsabs.harvard.edu/abs/1951PhRv...81..890D}
}

@ARTICLE{Lazarian&Hoang2007,
       author = {{Lazarian}, A. and {Hoang}, Thiem},
        title = "{Radiative torques: analytical model and basic properties}",
      journal = {\mnras},
         year = 2007,
        month = jul,
       volume = {378},
       number = {3},
        pages = {910-946},
          doi = {10.1111/j.1365-2966.2007.11817.x},
archivePrefix = {arXiv},
       eprint = {0707.0886},
 primaryClass = {astro-ph},
       adsurl = {https://ui.adsabs.harvard.edu/abs/2007MNRAS.378..910L}
}

@ARTICLE{Lazarian2007,
       author = {{Lazarian}, A.},
        title = "{Tracing magnetic fields with aligned grains}",
      journal = {\jqsrt},
         year = 2007,
        month = jul,
       volume = {106},
        pages = {225-256},
          doi = {10.1016/j.jqsrt.2007.01.038},
archivePrefix = {arXiv},
       eprint = {0707.0858},
 primaryClass = {astro-ph},
       adsurl = {https://ui.adsabs.harvard.edu/abs/2007JQSRT.106..225L}
}

@ARTICLE{Hull&Plambeck2015,
       author = {{Hull}, Charles L.~H. and {Plambeck}, Richard L.},
        title = "{The 1.3mm Full-Stokes Polarization System at CARMA}",
      journal = {Journal of Astronomical Instrumentation},
         year = 2015,
        month = jun,
       volume = {4},
          eid = {1550005},
        pages = {1550005},
          doi = {10.1142/S2251171715500051},
archivePrefix = {arXiv},
       eprint = {1506.04771},
 primaryClass = {astro-ph.IM},
       adsurl = {https://ui.adsabs.harvard.edu/abs/2015JAI.....450005H}
}

@INPROCEEDINGS{McMullin2007,
       author = {{McMullin}, J.~P. and {Waters}, B. and {Schiebel}, D. and {Young}, W. and {Golap}, K.},
        title = "{CASA Architecture and Applications}",
    booktitle = {Astronomical Data Analysis Software and Systems XVI},
         year = 2007,
       editor = {{Shaw}, R.~A. and {Hill}, F. and {Bell}, D.~J.},
       series = {Astronomical Society of the Pacific Conference Series},
       volume = {376},
        month = oct,
        pages = {127},
       adsurl = {https://ui.adsabs.harvard.edu/abs/2007ASPC..376..127M}
}

@ARTICLE{Valeille-Manet2025,
       author = {{Valeille-Manet}, M. and {Bontemps}, S. and {Csengeri}, T. and {Nony}, T. and {Motte}, F. and {Stutz}, A.~M. and {Gusdorf}, A. and {Ginsburg}, A. and {Galv{\'a}n-Madrid}, R. and {Sanhueza}, P. and {Bonfand}, M. and {Brouillet}, N. and {Dell'Ova}, P. and {Louvet}, F. and {Cunningham}, N. and {Fern{\'a}ndez-L{\'o}pez}, M. and {Herpin}, F. and {Wyrowski}, F. and {{\'A}lvarez-Guti{\'e}rrez}, R.~H. and {Armante}, M. and {Guzm{\'a}n}, A.~E. and {Kessler}, N. and {Koley}, A. and {Salinas}, J. and {Yoo}, T. and {Bronfman}, L. and {Le Nestour}, N.},
        title = "{ALMA-IMF: XVII. Census and lifetime of high-mass prestellar cores in 14 massive protoclusters}",
      journal = {\aap},
         year = 2025,
        month = apr,
       volume = {696},
          eid = {A11},
        pages = {A11},
          doi = {10.1051/0004-6361/202451291},
archivePrefix = {arXiv},
       eprint = {2502.09426},
 primaryClass = {astro-ph.GA},
       adsurl = {https://ui.adsabs.harvard.edu/abs/2025A&A...696A..11V}
}

@ARTICLE{Mignon-Risse2021a,
       author = {{Mignon-Risse}, R. and {Gonz{\'a}lez}, M. and {Commer{\c{c}}on}, B. and {Rosdahl}, J.},
        title = "{Collapse of turbulent massive cores with ambipolar diffusion and hybrid radiative transfer. I. Accretion and multiplicity}",
      journal = {\aap},
         year = 2021,
        month = aug,
       volume = {652},
          eid = {A69},
        pages = {A69},
          doi = {10.1051/0004-6361/202140617},
archivePrefix = {arXiv},
       eprint = {2105.14543},
 primaryClass = {astro-ph.SR},
       adsurl = {https://ui.adsabs.harvard.edu/abs/2021A&A...652A..69M}
}

@ARTICLE{Avison2021,
       author = {{Avison}, A. and {Fuller}, G.~A. and {Peretto}, N. and {Duarte-Cabral}, A. and {Rosen}, A.~L. and {Traficante}, A. and {Pineda}, J.~E. and {G{\"u}sten}, R. and {Cunningham}, N.},
        title = "{Continuity of accretion from clumps to Class 0 high-mass protostars in SDC335}",
      journal = {\aap},
         year = 2021,
        month = jan,
       volume = {645},
          eid = {A142},
        pages = {A142},
          doi = {10.1051/0004-6361/201936043},
archivePrefix = {arXiv},
       eprint = {2012.08948},
 primaryClass = {astro-ph.GA},
       adsurl = {https://ui.adsabs.harvard.edu/abs/2021A&A...645A.142A}
}

@ARTICLE{Matsushita2017,
       author = {{Matsushita}, Yuko and {Machida}, Masahiro N. and {Sakurai}, Yuya and {Hosokawa}, Takashi},
        title = "{Massive outflows driven by magnetic effects in star-forming clouds with high mass accretion rates}",
      journal = {\mnras},
         year = 2017,
        month = sep,
       volume = {470},
       number = {1},
        pages = {1026-1049},
          doi = {10.1093/mnras/stx893},
archivePrefix = {arXiv},
       eprint = {1704.03185},
 primaryClass = {astro-ph.SR},
       adsurl = {https://ui.adsabs.harvard.edu/abs/2017MNRAS.470.1026M}
}

@ARTICLE{Ferreira2006,
       author = {{Ferreira}, J. and {Dougados}, C. and {Cabrit}, S.},
        title = "{Which jet launching mechanism(s) in T Tauri stars?}",
      journal = {\aap},
         year = 2006,
        month = jul,
       volume = {453},
       number = {3},
        pages = {785-796},
          doi = {10.1051/0004-6361:20054231},
archivePrefix = {arXiv},
       eprint = {astro-ph/0604053},
 primaryClass = {astro-ph},
       adsurl = {https://ui.adsabs.harvard.edu/abs/2006A&A...453..785F}
}

@ARTICLE{Blandford&Payne1982,
       author = {{Blandford}, R.~D. and {Payne}, D.~G.},
        title = "{Hydromagnetic flows from accretion disks and the production of radio jets.}",
      journal = {\mnras},
         year = 1982,
        month = jun,
       volume = {199},
        pages = {883-903},
          doi = {10.1093/mnras/199.4.883},
       adsurl = {https://ui.adsabs.harvard.edu/abs/1982MNRAS.199..883B}
}

@ARTICLE{Padoan2020,
       author = {{Padoan}, Paolo and {Pan}, Liubin and {Juvela}, Mika and {Haugb{\o}lle}, Troels and {Nordlund}, {\r{A}}ke},
        title = "{The Origin of Massive Stars: The Inertial-inflow Model}",
      journal = {\apj},
         year = 2020,
        month = sep,
       volume = {900},
       number = {1},
          eid = {82},
        pages = {82},
          doi = {10.3847/1538-4357/abaa47},
archivePrefix = {arXiv},
       eprint = {1911.04465},
 primaryClass = {astro-ph.GA},
       adsurl = {https://ui.adsabs.harvard.edu/abs/2020ApJ...900...82P}
}

@ARTICLE{Commercon2022,
       author = {{Commer{\c{c}}on}, B. and {Gonz{\'a}lez}, M. and {Mignon-Risse}, R. and {Hennebelle}, P. and {Vaytet}, N.},
        title = "{Discs and outflows in the early phases of massive star formation: Influence of magnetic fields and ambipolar diffusion}",
      journal = {\aap},
         year = 2022,
        month = feb,
       volume = {658},
          eid = {A52},
        pages = {A52},
          doi = {10.1051/0004-6361/202037479},
archivePrefix = {arXiv},
       eprint = {2109.10580},
 primaryClass = {astro-ph.SR},
       adsurl = {https://ui.adsabs.harvard.edu/abs/2022A&A...658A..52C}
}

@ARTICLE{Peretto2013,
       author = {{Peretto}, N. and {Fuller}, G.~A. and {Duarte-Cabral}, A. and {Avison}, A. and {Hennebelle}, P. and {Pineda}, J.~E. and {Andr{\'e}}, Ph. and {Bontemps}, S. and {Motte}, F. and {Schneider}, N. and {Molinari}, S.},
        title = "{Global collapse of molecular clouds as a formation mechanism for the most massive stars}",
      journal = {\aap},
         year = 2013,
        month = jul,
       volume = {555},
          eid = {A112},
        pages = {A112},
          doi = {10.1051/0004-6361/201321318},
archivePrefix = {arXiv},
       eprint = {1307.2590},
 primaryClass = {astro-ph.GA},
       adsurl = {https://ui.adsabs.harvard.edu/abs/2013A&A...555A.112P}
}

@ARTICLE{Schneider2010,
       author = {{Schneider}, N. and {Csengeri}, T. and {Bontemps}, S. and {Motte}, F. and {Simon}, R. and {Hennebelle}, P. and {Federrath}, C. and {Klessen}, R.},
        title = "{Dynamic star formation in the massive DR21 filament}",
      journal = {\aap},
         year = 2010,
        month = sep,
       volume = {520},
          eid = {A49},
        pages = {A49},
          doi = {10.1051/0004-6361/201014481},
archivePrefix = {arXiv},
       eprint = {1003.4198},
 primaryClass = {astro-ph.GA},
       adsurl = {https://ui.adsabs.harvard.edu/abs/2010A&A...520A..49S}
}

@ARTICLE{Motte2022,
       author = {{Motte}, F. and {Bontemps}, S. and {Csengeri}, T. and {Pouteau}, Y. and {Louvet}, F. and {Stutz}, A.~M. and {Cunningham}, N. and {L{\'o}pez-Sepulcre}, A. and {Brouillet}, N. and {Galv{\'a}n-Madrid}, R. and {Ginsburg}, A. and {Maud}, L. and {Men'shchikov}, A. and {Nakamura}, F. and {Nony}, T. and {Sanhueza}, P. and {{\'A}lvarez-Guti{\'e}rrez}, R.~H. and {Armante}, M. and {Baug}, T. and {Bonfand}, M. and {Busquet}, G. and {Chapillon}, E. and {D{\'\i}az-Gonz{\'a}lez}, D. and {Fern{\'a}ndez-L{\'o}pez}, M. and {Guzm{\'a}n}, A.~E. and {Herpin}, F. and {Liu}, H. -L. and {Olguin}, F. and {Towner}, A.~P.~M. and {Bally}, J. and {Battersby}, C. and {Braine}, J. and {Bronfman}, L. and {Chen}, H. -R.~V. and {Dell'Ova}, P. and {Di Francesco}, J. and {Gonz{\'a}lez}, M. and {Gusdorf}, A. and {Hennebelle}, P. and {Izumi}, N. and {Joncour}, I. and {Lee}, Y. -N. and {Lefloch}, B. and {Lesaffre}, P. and {Lu}, X. and {Menten}, K.~M. and {Mignon-Risse}, R. and {Molet}, J. and {Moraux}, E. and {Mundy}, L. and {Nguyen Luong}, Q. and {Reyes}, N. and {Reyes Reyes}, S.~D. and {Robitaille}, J. -F. and {Rosolowsky}, E. and {Sandoval-Garrido}, N.~A. and {Schuller}, F. and {Svoboda}, B. and {Tatematsu}, K. and {Thomasson}, B. and {Walker}, D. and {Wu}, B. and {Whitworth}, A.~P. and {Wyrowski}, F.},
        title = "{ALMA-IMF. I. Investigating the origin of stellar masses: Introduction to the Large Program and first results}",
      journal = {\aap},
         year = 2022,
        month = jun,
       volume = {662},
          eid = {A8},
        pages = {A8},
          doi = {10.1051/0004-6361/202141677},
archivePrefix = {arXiv},
       eprint = {2112.08182},
 primaryClass = {astro-ph.GA},
       adsurl = {https://ui.adsabs.harvard.edu/abs/2022A&A...662A...8M},
    shorthand = {ALMA-IMF I}
}

@ARTICLE{Cunningham2023,
       author = {{Cunningham}, N. and {Ginsburg}, A. and {Galv{\'a}n-Madrid}, R. and {Motte}, F. and {Csengeri}, T. and {Stutz}, A.~M. and {Fern{\'a}ndez-L{\'o}pez}, M. and {{\'A}lvarez-Guti{\'e}rrez}, R.~H. and {Armante}, M. and {Baug}, T. and {Bonfand}, M. and {Bontemps}, S. and {Braine}, J. and {Brouillet}, N. and {Busquet}, G. and {D{\'\i}az-Gonz{\'a}lez}, D.~J. and {Di Francesco}, J. and {Gusdorf}, A. and {Herpin}, F. and {Liu}, H. and {L{\'o}pez-Sepulcre}, A. and {Louvet}, F. and {Lu}, X. and {Maud}, L. and {Nony}, T. and {Olguin}, F.~A. and {Pouteau}, Y. and {Rivera-Soto}, R. and {Sandoval-Garrido}, N.~A. and {Sanhueza}, P. and {Tatematsu}, K. and {Towner}, A.~P.~M. and {Valeille-Manet}, M.},
        title = "{ALMA-IMF. VII. First release of the full spectral line cubes: Core kinematics traced by DCN J = (3{\ensuremath{-}}2)}",
      journal = {\aap},
         year = 2023,
        month = oct,
       volume = {678},
          eid = {A194},
        pages = {A194},
          doi = {10.1051/0004-6361/202245429},
archivePrefix = {arXiv},
       eprint = {2306.14710},
 primaryClass = {astro-ph.GA},
       adsurl = {https://ui.adsabs.harvard.edu/abs/2023A&A...678A.194C}
}

@ARTICLE{Alvarez-Gutierrez2024,
       author = {{{\'A}lvarez-Guti{\'e}rrez}, R.~H. and {Stutz}, A.~M. and {Sandoval-Garrido}, N. and {Louvet}, F. and {Motte}, F. and {Galv{\'a}n-Madrid}, R. and {Cunningham}, N. and {Sanhueza}, P. and {Bonfand}, M. and {Bontemps}, S. and {Gusdorf}, A. and {Ginsburg}, A. and {Csengeri}, T. and {Reyes}, S.~D. and {Salinas}, J. and {Baug}, T. and {Bronfman}, L. and {Busquet}, G. and {D{\'\i}az-Gonz{\'a}lez}, D.~J. and {Fernandez-Lopez}, M. and {Guzm{\'a}n}, A. and {Koley}, A. and {Liu}, H. -L. and {Olguin}, F.~A. and {Valeille-Manet}, M. and {Wyrowski}, F.},
        title = "{ALMA-IMF: XIII. N$_{2}$H$^{+}$ kinematic analysis of the intermediate protocluster G353.41}",
      journal = {\aap},
         year = 2024,
        month = sep,
       volume = {689},
          eid = {A74},
        pages = {A74},
          doi = {10.1051/0004-6361/202450321},
archivePrefix = {arXiv},
       eprint = {2404.07363},
 primaryClass = {astro-ph.GA},
       adsurl = {https://ui.adsabs.harvard.edu/abs/2024A&A...689A..74A}
}

@ARTICLE{Bonfand2024,
       author = {{Bonfand}, M. and {Csengeri}, T. and {Bontemps}, S. and {Brouillet}, N. and {Motte}, F. and {Louvet}, F. and {Ginsburg}, A. and {Cunningham}, N. and {Galv{\'a}n-Madrid}, R. and {Herpin}, F. and {Wyrowski}, F. and {Valeille-Manet}, M. and {Stutz}, A.~M. and {Di Francesco}, J. and {Gusdorf}, A. and {Fern{\'a}ndez-L{\'o}pez}, M. and {Lefloch}, B. and {Liu}, H. -L. and {Sanhueza}, P. and {{\'A}lvarez-Guti{\'e}rrez}, R.~H. and {Olguin}, F. and {Nony}, T. and {Lopez-Sepulcre}, A. and {Dell'Ova}, P. and {Pouteau}, Y. and {Jeff}, D. and {Chen}, H. -R.~V. and {Armante}, M. and {Towner}, A. and {Bronfman}, L. and {Kessler}, N.},
        title = "{ALMA-IMF. XI. The sample of hot core candidates: A rich population of young high-mass protostars unveiled by the emission of methyl formate}",
      journal = {\aap},
         year = 2024,
        month = jul,
       volume = {687},
          eid = {A163},
        pages = {A163},
          doi = {10.1051/0004-6361/202347856},
archivePrefix = {arXiv},
       eprint = {2402.15023},
 primaryClass = {astro-ph.GA},
       adsurl = {https://ui.adsabs.harvard.edu/abs/2024A&A...687A.163B}
}

@ARTICLE{Brouillet2022,
       author = {{Brouillet}, N. and {Despois}, D. and {Molet}, J. and {Nony}, T. and {Motte}, F. and {Gusdorf}, A. and {Louvet}, F. and {Bontemps}, S. and {Herpin}, F. and {Bonfand}, M. and {Csengeri}, T. and {Ginsburg}, A. and {Cunningham}, N. and {Galv{\'a}n-Madrid}, R. and {Maud}, L. and {Busquet}, G. and {Bronfman}, L. and {Fern{\'a}ndez-L{\'o}pez}, M. and {Jeff}, D.~L. and {Lefloch}, B. and {Pouteau}, Y. and {Sanhueza}, P. and {Stutz}, A.~M. and {Valeille-Manet}, M.},
        title = "{ALMA-IMF. IV. A comparative study of the main hot cores in W43-MM1: Detection, temperature, and molecular composition}",
      journal = {\aap},
         year = 2022,
        month = sep,
       volume = {665},
          eid = {A140},
        pages = {A140},
          doi = {10.1051/0004-6361/202243669},
archivePrefix = {arXiv},
       eprint = {2207.03537},
 primaryClass = {astro-ph.GA},
       adsurl = {https://ui.adsabs.harvard.edu/abs/2022A&A...665A.140B}
}

@ARTICLE{Nony2020,
       author = {{Nony}, T. and {Motte}, F. and {Louvet}, F. and {Plunkett}, A. and {Gusdorf}, A. and {Fechtenbaum}, S. and {Pouteau}, Y. and {Lefloch}, B. and {Bontemps}, S. and {Molet}, J. and {Robitaille}, J. -F.},
        title = "{Episodic accretion constrained by a rich cluster of outflows}",
      journal = {\aap},
         year = 2020,
        month = apr,
       volume = {636},
          eid = {A38},
        pages = {A38},
          doi = {10.1051/0004-6361/201937046},
archivePrefix = {arXiv},
       eprint = {2002.05720},
 primaryClass = {astro-ph.GA},
       adsurl = {https://ui.adsabs.harvard.edu/abs/2020A&A...636A..38N}
}

@ARTICLE{Nony2018,
       author = {{Nony}, T. and {Louvet}, F. and {Motte}, F. and {Molet}, J. and {Marsh}, K. and {Chapillon}, E. and {Gusdorf}, A. and {Brouillet}, N. and {Bontemps}, S. and {Csengeri}, T. and {Despois}, D. and {Nguyen Luong}, Q. and {Duarte-Cabral}, A. and {Maury}, A.},
        title = "{Detection of a high-mass prestellar core candidate in W43-MM1}",
      journal = {\aap},
         year = 2018,
        month = oct,
       volume = {618},
          eid = {L5},
        pages = {L5},
          doi = {10.1051/0004-6361/201833863},
archivePrefix = {arXiv},
       eprint = {1810.01404},
 primaryClass = {astro-ph.GA},
       adsurl = {https://ui.adsabs.harvard.edu/abs/2018A&A...618L...5N}
}

@ARTICLE{Ossenkopf1994,
       author = {{Ossenkopf}, V. and {Henning}, Th.},
        title = "{Dust opacities for protostellar cores.}",
      journal = {\aap},
         year = 1994,
        month = nov,
       volume = {291},
        pages = {943-959},
       adsurl = {https://ui.adsabs.harvard.edu/abs/1994A&A...291..943O}
}

@ARTICLE{Nony2023,
       author = {{Nony}, T. and {Galv{\'a}n-Madrid}, R. and {Motte}, F. and {Pouteau}, Y. and {Cunningham}, N. and {Louvet}, F. and {Stutz}, A.~M. and {Lefloch}, B. and {Bontemps}, S. and {Brouillet}, N. and {Ginsburg}, A. and {Joncour}, I. and {Herpin}, F. and {Sanhueza}, P. and {Csengeri}, T. and {Towner}, A.~P.~M. and {Bonfand}, M. and {Fern{\'a}ndez-L{\'o}pez}, M. and {Baug}, T. and {Bronfman}, L. and {Busquet}, G. and {Di Francesco}, J. and {Gusdorf}, A. and {Lu}, X. and {Olguin}, F. and {Valeille-Manet}, M. and {Whitworth}, A.~P.},
        title = "{ALMA-IMF. V. Prestellar and protostellar core populations in the W43 cloud complex}",
      journal = {\aap},
         year = 2023,
        month = jun,
       volume = {674},
          eid = {A75},
        pages = {A75},
          doi = {10.1051/0004-6361/202244762},
archivePrefix = {arXiv},
       eprint = {2301.07238},
 primaryClass = {astro-ph.GA},
       adsurl = {https://ui.adsabs.harvard.edu/abs/2023A&A...674A..75N}
}

@ARTICLE{Morri2023,
       author = {{Morii}, Kaho and {Sanhueza}, Patricio and {Nakamura}, Fumitaka and {Zhang}, Qizhou and {Sabatini}, Giovanni and {Beuther}, Henrik and {Lu}, Xing and {Li}, Shanghuo and {Garay}, Guido and {Jackson}, James M. and {Olguin}, Fernando A. and {Tafoya}, Daniel and {Tatematsu}, Ken'ichi and {Izumi}, Natsuko and {Sakai}, Takeshi and {Silva}, Andrea},
        title = "{The ALMA Survey of 70 {\ensuremath{\mu}}m Dark High-mass Clumps in Early Stages (ASHES). IX. Physical Properties and Spatial Distribution of Cores in IRDCs}",
      journal = {\apj},
         year = 2023,
        month = jun,
       volume = {950},
       number = {2},
          eid = {148},
        pages = {148},
          doi = {10.3847/1538-4357/acccea},
archivePrefix = {arXiv},
       eprint = {2304.01757},
 primaryClass = {astro-ph.GA},
       adsurl = {https://ui.adsabs.harvard.edu/abs/2023ApJ...950..148M}
}

@ARTICLE{Csengeri2011,
       author = {{Csengeri}, T. and {Bontemps}, S. and {Schneider}, N. and {Motte}, F. and {Gueth}, F. and {Hora}, J.~L.},
        title = "{Convergent Flows and Low-velocity Shocks in DR21(OH)}",
      journal = {\apjl},
         year = 2011,
        month = oct,
       volume = {740},
       number = {1},
          eid = {L5},
        pages = {L5},
          doi = {10.1088/2041-8205/740/1/L5},
archivePrefix = {arXiv},
       eprint = {1108.4451},
 primaryClass = {astro-ph.GA},
       adsurl = {https://ui.adsabs.harvard.edu/abs/2011ApJ...740L...5C}
}

@ARTICLE{Sanhueza2019,
       author = {{Sanhueza}, Patricio and {Contreras}, Yanett and {Wu}, Benjamin and {Jackson}, James M. and {Guzm{\'a}n}, Andr{\'e}s E. and {Zhang}, Qizhou and {Li}, Shanghuo and {Lu}, Xing and {Silva}, Andrea and {Izumi}, Natsuko and {Liu}, Tie and {Miura}, Rie E. and {Tatematsu}, Ken'ichi and {Sakai}, Takeshi and {Beuther}, Henrik and {Garay}, Guido and {Ohashi}, Satoshi and {Saito}, Masao and {Nakamura}, Fumitaka and {Saigo}, Kazuya and {Veena}, V.~S. and {Nguyen-Luong}, Quang and {Tafoya}, Daniel},
        title = "{The ALMA Survey of 70 {\ensuremath{\mu}}m Dark High-mass Clumps in Early Stages (ASHES). I. Pilot Survey: Clump Fragmentation}",
      journal = {\apj},
         year = 2019,
        month = dec,
       volume = {886},
       number = {2},
          eid = {102},
        pages = {102},
          doi = {10.3847/1538-4357/ab45e9},
archivePrefix = {arXiv},
       eprint = {1909.07985},
 primaryClass = {astro-ph.GA},
       adsurl = {https://ui.adsabs.harvard.edu/abs/2019ApJ...886..102S}
}

@ARTICLE{Cunningham2016,
author = {{Cunningham}, Nichol and {Lumsden}, Stuart L. and {Cyganowski}, Claudia J. and {Maud}, Luke T. and {Purcell}, Cormac},
title = "{Submillimeter array observations of NGC 2264-C: molecular outflows and driving sources}",
journal = {\mnras},
year = 2016,
month = may,
volume = {458},
number = {2},
pages = {1742-1767},
doi = {10.1093/mnras/stw359},
archivePrefix = {arXiv},
eprint = {1602.06196},
primaryClass = {astro-ph.GA},
adsurl = {https://ui.adsabs.harvard.edu/abs/2016MNRAS.458.1742C}
}

@ARTICLE{Sanhueza2021,
       author = {{Sanhueza}, Patricio and {Girart}, Josep Miquel and {Padovani}, Marco and {Galli}, Daniele and {Hull}, Charles L.~H. and {Zhang}, Qizhou and {Cortes}, Paulo and {Stephens}, Ian W. and {Fern{\'a}ndez-L{\'o}pez}, Manuel and {Jackson}, James M. and {Frau}, Pau and {Kock}, Patrick M. and {Wu}, Benjamin and {Zapata}, Luis A. and {Olguin}, Fernando and {Lu}, Xing and {Silva}, Andrea and {Tang}, Ya-Wen and {Sakai}, Takeshi and {Guzm{\'a}n}, Andr{\'e}s E. and {Tatematsu}, Ken'ichi and {Nakamura}, Fumitaka and {Chen}, Huei-Ru Vivien},
        title = "{Gravity-driven Magnetic Field at  1000 au Scales in High-mass Star Formation}",
      journal = {\apjl},
         year = 2021,
        month = jul,
       volume = {915},
       number = {1},
          eid = {L10},
        pages = {L10},
          doi = {10.3847/2041-8213/ac081c},
archivePrefix = {arXiv},
       eprint = {2106.03866},
 primaryClass = {astro-ph.GA},
       adsurl = {https://ui.adsabs.harvard.edu/abs/2021ApJ...915L..10S}
}

@ARTICLE{Cortes2021,
       author = {{Cort{\'e}s}, Paulo C. and {Sanhueza}, Patricio and {Houde}, Martin and {Mart{\'\i}n}, Sergio and {Hull}, Charles L.~H. and {Girart}, Josep M. and {Zhang}, Qizhou and {Fernandez-Lopez}, Manuel and {Zapata}, Luis A. and {Stephens}, Ian W. and {Li}, Hua-bai and {Wu}, Benjamin and {Olguin}, Fernando and {Lu}, Xing and {Guzm{\'a}n}, Andres E. and {Nakamura}, Fumitaka},
        title = "{Magnetic Fields in Massive Star-forming Regions (MagMaR). II. Tomography through Dust and Molecular Line Polarization in NGC 6334I(N)}",
      journal = {\apj},
         year = 2021,
        month = dec,
       volume = {923},
       number = {2},
          eid = {204},
        pages = {204},
          doi = {10.3847/1538-4357/ac28a1},
archivePrefix = {arXiv},
       eprint = {2109.09270},
 primaryClass = {astro-ph.GA},
       adsurl = {https://ui.adsabs.harvard.edu/abs/2021ApJ...923..204C}
}

@ARTICLE{Motte2018,
       author = {{Motte}, F. and {Nony}, T. and {Louvet}, F. and {Marsh}, K.~A. and {Bontemps}, S. and {Whitworth}, A.~P. and {Men'shchikov}, A. and {Nguyen Luong}, Q. and {Csengeri}, T. and {Maury}, A.~J. and {Gusdorf}, A. and {Chapillon}, E. and {K{\"o}nyves}, V. and {Schilke}, P. and {Duarte-Cabral}, A. and {Didelon}, P. and {Gaudel}, M.},
        title = "{The unexpectedly large proportion of high-mass star-forming cores in a Galactic mini-starburst}",
      journal = {Nature Astronomy},
         year = 2018,
        month = apr,
       volume = {2},
        pages = {478-482},
          doi = {10.1038/s41550-018-0452-x},
archivePrefix = {arXiv},
       eprint = {1804.02392},
 primaryClass = {astro-ph.GA},
       adsurl = {https://ui.adsabs.harvard.edu/abs/2018NatAs...2..478M}
}

@INPROCEEDINGS{Pineda2023,
       author = {{Pineda}, J.~E. and {Arzoumanian}, D. and {Andre}, P. and {Friesen}, R.~K. and {Zavagno}, A. and {Clarke}, S.~D. and {Inoue}, T. and {Chen}, C. and {Lee}, Y. and {Soler}, J.~D. and {Kuffmeier}, M.},
        title = "{From Bubbles and Filaments to Cores and Disks: Gas Gathering and Growth of Structure Leading to the Formation of Stellar Systems}",
    booktitle = {Protostars and Planets VII},
         year = 2023,
       editor = {{Inutsuka}, S. and {Aikawa}, Y. and {Muto}, T. and {Tomida}, K. and {Tamura}, M.},
       series = {Astronomical Society of the Pacific Conference Series},
       volume = {534},
        month = jul,
        pages = {233},
          doi = {10.48550/arXiv.2205.03935},
archivePrefix = {arXiv},
       eprint = {2205.03935},
 primaryClass = {astro-ph.GA},
       adsurl = {https://ui.adsabs.harvard.edu/abs/2023ASPC..534..233P}
}

@ARTICLE{Louvet2016,
       author = {{Louvet}, F. and {Motte}, F. and {Gusdorf}, A. and {Nguy{\^e}n Luong}, Q. and {Lesaffre}, P. and {Duarte-Cabral}, A. and {Maury}, A. and {Schneider}, N. and {Hill}, T. and {Schilke}, P. and {Gueth}, F.},
        title = "{Tracing extended low-velocity shocks through SiO emission. Case study of the W43-MM1 ridge}",
      journal = {\aap},
         year = 2016,
        month = nov,
       volume = {595},
          eid = {A122},
        pages = {A122},
          doi = {10.1051/0004-6361/201629077},
archivePrefix = {arXiv},
       eprint = {1607.08668},
 primaryClass = {astro-ph.GA},
       adsurl = {https://ui.adsabs.harvard.edu/abs/2016A&A...595A.122L}
}

@ARTICLE{Cortes2024,
       author = {{Cort{\'e}s}, Paulo C. and {Girart}, Josep M. and {Sanhueza}, Patricio and {Liu}, Junhao and {Mart{\'\i}n}, Sergio and {Stephens}, Ian W. and {Beuther}, Henrik and {Koch}, Patrick M. and {Fern{\'a}ndez-L{\'o}pez}, M. and {S{\'a}nchez-Monge}, {\'A}lvaro and {Wang}, Jia-Wei and {Morii}, Kaho and {Li}, Shanghuo and {Saha}, Piyali and {Zhang}, Qizhou and {Rebolledo}, David and {Zapata}, Luis A. and {Kang}, Ji-hyun and {Jiao}, Wenyu and {Kim}, Jongsoo and {Cheng}, Yu and {Hwang}, Jihye and {Chung}, Eun Jung and {Choudhury}, Spandan and {Lyo}, A. -Ran and {Olguin}, Fernando},
        title = "{MagMaR III{\textemdash}Resisting the Pressure, Is the Magnetic Field Overwhelmed in NGC6334I?}",
      journal = {\apj},
         year = 2024,
        month = sep,
       volume = {972},
       number = {1},
          eid = {115},
        pages = {115},
          doi = {10.3847/1538-4357/ad59a7},
archivePrefix = {arXiv},
       eprint = {2406.14663},
 primaryClass = {astro-ph.GA},
       adsurl = {https://ui.adsabs.harvard.edu/abs/2024ApJ...972..115C}
}

@ARTICLE{Sanhueza2025,
       author = {{Sanhueza}, Patricio and {Liu}, Junhao and {Morii}, Kaho and {Girart}, Josep Miquel and {Zhang}, Qizhou and {Stephens}, Ian W. and {Jackson}, James M. and {Cort{\'e}s}, Paulo C. and {Koch}, Patrick M. and {Cyganowski}, Claudia J. and {Saha}, Piyali and {Beuther}, Henrik and {Zhang}, Suinan and {Beltr{\'a}n}, Maria T. and {Cheng}, Yu and {Olguin}, Fernando A. and {Lu}, Xing and {Choudhury}, Spandan and {Pattle}, Kate and {Fern{\'a}ndez-L{\'o}pez}, Manuel and {Hwang}, Jihye and {Kang}, Ji-hyun and {Karoly}, Janik and {Ginsburg}, Adam and {Lyo}, A. -Ran and {Taniguchi}, Kotomi and {Jiao}, Wenyu and {Eswaraiah}, Chakali and {Luo}, Qiu-yi and {Wang}, Jia-Wei and {Commer{\c{c}}on}, Beno{\^\i}t and {Li}, Shanghuo and {Xu}, Fengwei and {Chen}, Huei-Ru Vivien and {Zapata}, Luis A. and {Chung}, Eun Jung and {Nakamura}, Fumitaka and {Panigrahy}, Sandhyarani and {Sakai}, Takeshi},
        title = "{Magnetic Fields in Massive Star-forming Regions (MagMaR). V. The Magnetic Field at the Onset of High-mass Star Formation}",
      journal = {\apj},
         year = 2025,
        month = feb,
       volume = {980},
       number = {1},
          eid = {87},
        pages = {87},
          doi = {10.3847/1538-4357/ad9d40},
archivePrefix = {arXiv},
       eprint = {2412.08790},
 primaryClass = {astro-ph.GA},
       adsurl = {https://ui.adsabs.harvard.edu/abs/2025ApJ...980...87S}
}

@ARTICLE{Crutcher2012,
       author = {{Crutcher}, Richard M.},
        title = "{Magnetic Fields in Molecular Clouds}",
      journal = {\araa},
         year = 2012,
        month = sep,
       volume = {50},
        pages = {29-63},
          doi = {10.1146/annurev-astro-081811-125514},
       adsurl = {https://ui.adsabs.harvard.edu/abs/2012ARA&A..50...29C}
}

@ARTICLE{LeGouellec2024,
       author = {{Le Gouellec}, Valentin J.~M. and {Greene}, Thomas P. and {Hillenbrand}, Lynne A. and {Yates}, Zoe},
        title = "{New Insights on the Accretion Properties of Class 0 Protostars from 2 {\ensuremath{\mu}}m Spectroscopy}",
      journal = {\apj},
         year = 2024,
        month = may,
       volume = {966},
       number = {1},
          eid = {91},
        pages = {91},
          doi = {10.3847/1538-4357/ad2935},
archivePrefix = {arXiv},
       eprint = {2401.16532},
 primaryClass = {astro-ph.SR},
       adsurl = {https://ui.adsabs.harvard.edu/abs/2024ApJ...966...91L}
}

@ARTICLE{Nguyen2013,
       author = {{Nguyen-Lu'o'ng}, Q. and {Motte}, F. and {Carlhoff}, P. and {Louvet}, F. and {Lesaffre}, P. and {Schilke}, P. and {Hill}, T. and {Hennemann}, M. and {Gusdorf}, A. and {Didelon}, P. and {Schneider}, N. and {Bontemps}, S. and {Duarte-Cabral}, A. and {Menten}, K.~M. and {Martin}, P.~G. and {Wyrowski}, F. and {Bendo}, G. and {Roussel}, H. and {Bernard}, J. -P. and {Bronfman}, L. and {Henning}, T. and {Kramer}, C. and {Heitsch}, F.},
        title = "{Low-velocity Shocks Traced by Extended SiO Emission along the W43 Ridges: Witnessing the Formation of Young Massive Clusters}",
      journal = {\apj},
         year = 2013,
        month = oct,
       volume = {775},
       number = {2},
          eid = {88},
        pages = {88},
          doi = {10.1088/0004-637X/775/2/88},
archivePrefix = {arXiv},
       eprint = {1306.0547},
 primaryClass = {astro-ph.GA},
       adsurl = {https://ui.adsabs.harvard.edu/abs/2013ApJ...775...88N}
}

@ARTICLE{Cortes2016,
       author = {{Cortes}, Paulo C. and {Girart}, Josep M. and {Hull}, Charles L.~H. and {Sridharan}, Tirupati K. and {Louvet}, Fabien and {Plambeck}, Richard and {Li}, Zhi-Yun and {Crutcher}, Richard M. and {Lai}, Shih-Ping},
        title = "{Interferometric Mapping of Magnetic Fields: The ALMA View of the Massive Star-forming Clump W43-MM1}",
      journal = {\apjl},
         year = 2016,
        month = jul,
       volume = {825},
       number = {1},
          eid = {L15},
        pages = {L15},
          doi = {10.3847/2041-8205/825/1/L15},
archivePrefix = {arXiv},
       eprint = {1605.08037},
 primaryClass = {astro-ph.GA},
       adsurl = {https://ui.adsabs.harvard.edu/abs/2016ApJ...825L..15C}
}

@ARTICLE{Arce-Tord2020,
       author = {{Arce-Tord}, C. and {Louvet}, F. and {Cortes}, P.~C. and {Motte}, F. and {Hull}, C.~L.~H. and {Le Gouellec}, V.~J.~M. and {Garay}, G. and {Nony}, T. and {Didelon}, P. and {Bronfman}, L.},
        title = "{Outflows, cores, and magnetic field orientations in W43-MM1 as seen by ALMA}",
      journal = {\aap},
         year = 2020,
        month = aug,
       volume = {640},
          eid = {A111},
        pages = {A111},
          doi = {10.1051/0004-6361/202038024},
archivePrefix = {arXiv},
       eprint = {2005.12921},
 primaryClass = {astro-ph.GA},
       adsurl = {https://ui.adsabs.harvard.edu/abs/2020A&A...640A.111A}
}

@ARTICLE{Saha2022,
       author = {{Saha}, Anindya and {Tej}, Anandmayee and {Liu}, Hong-Li and {Liu}, Tie and {Issac}, Namitha and {Lee}, Chang Won and {Garay}, Guido and {Goldsmith}, Paul F. and {Juvela}, Mika and {Qin}, Sheng-Li and {Stutz}, Amelia and {Li}, Shanghuo and {Wang}, Ke and {Baug}, Tapas and {Bronfman}, Leonardo and {Xu}, Feng-Wei and {Zhang}, Yong and {Eswaraiah}, Chakali},
        title = "{ATOMS: ALMA three-millimeter observations of massive star-forming regions - XII: Fragmentation and multiscale gas kinematics in protoclusters G12.42+0.50 and G19.88-0.53}",
      journal = {\mnras},
         year = 2022,
        month = oct,
       volume = {516},
       number = {2},
        pages = {1983-2005},
          doi = {10.1093/mnras/stac2353},
archivePrefix = {arXiv},
       eprint = {2208.09877},
 primaryClass = {astro-ph.GA},
       adsurl = {https://ui.adsabs.harvard.edu/abs/2022MNRAS.516.1983S}
}

@ARTICLE{Li2023,
       author = {{Li}, Shanghuo and {Sanhueza}, Patricio and {Zhang}, Qizhou and {Guido}, Garay and {Sabatini}, Giovanni and {Morii}, Kaho and {Lu}, Xing and {Tafoya}, Daniel and {Nakamura}, Fumitaka and {Izumi}, Natsuko and {Tatematsu}, Ken'ichi and {Li}, Fei},
        title = "{The ALMA Survey of 70 {\ensuremath{\mu}}m Dark High-mass Clumps in Early Stages (ASHES). VIII. Dynamics of Embedded Dense Cores}",
      journal = {\apj},
         year = 2023,
        month = jun,
       volume = {949},
       number = {2},
          eid = {109},
        pages = {109},
          doi = {10.3847/1538-4357/acc58f},
archivePrefix = {arXiv},
       eprint = {2304.01718},
 primaryClass = {astro-ph.GA},
       adsurl = {https://ui.adsabs.harvard.edu/abs/2023ApJ...949..109L}
}

@ARTICLE{Cortes2019,
       author = {{Cortes}, Paulo C. and {Hull}, Charles L.~H. and {Girart}, Josep M. and {Orquera-Rojas}, Carlos and {Sridharan}, Tirupati K. and {Li}, Zhi-Yun and {Louvet}, Fabien and {Cortes}, Juan R. and {Le Gouellec}, Valentin J.~M. and {Crutcher}, Richard M. and {Lai}, Shih-Ping},
        title = "{The Seven Most Massive Clumps in W43-Main as Seen by ALMA: Dynamical Equilibrium and Magnetic Fields}",
      journal = {\apj},
         year = 2019,
        month = oct,
       volume = {884},
       number = {1},
          eid = {48},
        pages = {48},
          doi = {10.3847/1538-4357/ab378d},
archivePrefix = {arXiv},
       eprint = {1907.12994},
 primaryClass = {astro-ph.GA},
       adsurl = {https://ui.adsabs.harvard.edu/abs/2019ApJ...884...48C}
}

@ARTICLE{Liu2020,
       author = {{Liu}, Junhao and {Zhang}, Qizhou and {Qiu}, Keping and {Liu}, Hauyu Baobab and {Pillai}, Thushara and {Girart}, Josep Miquel and {Li}, Zhi-Yun and {Wang}, Ke},
        title = "{Magnetic Fields in the Early Stages of Massive Star Formation as Revealed by ALMA}",
      journal = {\apj},
         year = 2020,
        month = jun,
       volume = {895},
       number = {2},
          eid = {142},
        pages = {142},
          doi = {10.3847/1538-4357/ab9087},
archivePrefix = {arXiv},
       eprint = {2005.01705},
 primaryClass = {astro-ph.GA},
       adsurl = {https://ui.adsabs.harvard.edu/abs/2020ApJ...895..142L}
}

@ARTICLE{Pillai2015,
       author = {{Pillai}, T. and {Kauffmann}, J. and {Tan}, J.~C. and {Goldsmith}, P.~F. and {Carey}, S.~J. and {Menten}, K.~M.},
        title = "{Magnetic Fields in High-mass Infrared Dark Clouds}",
      journal = {\apj},
         year = 2015,
        month = jan,
       volume = {799},
       number = {1},
          eid = {74},
        pages = {74},
          doi = {10.1088/0004-637X/799/1/74},
archivePrefix = {arXiv},
       eprint = {1410.7390},
 primaryClass = {astro-ph.GA},
       adsurl = {https://ui.adsabs.harvard.edu/abs/2015ApJ...799...74P}
}

@ARTICLE{McKee&Ostriker2007,
       author = {{McKee}, Christopher F. and {Ostriker}, Eve C.},
        title = "{Theory of Star Formation}",
      journal = {\araa},
         year = 2007,
        month = sep,
       volume = {45},
       number = {1},
        pages = {565-687},
          doi = {10.1146/annurev.astro.45.051806.110602},
archivePrefix = {arXiv},
       eprint = {0707.3514},
 primaryClass = {astro-ph},
       adsurl = {https://ui.adsabs.harvard.edu/abs/2007ARA&A..45..565M}
}

@ARTICLE{Hacar2016,
       author = {{Hacar}, A. and {Alves}, J. and {Burkert}, A. and {Goldsmith}, P.},
        title = "{Opacity broadening and interpretation of suprathermal CO linewidths: Macroscopic turbulence and tangled molecular clouds}",
      journal = {\aap},
         year = 2016,
        month = jun,
       volume = {591},
          eid = {A104},
        pages = {A104},
          doi = {10.1051/0004-6361/201527319},
archivePrefix = {arXiv},
       eprint = {1603.08521},
 primaryClass = {astro-ph.GA},
       adsurl = {https://ui.adsabs.harvard.edu/abs/2016A&A...591A.104H}
}

@ARTICLE{Morii2025,
       author = {{Morii}, Kaho and {Sanhueza}, Patricio and {Csengeri}, Timea and {Nakamura}, Fumitaka and {Bontemps}, Sylvain and {Garay}, Guido and {Zhang}, Qizhou},
        title = "{Global and Local Infall in the ASHES Sample (GLASHES). I. Pilot Study in G337.541}",
      journal = {\apj},
         year = 2025,
        month = feb,
       volume = {979},
       number = {2},
          eid = {233},
        pages = {233},
          doi = {10.3847/1538-4357/ada27f},
archivePrefix = {arXiv},
       eprint = {2412.17901},
 primaryClass = {astro-ph.GA},
       adsurl = {https://ui.adsabs.harvard.edu/abs/2025ApJ...979..233M}
}

@ARTICLE{Traficante2018,
       author = {{Traficante}, A. and {Fuller}, G.~A. and {Smith}, R.~J. and {Billot}, N. and {Duarte-Cabral}, A. and {Peretto}, N. and {Molinari}, S. and {Pineda}, J.~E.},
        title = "{Massive 70 {\ensuremath{\mu}}m quiet clumps - II. Non-thermal motions driven by gravity in massive star formation?}",
      journal = {\mnras},
         year = 2018,
        month = feb,
       volume = {473},
       number = {4},
        pages = {4975-4985},
          doi = {10.1093/mnras/stx2672},
archivePrefix = {arXiv},
       eprint = {1710.04904},
 primaryClass = {astro-ph.GA},
       adsurl = {https://ui.adsabs.harvard.edu/abs/2018MNRAS.473.4975T}
}

@ARTICLE{Pineda2010,
       author = {{Pineda}, Jaime E. and {Goodman}, Alyssa A. and {Arce}, H{\'e}ctor G. and {Caselli}, Paola and {Foster}, Jonathan B. and {Myers}, Philip C. and {Rosolowsky}, Erik W.},
        title = "{Direct Observation of a Sharp Transition to Coherence in Dense Cores}",
      journal = {\apjl},
         year = 2010,
        month = mar,
       volume = {712},
       number = {1},
        pages = {L116-L121},
          doi = {10.1088/2041-8205/712/1/L116},
archivePrefix = {arXiv},
       eprint = {1002.2946},
 primaryClass = {astro-ph.GA},
       adsurl = {https://ui.adsabs.harvard.edu/abs/2010ApJ...712L.116P}
}

@INPROCEEDINGS{Pattle2023,
       author = {{Pattle}, K. and {Fissel}, L. and {Tahani}, M. and {Liu}, T. and {Ntormousi}, E.},
        title = "{Magnetic Fields in Star Formation: from Clouds to Cores}",
    booktitle = {Protostars and Planets VII},
         year = 2023,
       editor = {{Inutsuka}, S. and {Aikawa}, Y. and {Muto}, T. and {Tomida}, K. and {Tamura}, M.},
       series = {Astronomical Society of the Pacific Conference Series},
       volume = {534},
        month = jul,
        pages = {193},
          doi = {10.48550/arXiv.2203.11179},
archivePrefix = {arXiv},
       eprint = {2203.11179},
 primaryClass = {astro-ph.GA},
       adsurl = {https://ui.adsabs.harvard.edu/abs/2023ASPC..534..193P}
}

@ARTICLE{Rosen2020,
       author = {{Rosen}, Anna L. and {Krumholz}, Mark R.},
        title = "{The Role of Outflows, Radiation Pressure, and Magnetic Fields in Massive Star Formation}",
      journal = {\aj},
         year = 2020,
        month = aug,
       volume = {160},
       number = {2},
          eid = {78},
        pages = {78},
          doi = {10.3847/1538-3881/ab9abf},
archivePrefix = {arXiv},
       eprint = {2006.04829},
 primaryClass = {astro-ph.SR},
       adsurl = {https://ui.adsabs.harvard.edu/abs/2020AJ....160...78R}
}

@ARTICLE{Commercon2011,
       author = {{Commer{\c{c}}on}, Beno{\^\i}t and {Hennebelle}, Patrick and {Henning}, Thomas},
        title = "{Collapse of Massive Magnetized Dense Cores Using Radiation Magnetohydrodynamics: Early Fragmentation Inhibition}",
      journal = {\apjl},
         year = 2011,
        month = nov,
       volume = {742},
       number = {1},
          eid = {L9},
        pages = {L9},
          doi = {10.1088/2041-8205/742/1/L9},
archivePrefix = {arXiv},
       eprint = {1110.2955},
 primaryClass = {astro-ph.SR},
       adsurl = {https://ui.adsabs.harvard.edu/abs/2011ApJ...742L...9C}
}

@ARTICLE{Mestel1966,
       author = {{Mestel}, L.},
        title = "{The magnetic field of a contracting gas cloud. I,Strict flux-freezing}",
      journal = {\mnras},
         year = 1966,
        month = jan,
       volume = {133},
        pages = {265},
          doi = {10.1093/mnras/133.2.265},
       adsurl = {https://ui.adsabs.harvard.edu/abs/1966MNRAS.133..265M}
}

@INPROCEEDINGS{Mouschovias1999,
       author = {{Mouschovias}, Telemachos Ch. and {Ciolek}, Glenn E.},
        title = "{Magnetic Fields and Star Formation: A Theory Reaching Adulthood}",
    booktitle = {The Origin of Stars and Planetary Systems},
         year = 1999,
       editor = {{Lada}, Charles J. and {Kylafis}, Nikolaos D.},
       series = {NATO Advanced Study Institute (ASI) Series C},
       volume = {540},
        month = jan,
        pages = {305},
       adsurl = {https://ui.adsabs.harvard.edu/abs/1999ASIC..540..305M}
}

@ARTICLE{Beuther2025,
       author = {{Beuther}, H. and {Kuiper}, R. and {Tafalla}, M.},
        title = "{Star Formation from Low to High Mass: A Comparative View}",
      journal = {\araa},
         year = 2025,
        month = aug,
       volume = {63},
       number = {1},
        pages = {1-44},
          doi = {10.1146/annurev-astro-013125-122023},
archivePrefix = {arXiv},
       eprint = {2501.16866},
 primaryClass = {astro-ph.GA},
       adsurl = {https://ui.adsabs.harvard.edu/abs/2025ARA&A..63....1B}
}

@ARTICLE{Dib2007,
       author = {{Dib}, Sami and {Kim}, Jongsoo and {V{\'a}zquez-Semadeni}, Enrique and {Burkert}, Andreas and {Shadmehri}, Mohsen},
        title = "{The Virial Balance of Clumps and Cores in Molecular Clouds}",
      journal = {\apj},
         year = 2007,
        month = may,
       volume = {661},
       number = {1},
        pages = {262-284},
          doi = {10.1086/513708},
archivePrefix = {arXiv},
       eprint = {astro-ph/0607362},
 primaryClass = {astro-ph},
       adsurl = {https://ui.adsabs.harvard.edu/abs/2007ApJ...661..262D}
}

@INPROCEEDINGS{Li2014,
       author = {{Li}, Z.-Y. and {Banerjee}, R. and {Pudritz}, R.~E. and {J{\o}rgensen}, J.~K. and {Shang}, H. and {Krasnopolsky}, R. and {Maury}, A.},
        title = "{The Earliest Stages of Star and Planet Formation: Core Collapse, and the Formation of Disks and Outflows}",
    booktitle = {Protostars and Planets VI},
         year = 2014,
       editor = {{Beuther}, Henrik and {Klessen}, Ralf S. and {Dullemond}, Cornelis P. and {Henning}, Thomas},
        month = jan,
        pages = {173-194},
          doi = {10.2458/azu_uapress_9780816531240-ch008},
archivePrefix = {arXiv},
       eprint = {1401.2219},
 primaryClass = {astro-ph.SR},
       adsurl = {https://ui.adsabs.harvard.edu/abs/2014prpl.conf..173L}
}

@INPROCEEDINGS{Shadmehri2002,
       author = {{Shadmehri}, M. and {V{\'a}zquez-Semadeni}, E. and {Ballesteros-Paredes}, J.},
        title = "{Virial Theorem Analysis of 3D Numerical Simulations of MHD Self-Gravitating Turbulence}",
    booktitle = {Seeing Through the Dust: The Detection of HI and the Exploration of the ISM in Galaxies},
         year = 2002,
       editor = {{Taylor}, A.~R. and {Landecker}, T.~L. and {Willis}, A.~G.},
       series = {Astronomical Society of the Pacific Conference Series},
       volume = {276},
        month = dec,
        pages = {190},
          doi = {10.48550/arXiv.astro-ph/0111574},
archivePrefix = {arXiv},
       eprint = {astro-ph/0111574},
 primaryClass = {astro-ph},
       adsurl = {https://ui.adsabs.harvard.edu/abs/2002ASPC..276..190S}
}

@ARTICLE{Mininni2025,
       author = {{Mininni}, C. and {Molinari}, S. and {Soler}, J.~D. and {S{\'a}nchez-Monge}, {\'A}. and {Coletta}, A. and {Benedettini}, M. and {Traficante}, A. and {Schisano}, E. and {Elia}, D. and {Pezzuto}, S. and {Nucara}, A. and {Schilke}, P. and {Battersby}, C. and {Ho}, P.~T.~P. and {Beltr{\'a}n}, M.~T. and {Beuther}, H. and {Fuller}, G.~A. and {Jones}, B. and {Klessen}, R.~S. and {Zhang}, Q. and {Walch}, S. and {Tang}, Y. and {Ahmadi}, A. and {Allande}, J. and {Avison}, A. and {Brogan}, C.~L. and {De Angelis}, F. and {Fontani}, F. and {Hennebelle}, P. and {Hunter}, T.~R. and {Johnston}, K.~G. and {Koch}, P. and {Kuiper}, R. and {Law}, C.-Y. and {Lis}, D.~C. and {Liu}, S. and {Liu}, T. and {Liu}, S.-Y. and {Moscadelli}, L. and {M{\"o}ller}, T. and {Rigby}, A.~J. and {Rygl}, K.~L.~J. and {Sanhueza}, P. and {Testi}, L. and {Su}, Y.-N. and {van der Tak}, F.~F.~S. and {Wells}, M.~R.~A. and {Bronfman}, L. and {Zhang}, T. and {Zinnecker}, H.},
        title = "{ALMAGAL: IV. Morphological comparison of molecular and thermal dust emission using the histogram of oriented gradients method}",
      journal = {\aap},
         year = 2025,
        month = jun,
       volume = {699},
          eid = {A34},
        pages = {A34},
          doi = {10.1051/0004-6361/202452700},
archivePrefix = {arXiv},
       eprint = {2504.12963},
 primaryClass = {astro-ph.GA},
       adsurl = {https://ui.adsabs.harvard.edu/abs/2025A&A...699A..34M}
}

@ARTICLE{Saha2024,
       author = {{Saha}, Piyali and {Sanhueza}, Patricio and {Padovani}, Marco and {Girart}, Josep M. and {Cort{\'e}s}, Paulo C. and {Morii}, Kaho and {Liu}, Junhao and {S{\'a}nchez-Monge}, {\'A}. and {Galli}, Daniele and {Basu}, Shantanu and {Koch}, Patrick M. and {Beltr{\'a}n}, Maria T. and {Li}, Shanghuo and {Beuther}, Henrik and {Stephens}, Ian W. and {Nakamura}, Fumitaka and {Zhang}, Qizhou and {Jiao}, Wenyu and {Fern{\'a}ndez-L{\'o}pez}, M. and {Hwang}, Jihye and {Chung}, Eun Jung and {Pattle}, Kate and {Zapata}, Luis A. and {Xu}, Fengwei and {Olguin}, Fernando A. and {Kang}, Ji-hyun and {Karoly}, Janik and {Law}, Chi-Yan and {Wang}, Jia-Wei and {Csengeri}, Timea and {Lu}, Xing and {Cheng}, Yu and {Kim}, Jongsoo and {Choudhury}, Spandan and {Chen}, Huei-Ru Vivien and {Hull}, Charles L.~H.},
        title = "{Magnetic Fields in Massive Star-forming Regions (MagMaR): Unveiling an Hourglass Magnetic Field in G333.46{\textendash}0.16 Using ALMA}",
      journal = {\apjl},
         year = 2024,
        month = sep,
       volume = {972},
       number = {1},
          eid = {L6},
        pages = {L6},
          doi = {10.3847/2041-8213/ad660c},
archivePrefix = {arXiv},
       eprint = {2407.16654},
 primaryClass = {astro-ph.GA},
       adsurl = {https://ui.adsabs.harvard.edu/abs/2024ApJ...972L...6S}
}

@ARTICLE{Koley2025,
       author = {{Koley}, A. and {Stutz}, A.~M. and {Louvet}, F. and {Motte}, F. and {Ginsburg}, A. and {Galv{\'a}n-Madrid}, R. and {{\'A}lvarez-Guti{\'e}rrez}, R.~H. and {Sanhueza}, P. and {Baug}, T. and {Sandoval-Garrido}, N. and {Salinas}, J. and {Busquet}, G. and {Braine}, J. and {Liu}, H.-L. and {Csengeri}, T. and {Gusdorf}, A. and {Fern{\'a}ndez-L{\'o}pez}, M. and {Cunningham}, N. and {Bronfman}, L. and {Bonfand}, M.},
        title = "{ALMA-IMF: XIX. C$^{18}$O (J = 2─1): Measurements of turbulence in 15 massive protoclusters}",
      journal = {\aap},
         year = 2025,
        month = oct,
       volume = {702},
          eid = {A133},
        pages = {A133},
          doi = {10.1051/0004-6361/202553830},
archivePrefix = {arXiv},
       eprint = {2507.14502},
 primaryClass = {astro-ph.GA},
       adsurl = {https://ui.adsabs.harvard.edu/abs/2025A&A...702A.133K}
}

@ARTICLE{Ballesteros-Paredes2018,
       author = {{Ballesteros-Paredes}, Javier and {V{\'a}zquez-Semadeni}, Enrique and {Palau}, Aina and {Klessen}, Ralf S.},
        title = "{Gravity or turbulence? - IV. Collapsing cores in out-of-virial disguise}",
      journal = {\mnras},
         year = 2018,
        month = sep,
       volume = {479},
       number = {2},
        pages = {2112-2125},
          doi = {10.1093/mnras/sty1515},
archivePrefix = {arXiv},
       eprint = {1710.07384},
 primaryClass = {astro-ph.GA},
       adsurl = {https://ui.adsabs.harvard.edu/abs/2018MNRAS.479.2112B}
}

@ARTICLE{Liu2019,
       author = {{Liu}, Junhao and {Zhang}, Qizhou and {Commer{\c{c}}on}, Beno{\^\i}t and {Valdivia}, Valeska and {Maury}, Ana{\"e}lle and {Qiu}, Keping},
        title = "{Calibrating the Davis-Chandrasekhar-Fermi Method with Numerical Simulations: Uncertainties in Estimating the Magnetic Field Strength from Statistics of Field Orientations}",
      journal = {\apj},
         year = 2021,
        month = oct,
       volume = {919},
       number = {2},
          eid = {79},
        pages = {79},
          doi = {10.3847/1538-4357/ac0cec},
archivePrefix = {arXiv},
       eprint = {2106.09934},
 primaryClass = {astro-ph.GA},
       adsurl = {https://ui.adsabs.harvard.edu/abs/2021ApJ...919...79L}
}

@ARTICLE{Zapata2024,
       author = {{Zapata}, Luis A. and {Fern{\'a}ndez-L{\'o}pez}, Manuel and {Sanhueza}, Patricio and {Girart}, Josep M. and {Rodr{\'\i}guez}, Luis F. and {Cort{\'e}s}, Paulo and {Koch}, Patrick and {Beltr{\'a}n}, Maria T. and {Pattle}, Kate and {Beuther}, Henrik and {Saha}, Piyali and {Jiao}, Wenyu and {Xu}, Fengwei and {Lu}, Xing Walker and {Olguin}, Fernando and {Li}, Shanghuo and {Stephens}, Ian W. and {Kang}, Ji-hyun and {Cheng}, Yu and {Choudhury}, Spandan and {Morii}, Kaho and {Chung}, Eun Jung and {Wang}, Jia-Wei and {Hwang}, Jihye and {Lyo}, A.-Ran and {Zhang}, Q. and {Chen}, Huei-Ru Vivien},
        title = "{Magnetic Fields in Massive Star-forming Regions (MagMaR). IV. Tracing the Magnetic Fields in the O-type Protostellar System IRAS 16547{\textendash}4247}",
      journal = {\apj},
         year = 2024,
        month = oct,
       volume = {974},
       number = {2},
          eid = {257},
        pages = {257},
          doi = {10.3847/1538-4357/ad701d},
archivePrefix = {arXiv},
       eprint = {2408.10199},
 primaryClass = {astro-ph.SR},
       adsurl = {https://ui.adsabs.harvard.edu/abs/2024ApJ...974..257Z}
}

@ARTICLE{Sandoval-Garrido2025,
       author = {{Sandoval-Garrido}, N.~A. and {Stutz}, A.~M. and {{\'A}lvarez-Guti{\'e}rrez}, R.~H. and {Galv{\'a}n-Madrid}, R. and {Motte}, F. and {Ginsburg}, A. and {Cunningham}, N. and {Reyes-Reyes}, S. and {Redaelli}, E. and {Bonfand}, M. and {Salinas}, J. and {Koley}, A. and {Bernal-Mesina}, G. and {Braine}, J. and {Bronfman}, L. and {Busquet}, G. and {Csengeri}, T. and {Di Francesco}, J. and {Fern{\'a}ndez-L{\'o}pez}, M. and {Garcia}, P. and {Gusdorf}, A. and {Liu}, H.-L. and {Sanhueza}, P.},
        title = "{ALMA-IMF: XVIII. The assembly of a star cluster: Dense N$_{2}$H$^{+}$ (1─0) kinematics in the massive G351.77 protocluster}",
      journal = {\aap},
         year = 2025,
        month = apr,
       volume = {696},
          eid = {A202},
        pages = {A202},
          doi = {10.1051/0004-6361/202452589},
archivePrefix = {arXiv},
       eprint = {2410.09843},
 primaryClass = {astro-ph.GA},
       adsurl = {https://ui.adsabs.harvard.edu/abs/2025A&A...696A.202S}
}

@ARTICLE{Kauffman2013,
       author = {{Kauffmann}, Jens and {Pillai}, Thushara and {Goldsmith}, Paul F.},
        title = "{Low Virial Parameters in Molecular Clouds: Implications for High-mass Star Formation and Magnetic Fields}",
      journal = {\apj},
         year = 2013,
        month = dec,
       volume = {779},
       number = {2},
          eid = {185},
        pages = {185},
          doi = {10.1088/0004-637X/779/2/185},
archivePrefix = {arXiv},
       eprint = {1308.5679},
 primaryClass = {astro-ph.GA},
       adsurl = {https://ui.adsabs.harvard.edu/abs/2013ApJ...779..185K}
}

@ARTICLE{Myers&Goodman1988,
       author = {{Myers}, P.~C. and {Goodman}, A.~A.},
        title = "{Evidence for Magnetic and Virial Equilibrium in Molecular Clouds}",
      journal = {\apjl},
         year = 1988,
        month = mar,
       volume = {326},
        pages = {L27},
          doi = {10.1086/185116},
       adsurl = {https://ui.adsabs.harvard.edu/abs/1988ApJ...326L..27M}
}

@ARTICLE{Bertoldi&McKee1992,
       author = {{Bertoldi}, Frank and {McKee}, Christopher F.},
        title = "{Pressure-confined Clumps in Magnetized Molecular Clouds}",
      journal = {\apj},
         year = 1992,
        month = aug,
       volume = {395},
        pages = {140},
          doi = {10.1086/171638},
       adsurl = {https://ui.adsabs.harvard.edu/abs/1992ApJ...395..140B}
}

@ARTICLE{Wang2024,
       author = {{Wang}, Chao and {Wang}, Ke and {Xu}, Feng-Wei and {Sanhueza}, Patricio and {Liu}, Hauyu Baobab and {Zhang}, Qizhou and {Lu}, Xing and {Fontani}, F. and {Caselli}, Paola and {Busquet}, Gemma and {Tan}, Jonathan C. and {Li}, Di and {Jackson}, J.~M. and {Pillai}, Thushara and {Ho}, Paul T.~P. and {Guzm{\'a}n}, Andr{\'e}s E. and {Yue}, Nannan},
        title = "{The role of turbulence in high-mass star formation: Subsonic and transonic turbulence are ubiquitously found at early stages}",
      journal = {\aap},
         year = 2024,
        month = jan,
       volume = {681},
          eid = {A51},
        pages = {A51},
          doi = {10.1051/0004-6361/202347024},
archivePrefix = {arXiv},
       eprint = {2310.17970},
 primaryClass = {astro-ph.GA},
       adsurl = {https://ui.adsabs.harvard.edu/abs/2024A&A...681A..51W}
}

@ARTICLE{Larson1981,
       author = {{Larson}, R.~B.},
        title = "{Turbulence and star formation in molecular clouds.}",
      journal = {\mnras},
         year = 1981,
        month = mar,
       volume = {194},
        pages = {809-826},
          doi = {10.1093/mnras/194.4.809},
       adsurl = {https://ui.adsabs.harvard.edu/abs/1981MNRAS.194..809L}
}

@ARTICLE{Ballesteros-Paredes2006,
       author = {{Ballesteros-Paredes}, Javier},
        title = "{Six myths on the virial theorem for interstellar clouds}",
      journal = {\mnras},
         year = 2006,
        month = oct,
       volume = {372},
       number = {1},
        pages = {443-449},
          doi = {10.1111/j.1365-2966.2006.10880.x},
archivePrefix = {arXiv},
       eprint = {astro-ph/0606071},
 primaryClass = {astro-ph},
       adsurl = {https://ui.adsabs.harvard.edu/abs/2006MNRAS.372..443B}
}

@ARTICLE{Li2022,
       author = {{Li}, Shanghuo and {Sanhueza}, Patricio and {Lee}, Chang Won and {Zhang}, Qizhou and {Beuther}, Henrik and {Palau}, Aina and {Liu}, Hong-Li and {Smith}, Howard A. and {Liu}, Hauyu Baobab and {Jim{\'e}nez-Serra}, Izaskun and {Kim}, Kee-Tae and {Feng}, Siyi and {Liu}, Tie and {Wang}, Junzhi and {Li}, Di and {Qiu}, Keping and {Lu}, Xing and {Girart}, Josep Miquel and {Wang}, Ke and {Li}, Fei and {Li}, Juan and {Cao}, Yue and {Kim}, Shinyoung and {Strom}, Shaye},
        title = "{ALMA Observations of NGC 6334S. II. Subsonic and Transonic Narrow Filaments in a High-mass Star Formation Cloud}",
      journal = {\apj},
         year = 2022,
        month = feb,
       volume = {926},
       number = {2},
          eid = {165},
        pages = {165},
          doi = {10.3847/1538-4357/ac3df8},
archivePrefix = {arXiv},
       eprint = {2111.12593},
 primaryClass = {astro-ph.GA},
       adsurl = {https://ui.adsabs.harvard.edu/abs/2022ApJ...926..165L}
}

@ARTICLE{Hwang2026,
       author = {{Hwang}, Jihye and {Sanhueza}, Patricio and {Girart}, Josep Miquel and {Stephens}, Ian W. and {Beltr{\'a}n}, Maria T. and {Law}, Chi Yan and {Zhang}, Qizhou and {Liu}, Junhao and {Cort{\'e}s}, Paulo and {Olguin}, Fernando A. and {Koch}, Patrick M. and {Nakamura}, Fumitaka and {Saha}, Piyali and {Wang}, Jia-Wei and {Xu}, Fengwei and {Beuther}, Henrik and {Morii}, Kaho and {Fern{\'a}ndez L{\'o}pez}, Manuel and {Jiao}, Wenyu and {Kim}, Kee-Tae and {Li}, Shanghuo and {Zapata}, Luis A. and {Kim}, Jongsoo and {Choudhury}, Spandan and {Cheng}, Yu and {Pattle}, Kate and {Eswaraiah}, Chakali and {Sandhyarani}, Panigrahy and {Dewangan}, L.~K. and {Jadhav}, O.~R.},
        title = "{Magnetic Fields in Massive Star-forming Regions (MagMaR). VI. Magnetic Field Dragging in the Filamentary High-mass Star-forming Region G35.20─0.74N Due to Gravity}",
      journal = {\aj},
         year = 2026,
        month = jan,
       volume = {171},
       number = {1},
          eid = {50},
        pages = {50},
          doi = {10.3847/1538-3881/ae18c9},
archivePrefix = {arXiv},
       eprint = {2510.25078},
 primaryClass = {astro-ph.GA},
       adsurl = {https://ui.adsabs.harvard.edu/abs/2026AJ....171...50H}
}

@ARTICLE{Louvet2014,
       author = {{Louvet}, F. and {Motte}, F. and {Hennebelle}, P. and {Maury}, A. and {Bonnell}, I. and {Bontemps}, S. and {Gusdorf}, A. and {Hill}, T. and {Gueth}, F. and {Peretto}, N. and {Duarte-Cabral}, A. and {Stephan}, G. and {Schilke}, P. and {Csengeri}, T. and {Nguyen Luong}, Q. and {Lis}, D.~C.},
        title = "{The W43-MM1 mini-starburst ridge, a test for star formation efficiency models}",
      journal = {\aap},
         year = 2014,
        month = oct,
       volume = {570},
          eid = {A15},
        pages = {A15},
          doi = {10.1051/0004-6361/201423603},
archivePrefix = {arXiv},
       eprint = {1404.4843},
 primaryClass = {astro-ph.SR},
       adsurl = {https://ui.adsabs.harvard.edu/abs/2014A&A...570A..15L}
}

@ARTICLE{Beuther2025_noreview,
       author = {{Beuther}, H. and {Olguin}, F.~A. and {Sanhueza}, P. and {Cunningham}, N. and {Ginsburg}, A.},
        title = "{Hierarchical accretion flow from the G351 infrared dark filament to its central cores}",
      journal = {\aap},
         year = 2025,
        month = mar,
       volume = {695},
          eid = {A51},
        pages = {A51},
          doi = {10.1051/0004-6361/202452754},
archivePrefix = {arXiv},
       eprint = {2502.13866},
 primaryClass = {astro-ph.GA},
       adsurl = {https://ui.adsabs.harvard.edu/abs/2025A&A...695A..51B}
}

@ARTICLE{Tritsis2026,
       author = {{Tritsis}, Aris},
        title = "{The mass-to-flux ratio in molecular clouds: What are we really measuring?}",
      journal = {\aap},
         year = 2026,
        month = feb,
       volume = {706},
          eid = {A60},
        pages = {A60},
          doi = {10.1051/0004-6361/202555979},
archivePrefix = {arXiv},
       eprint = {2512.07943},
 primaryClass = {astro-ph.GA},
       adsurl = {https://ui.adsabs.harvard.edu/abs/2026A&A...706A..60T}
}

@ARTICLE{Palau2021,
       author = {{Palau}, Aina and {Zhang}, Qizhou and {Girart}, Josep M. and {Liu}, Junhao and {Rao}, Ramprasad and {Koch}, Patrick M. and {Estalella}, Robert and {Chen}, Huei-Ru Vivien and {Liu}, Hauyu Baobab and {Qiu}, Keping and {Li}, Zhi-Yun and {Zapata}, Luis A. and {Bontemps}, Sylvain and {Ho}, Paul T.~P. and {Beuther}, Henrik and {Ching}, Tao-Chung and {Shinnaga}, Hiroko and {Ahmadi}, Aida},
        title = "{Does the Magnetic Field Suppress Fragmentation in Massive Dense Cores?}",
      journal = {\apj},
         year = 2021,
        month = may,
       volume = {912},
       number = {2},
          eid = {159},
        pages = {159},
          doi = {10.3847/1538-4357/abee1e},
archivePrefix = {arXiv},
       eprint = {2010.12099},
 primaryClass = {astro-ph.GA},
       adsurl = {https://ui.adsabs.harvard.edu/abs/2021ApJ...912..159P}
}

@ARTICLE{Polychronakis2025,
       author = {{Polychronakis}, A. and {Tritsis}, A. and {Skalidis}, R. and {Tassis}, K.},
        title = "{A three-step approach to reliably estimate magnetic field strengths in star-forming regions}",
      journal = {\aap},
         year = 2025,
        month = aug,
       volume = {700},
          eid = {A256},
        pages = {A256},
          doi = {10.1051/0004-6361/202553774},
archivePrefix = {arXiv},
       eprint = {2507.06297},
 primaryClass = {astro-ph.GA},
       adsurl = {https://ui.adsabs.harvard.edu/abs/2025A&A...700A.256P}
}

@ARTICLE{Huang2025,
       author = {{Huang}, Bo and {Girart}, Josep M. and {Stephens}, Ian W. and {Myers}, Philip C. and {Zhang}, Qizhou and {Cortes}, Paulo and {S{\'a}nchez-Monge}, {\'A}lvaro and {Fern{\'a}ndez L{\'o}pez}, Manuel and {Le Gouellec}, Valentin J.~M. and {Megeath}, Tom and {Murillo}, Nadia M. and {Carpenter}, John M. and {Li}, Zhi-Yun and {Liu}, Junhao and {Looney}, Leslie W. and {Sadavoy}, Sarah and {Karnath}, Nicole and {Kwon}, Woojin},
        title = "{Characterizing Magnetic Properties of Young Protostars in Orion}",
      journal = {\apj},
         year = 2025,
        month = may,
       volume = {984},
       number = {1},
          eid = {29},
        pages = {29},
          doi = {10.3847/1538-4357/adc30b},
archivePrefix = {arXiv},
       eprint = {2503.14726},
 primaryClass = {astro-ph.GA},
       adsurl = {https://ui.adsabs.harvard.edu/abs/2025ApJ...984...29H}
}

@ARTICLE{Falceta-Goncalves2008,
       author = {{Falceta-Gon{\c{c}}alves}, Diego and {Lazarian}, Alex and {Kowal}, Grzegorz},
        title = "{Studies of Regular and Random Magnetic Fields in the ISM: Statistics of Polarization Vectors and the Chandrasekhar-Fermi Technique}",
      journal = {\apj},
         year = 2008,
        month = may,
       volume = {679},
       number = {1},
        pages = {537-551},
          doi = {10.1086/587479},
archivePrefix = {arXiv},
       eprint = {0801.0279},
 primaryClass = {astro-ph},
       adsurl = {https://ui.adsabs.harvard.edu/abs/2008ApJ...679..537F}
}

@ARTICLE{Houde2009,
       author = {{Houde}, Martin and {Vaillancourt}, John E. and {Hildebrand}, Roger H. and {Chitsazzadeh}, Shadi and {Kirby}, Larry},
        title = "{Dispersion of Magnetic Fields in Molecular Clouds. II.}",
      journal = {\apj},
         year = 2009,
        month = dec,
       volume = {706},
       number = {2},
        pages = {1504-1516},
          doi = {10.1088/0004-637X/706/2/1504},
archivePrefix = {arXiv},
       eprint = {0909.5227},
 primaryClass = {astro-ph.GA},
       adsurl = {https://ui.adsabs.harvard.edu/abs/2009ApJ...706.1504H}
}

@ARTICLE{Skalidis_Tassis2021,
       author = {{Skalidis}, Raphael and {Tassis}, Konstantinos},
        title = "{High-accuracy estimation of magnetic field strength in the interstellar medium from dust polarization}",
      journal = {\aap},
         year = 2021,
        month = mar,
       volume = {647},
          eid = {A186},
        pages = {A186},
          doi = {10.1051/0004-6361/202039779},
archivePrefix = {arXiv},
       eprint = {2010.15141},
 primaryClass = {astro-ph.GA},
       adsurl = {https://ui.adsabs.harvard.edu/abs/2021A&A...647A.186S}
}

@ARTICLE{Tritsis2015,
       author = {{Tritsis}, A. and {Panopoulou}, G.~V. and {Mouschovias}, T. Ch. and {Tassis}, K. and {Pavlidou}, V.},
        title = "{Magnetic field-gas density relation and observational implications revisited}",
      journal = {\mnras},
         year = 2015,
        month = aug,
       volume = {451},
       number = {4},
        pages = {4384-4396},
          doi = {10.1093/mnras/stv1133},
archivePrefix = {arXiv},
       eprint = {1505.05508},
 primaryClass = {astro-ph.GA},
       adsurl = {https://ui.adsabs.harvard.edu/abs/2015MNRAS.451.4384T}
}

@ARTICLE{Vaillancourt2006,
       author = {{Vaillancourt}, John E.},
        title = "{Placing Confidence Limits on Polarization Measurements}",
      journal = {\pasp},
         year = 2006,
        month = sep,
       volume = {118},
       number = {847},
        pages = {1340-1343},
          doi = {10.1086/507472},
archivePrefix = {arXiv},
       eprint = {astro-ph/0603110},
 primaryClass = {astro-ph},
       adsurl = {https://ui.adsabs.harvard.edu/abs/2006PASP..118.1340V}
}

@ARTICLE{Padoan2011,
       author = {{Padoan}, Paolo and {Nordlund}, {\r{A}}ke},
        title = "{The Star Formation Rate of Supersonic Magnetohydrodynamic Turbulence}",
      journal = {\apj},
         year = 2011,
        month = mar,
       volume = {730},
       number = {1},
          eid = {40},
        pages = {40},
          doi = {10.1088/0004-637X/730/1/40},
archivePrefix = {arXiv},
       eprint = {0907.0248},
 primaryClass = {astro-ph.GA},
       adsurl = {https://ui.adsabs.harvard.edu/abs/2011ApJ...730...40P}
}

@ARTICLE{Sakai2022,
       author = {{Sakai}, Takeshi and {Sanhueza}, Patricio and {Furuya}, Kenji and {Tatematsu}, Ken'ichi and {Li}, Shanghuo and {Aikawa}, Yuri and {Lu}, Xing and {Zhang}, Qizhou and {Morii}, Kaho and {Nakamura}, Fumitaka and {Takemura}, Hideaki and {Izumi}, Natsuko and {Hirota}, Tomoya and {Silva}, Andrea and {Guzman}, Andres E. and {Sakai}, Nami and {Yamamoto}, Satoshi},
        title = "{The ALMA Survey of 70 {\ensuremath{\mu}}m Dark High-mass Clumps in Early Stages (ASHES). V. Deuterated Molecules in the 70 {\ensuremath{\mu}}m Dark IRDC G14.492-00.139}",
      journal = {\apj},
         year = 2022,
        month = feb,
       volume = {925},
       number = {2},
          eid = {144},
        pages = {144},
          doi = {10.3847/1538-4357/ac3d2e},
archivePrefix = {arXiv},
       eprint = {2111.13325},
 primaryClass = {astro-ph.GA},
       adsurl = {https://ui.adsabs.harvard.edu/abs/2022ApJ...925..144S}
}

@ARTICLE{Myers2018,
       author = {{Myers}, Philip C. and {Basu}, Shantanu and {Auddy}, Sayantan},
        title = "{Magnetic Field Structure in Spheroidal Star-forming Clouds}",
      journal = {\apj},
         year = 2018,
        month = nov,
       volume = {868},
       number = {1},
          eid = {51},
        pages = {51},
          doi = {10.3847/1538-4357/aae695},
archivePrefix = {arXiv},
       eprint = {1810.04132},
 primaryClass = {astro-ph.GA},
       adsurl = {https://ui.adsabs.harvard.edu/abs/2018ApJ...868...51M}
}

@ARTICLE{Skalidis2021,
       author = {{Skalidis}, R. and {Sternberg}, J. and {Beattie}, J.~R. and {Pavlidou}, V. and {Tassis}, K.},
        title = "{Why take the square root? An assessment of interstellar magnetic field strength estimation methods}",
      journal = {\aap},
         year = 2021,
        month = dec,
       volume = {656},
          eid = {A118},
        pages = {A118},
          doi = {10.1051/0004-6361/202142045},
archivePrefix = {arXiv},
       eprint = {2109.10925},
 primaryClass = {astro-ph.GA},
       adsurl = {https://ui.adsabs.harvard.edu/abs/2021A&A...656A.118S}
}

\begin{appendix}
\section{Test on the incompleteness sampling of polarization semi-vectors}

Here we present two tests assessing the impact of the number of retrieved semi-vectors within the three-beam area on the dispersion of polarization angles.
These tests are performed for cores \#1 and \#14, both of which contain 45 semi-vectors detected above $3\,\sigma_{\mathrm{P}}$ at this scale.
As discussed in Sect.\,\ref{subsec:mag_field_estimate}, reducing the number of semi-vectors has only a minor impact on the magnetic field strength estimation, leading to variations of less than about 10\,\% in the computed angle dispersion.

\begin{figure}[h]
\centering
\includegraphics[width=0.49\textwidth]{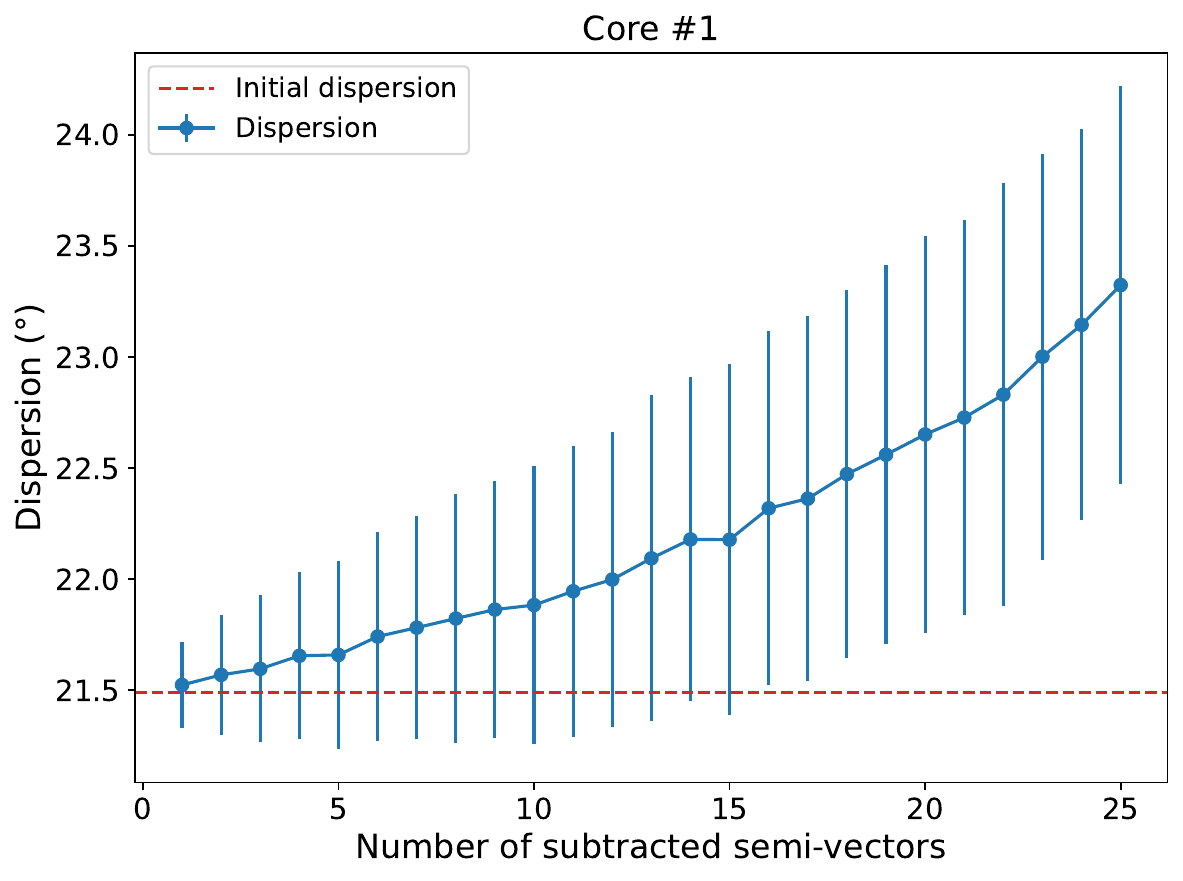}
\includegraphics[width=0.49\textwidth]{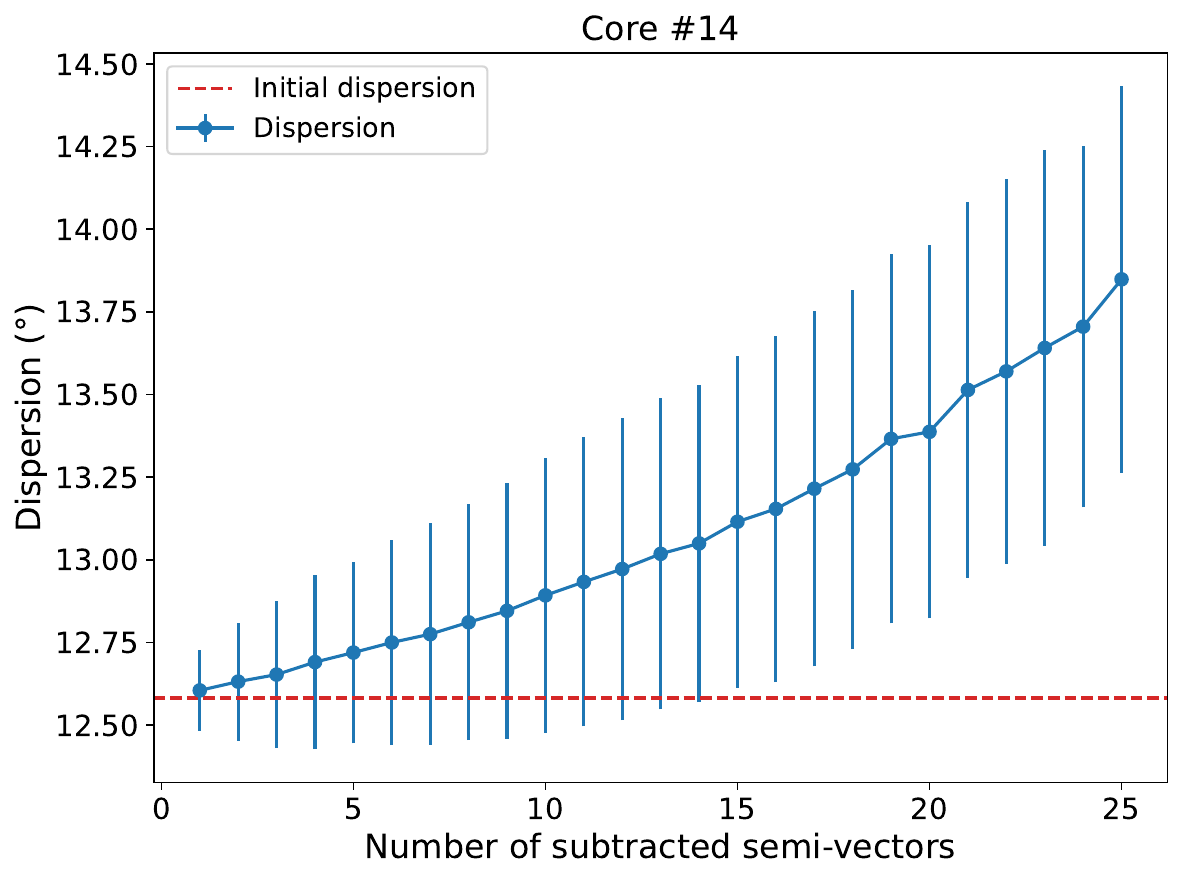}
\caption{Test of the incompleteness sampling of polarization semi-vectors for cores \#1 and \#14.}
\label{fig:app_sampling_effect}  

\end{figure}

\section{Core properties}

We present here Table\,\ref{tab:Core_properties} compiling all core properties, including their identification number, classification, mass, size, density, velocity dispersion at both the core and three-beam scales, dispersion of polarization angles, and magnetic field strengths measured at the three-beam scale and extrapolated to the core-scale.

\begin{table*} [htbp]
    \centering     
    \begin{threeparttable}[c]
    \caption{Properties of the sample of cores studied in W43-MM1.}
    
    \begin{tabular}{llcccccccccl}
    \hline \noalign {\smallskip}
     \col{Core ID\tnote{1}} &  \col{Class\tnote{1}} & \col{Mass\tnote{2}}  &  \col{Size\tnote{2}}  & \col{Density\tnote{2}} & \col{$\sigma_{\mathrm{v}}^{\mathrm{core}}$} & \col{$\sigma_{\mathrm{v}}^{\mathrm{3beam}}$} & \col{$\sigma_{\theta}$}  & \col{Semi-vectors} & \col{$\mathrm{B_{3beam}}$} & \col{$\mathrm{B_{core}}$\tnote{3}}\\ \noalign {\smallskip}
        &  &  \col{$\left[M_\odot \right]$}  & \col{$\left[\mathrm{au} \right]$}  & \col{$\left[\times 10^6 \rm \, cm^{-3} \right]$}  &  \col{$\left[\mathrm{km\,s^{-1}}\right]$} & \col{$\left[\mathrm{km\,s^{-1}}\right]$} & \col{$\left[ ^{\circ}\right]$} & \col{number} & \col{$\left[\mathrm{mG} \right]$}  & \col{$\left[\mathrm{mG} \right]$} {\smallskip}\\  
    \hline \noalign {\smallskip}
1 & Proto & $115.2 \pm 49.6$ & 2520 & $218.0 \pm 93.8$ & 4.48 & 3.12 & 21.5 & 45 & $17.6 \pm {0.5}$ & $48.6 \pm {7.6}$ \\
2 & Proto & $50.8 \pm 2.6$ & 1830 & $251.0 \pm 12.8$ & 1.76 & 1.93 & 18.2 & 41 & $9.6 \pm {0.2}$ & $39.0 \pm {8.3}$ \\
3 & Proto & $49.1 \pm 3.0$ & 2520 & $92.9 \pm 5.7$ & 1.61 & 1.62 & 41.1 & 20 & $2.8 \pm {0.1}$ & $8.7 \pm {1.5}$ \\
4 & Proto & $54.4 \pm 2.4$ & 3900 & $27.8 \pm 1.2$ & 3.15 & 2.98 & 27.7 & 43 & $12.3 \pm {0.3}$ & $12.2 \pm {0.3}$ \\
5 & Proto & $11.6 \pm 0.7$ & 1850 & $55.5 \pm 3.4$ & 2.91 & 1.80 & 35.7 & 23 & $3.2 \pm {0.1}$ & $8.5 \pm {1.3}$ \\
6 & Pre & $45.2 \pm 1.1$ & 1930 & $190.0 \pm 4.6$ & 1.16 & - & - & - & - & - \\
7 & Proto & $9.8 \pm 0.5$ & 1630 & $68.5 \pm 3.5$ & 0.60 & - & - & - & - & - \\
$8^{\dagger}$ & Proto & $11.5 \pm 0.6$ & 1920 & $49.2 \pm 2.6$ & 0.86 & 1.16 & 23.5 & 35 & $2.4 \pm {0.1}$ & $7.9 \pm {1.4}$ \\
9 & Proto & $6.9 \pm 0.6$ & 1710 & $41.8 \pm 3.6$ & 1.48 & 1.34 & 32.6 & 25 & $2.6 \pm {0.1}$ & $6.0 \pm {0.8}$ \\
$10^{\dagger}$ & Proto & $7.5 \pm 0.8$ & 2670 & $11.9 \pm 1.3$ & 2.31 & 1.77 & 16.7 & 45 & $9.5 \pm {0.2}$ & $7.8 \pm {0.3}$ \\
11 & Proto & $2.4 \pm 0.2$ & 1510 & $21.1 \pm 1.8$ & 2.41 & 1.06 & 20.3 & 29 & $2.3 \pm {0.1}$ & $5.4 \pm {0.7}$ \\
12 & Proto & $4.8 \pm 0.6$ & 1820 & $24.1 \pm 3.0$ & 0.50 & 1.53 & 13.7 & 29 & $7.1 \pm {0.2}$ & $12.1 \pm {1.0}$ \\
$13^{\dagger}$ & Proto & $3.5 \pm 0.7$ & 1160 & $67.9 \pm 13.6$ & 1.29 & 1.76 & 15.3 & 43 & $10.2 \pm {0.3}$ & $21.2 \pm {2.4}$ \\
$14^{*}$ & Pre & $11.1 \pm 1.4$ & 1170 & $210.0 \pm 26.5$ & 0.50 & 1.44 & 12.6 & 45 & $8.3 \pm {0.2}$ & $38.7 \pm {9.0}$ \\
15 & Proto & $3.8 \pm 0.3$ & 2070 & $13.0 \pm 1.0$ & 0.83 & - & - & - & - & - \\
16 & Proto & $12.1 \pm 1.7$ & 2680 & $19.0 \pm 2.7$ & 1.04 & 1.52 & 8.9 & 45 & $12.1 \pm {0.3}$ & $16.2 \pm {0.8}$ \\
17 & Pre & $18.2 \pm 1.7$ & 1820 & $91.4 \pm 8.5$ & 1.01 & 1.73 & 12.1 & 45 & $15.2 \pm {0.4}$ & $30.4 \pm {3.3}$ \\
18 & Proto & $4.6 \pm 0.4$ & 2100 & $15.0 \pm 1.3$ & 0.57 & - & - & - & - & - \\
$19^{\dagger}$ & Proto & $3.5 \pm 0.8$ & 1410 & $37.8 \pm 8.6$ & 1.16 & 1.77 & 8.6 & 29 & $11.7 \pm {0.3}$ & $28.4 \pm {3.9}$ \\
20 & Pre & $10.3 \pm 0.3$ & 1410 & $111.0 \pm 3.2$ & 0.46 & - & - & - & - & - \\
21 & Pre & $12.2 \pm 0.5$ & 1380 & $141.0 \pm 5.8$ & - & - & - & - & - & - \\
$22^{\dagger}$ & Proto & $2.4 \pm 0.5$ & 1160 & $46.6 \pm 9.7$ & 0.87 & 0.63 & 4.1 & 32 & $7.4 \pm {0.2}$ & $23.8 \pm {4.2}$ \\
23 & Proto & $3.8 \pm 0.6$ & 1660 & $25.2 \pm 4.0$ & 0.35 & - & - & - & - & - \\
24 & Pre & $6.4 \pm 1.2$ & 1450 & $63.6 \pm 11.9$ & 0.71 & 1.45 & 5.4 & 42 & $15.4 \pm {0.4}$ & $49.3 \pm {8.8}$ \\
26 & Proto & $2.9 \pm 0.3$ & 1920 & $12.4 \pm 1.3$ & - & - & - & - & - & - \\
$28^{\dagger}$ & Pre & $2.4 \pm 0.4$ & 1200 & $42.1 \pm 7.0$ & 0.80 & - & - & - & - & - \\
$29^{\dagger}$ & Proto & $1.3 \pm 0.1$ & 1410 & $14.0 \pm 1.1$ & 1.05 & - & - & - & - & - \\
31 & Proto & $1.2 \pm 0.1$ & 1460 & $11.7 \pm 1.0$ & - & - & - & - & - & - \\
$32^{\dagger}$ & Pre & $4.7 \pm 0.4$ & 1490 & $43.0 \pm 3.7$ & 0.62 & - & - & - & - & - \\
$34^{\dagger}$ & Pre & $3.8 \pm 0.3$ & 1670 & $24.7 \pm 1.9$ & 1.24 & - & - & - & - & - \\
$36^{\dagger}$ & Proto & $1.1 \pm 0.2$ & 1470 & $10.5 \pm 1.9$ & 0.54 & - & - & - & - & - \\
39 & Proto & $0.7 \pm 0.1$ & 1660 & $4.6 \pm 0.7$ & - & - & - & - & - & - \\
$40^{\dagger}$ & Pre & $4.4 \pm 0.3$ & 2140 & $13.6 \pm 0.9$ & 0.64 & - & - & - & - & - \\
44 & Proto & $1.2 \pm 0.2$ & 2080 & $4.0 \pm 0.7$ & 0.38 & - & - & - & - & - \\
46 & Pre & $2.6 \pm 0.3$ & 2030 & $9.4 \pm 1.1$ & 0.82 & - & - & - & - & - \\
49 & Proto & $1.3 \pm 0.2$ & 1450 & $12.9 \pm 2.0$ & 0.65 & - & - & - & - & - \\
$51^{\dagger}$ & Proto & $2.8 \pm 0.3$ & 3730 & $1.6 \pm 0.2$ & 0.55 & - & - & - & - & - \\
54 & Pre & $2.1 \pm 0.3$ & 1510 & $18.5 \pm 2.6$ & 0.39 & - & - & - & - & - \\
59 & Proto & $0.5 \pm 0.2$ & 1550 & $4.1 \pm 1.6$ & 0.34 & - & - & - & - & - \\
64 & Pre & $3.0 \pm 0.4$ & 1430 & $31.1 \pm 4.1$ & - & - & - & - & - & - \\
67 & Proto & $0.7 \pm 0.1$ & 1900 & $3.1 \pm 0.4$ & - & - & - & - & - & - \\
$73^{\dagger}$ & Pre & $2.0 \pm 0.3$ & 2570 & $3.6 \pm 0.5$ & 0.54 & - & - & - & - & - \\
$74^{\dagger}$ & Pre & $2.0 \pm 0.3$ & 1700 & $12.3 \pm 1.9$ & 0.57 & - & - & - & - & - \\
99 & Pre & $2.2 \pm 0.2$ & 3930 & $1.1 \pm 0.1$ & - & - & - & - & - & - \\
$134^{\dagger}$ & Pre & $21.1 \pm 2.0$ & 2480 & $41.9 \pm 4.0$ & 1.93 & 1.57 & 18.4 & 44 & $8.7 \pm {0.2}$ & $12.1 \pm {0.7}$ \\
$136^{\dagger}$ & Pre & $12.2 \pm 1.9$ & 1850 & $58.3 \pm 9.1$ & 2.30 & 3.04 & 12.3 & 45 & $22.5 \pm {0.6}$ & $42.0 \pm {4.1}$ \\
$141^{\dagger}$ & Pre & $2.2 \pm 0.3$ & 3250 & $1.9 \pm 0.3$ & 0.39 & - & - & - & - & - \\
143 & Pre & $1.6 \pm 0.2$ & 5640 & $0.3 \pm 0.0$ & - & - & - & - & - & - \\
146 & Pre & $1.8 \pm 0.3$ & 3560 & $1.2 \pm 0.2$ & 0.74 & 0.81 & 17.8 & 26 & $1.1 \pm {0.1}$ & $1.1 \pm {0.1}$ \\
148 & Pre & $2.4 \pm 0.5$ & 2930 & $2.9 \pm 0.6$ & 0.61 & - & - & - & - & - \\
151 & Pre & $1.5 \pm 0.2$ & 3510 & $1.1 \pm 0.1$ & 0.55 & - & - & - & - & - \\
$154^{\dagger}$ & Pre & $1.9 \pm 0.3$ & 3300 & $1.6 \pm 0.3$ & 0.48 & - & - & - & - & - \\
155 & Pre & $1.8 \pm 0.3$ & 4500 & $0.6 \pm 0.1$ & - & - & - & - & - & - \\
162 & Pre & $2.0 \pm 0.4$ & 2750 & $2.9 \pm 0.6$ & - & - & 8.7 & 27 & - & - \\
174 & Proto & $0.3 \pm 0.1$ & 3090 & $0.3 \pm 0.1$ & - & - & - & - & - & - \\
177 & Pre & $6.2 \pm 1.5$ & 2140 & $19.2 \pm 4.6$ & 1.52 & 1.69 & 11.3 & 45 & $11.4 \pm {0.3}$ & $14.2 \pm {0.6}$ \\
    \hline
    \end{tabular}
    \label{tab:Core_properties}
\begin{tablenotes}
\item[1] The cores ID and the classification of the cores are the one published in \cite{Nony2020} and \cite{Nony2023}. 
\item[2] The masses and sizes are extracted from \cite{Motte2025} and the densities are derived assuming a spherical symmetry.
\item[3] The uncertainties are derived following a mean uncertainty of 0.08 on the density power-law index of the density-field relation.
\item[*] This core was classified as protostellar in \cite{Nony2020} and as prestellar in \cite{Valeille-Manet2025} (see text).
\item[$\dagger$] These cores have their core velocity dispersion measured with their On spectra.

\end{tablenotes}    
\end{threeparttable}
\end{table*}

\section{Examples of Gaussian fits and moment maps} \label{app:fits+mom_maps}
We present here the Gaussian fits and moment maps for cores \#1, \#3, \#13 and \#17 of the line tracers used to estimate the velocity dispersions at the core and three-beam scales. In the moment-1 maps, the dark-green region shows the mask used to extract the three-beam spectra and densities. 

\begin{figure*}[h]
\centering
\includegraphics[width=0.49\textwidth]{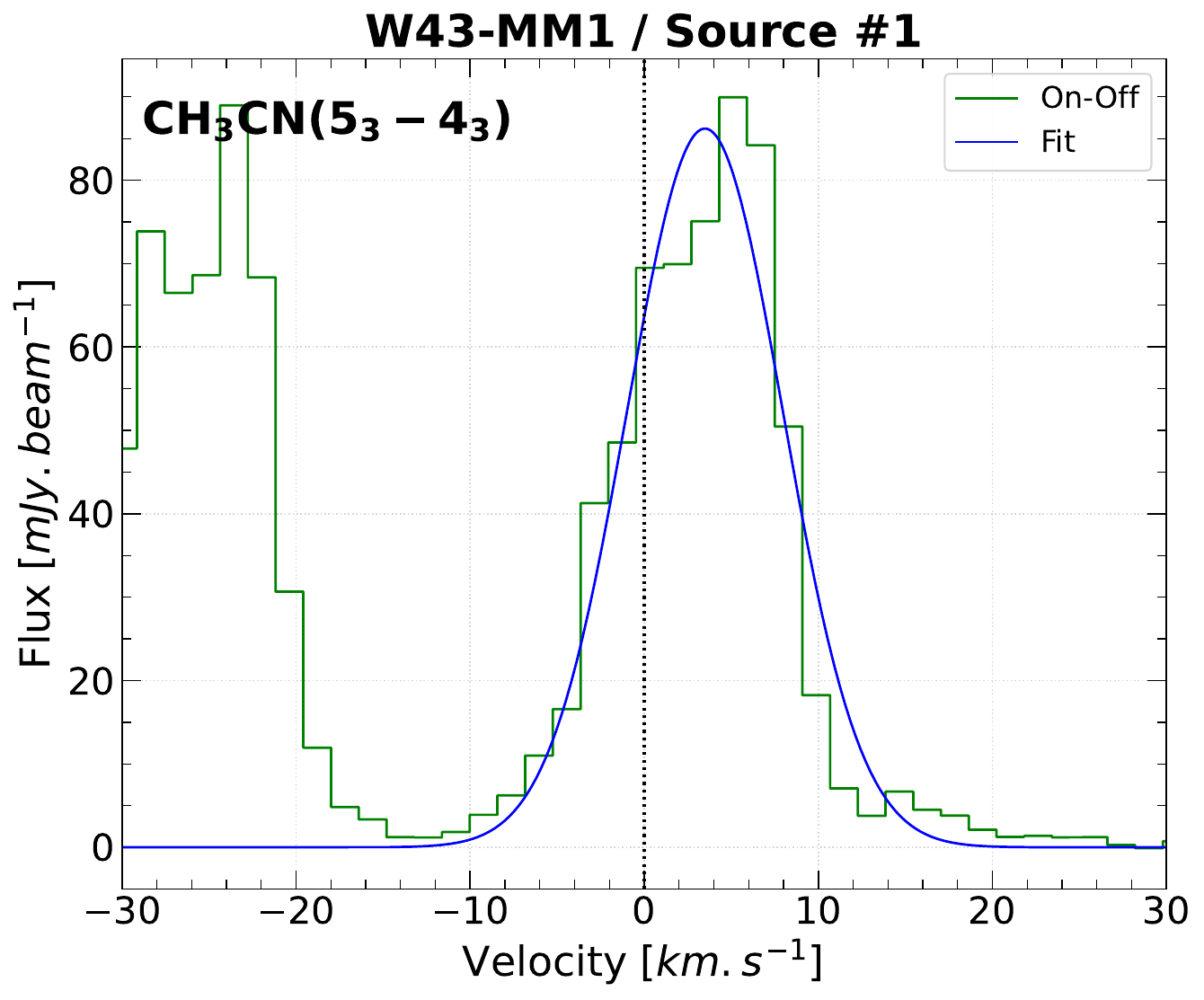}
\includegraphics[width=0.49\textwidth]{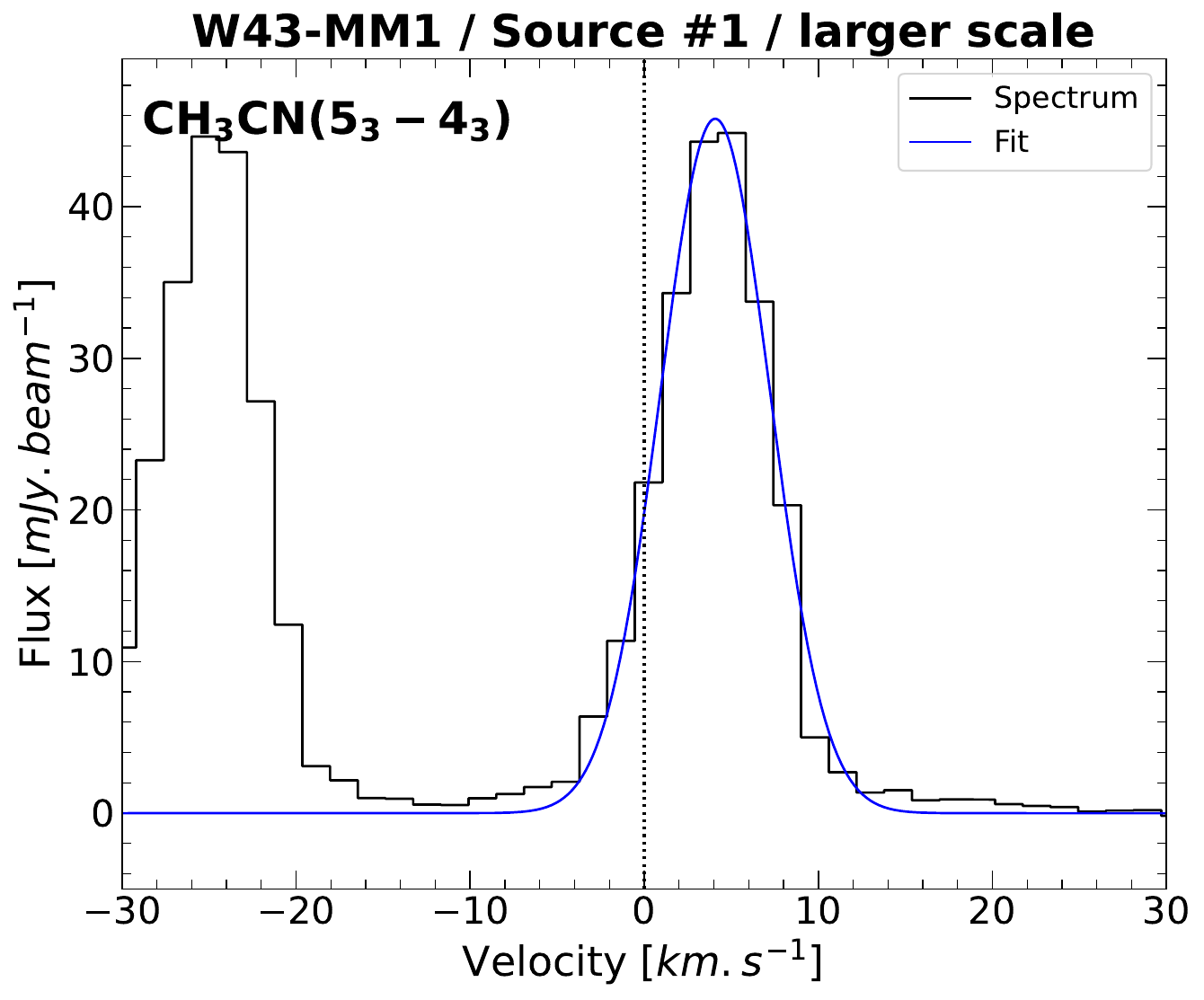}
\includegraphics[width=0.49\textwidth]{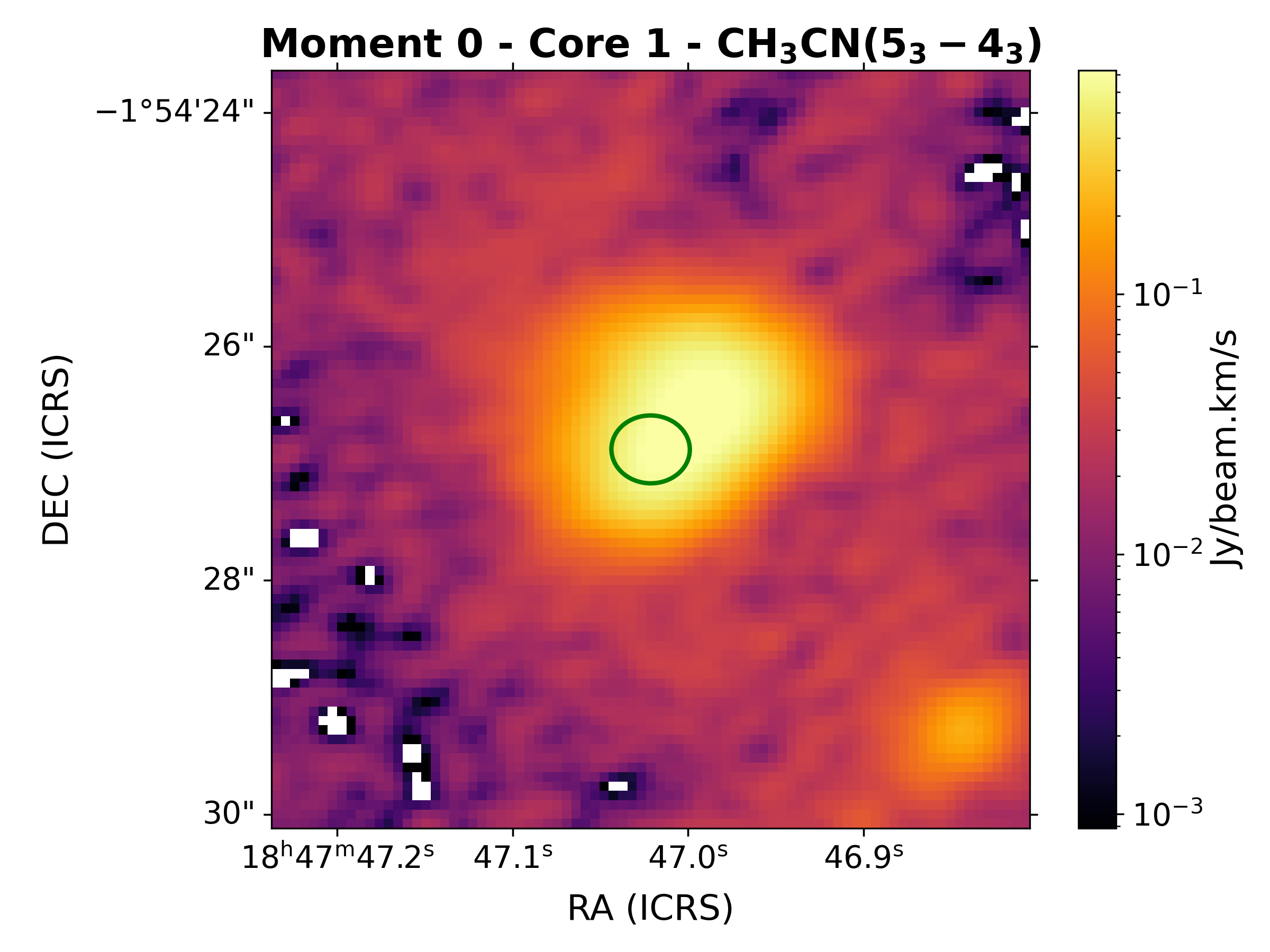}
\includegraphics[width=0.49\textwidth]{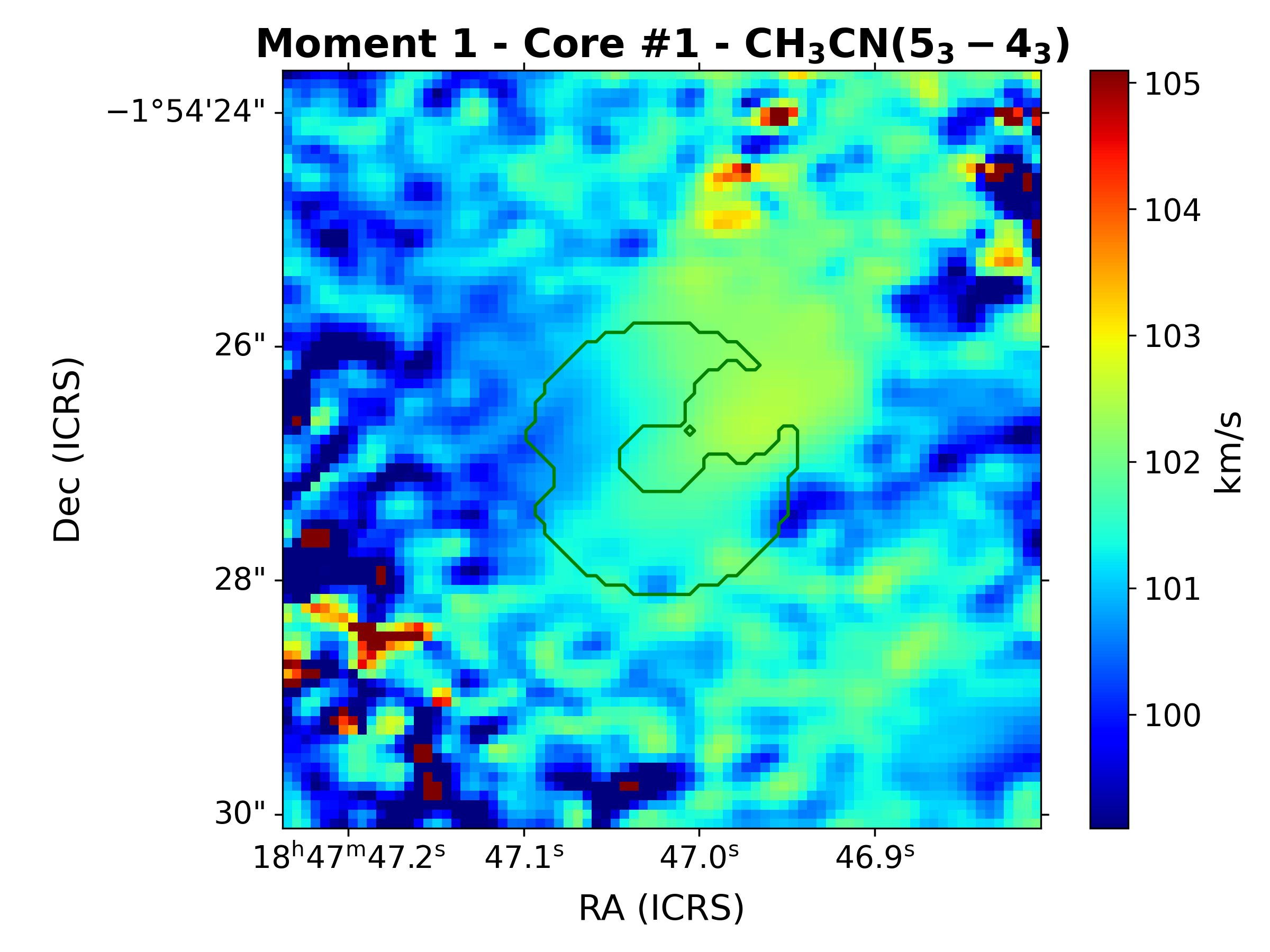}
\caption{Same as Fig.\,\ref{fig:fits_maps_core2} for core \#1.}
\label{fig:fits_maps_core1}  
\end{figure*}

\begin{figure*}[h]
\centering
\includegraphics[width=0.49\textwidth]{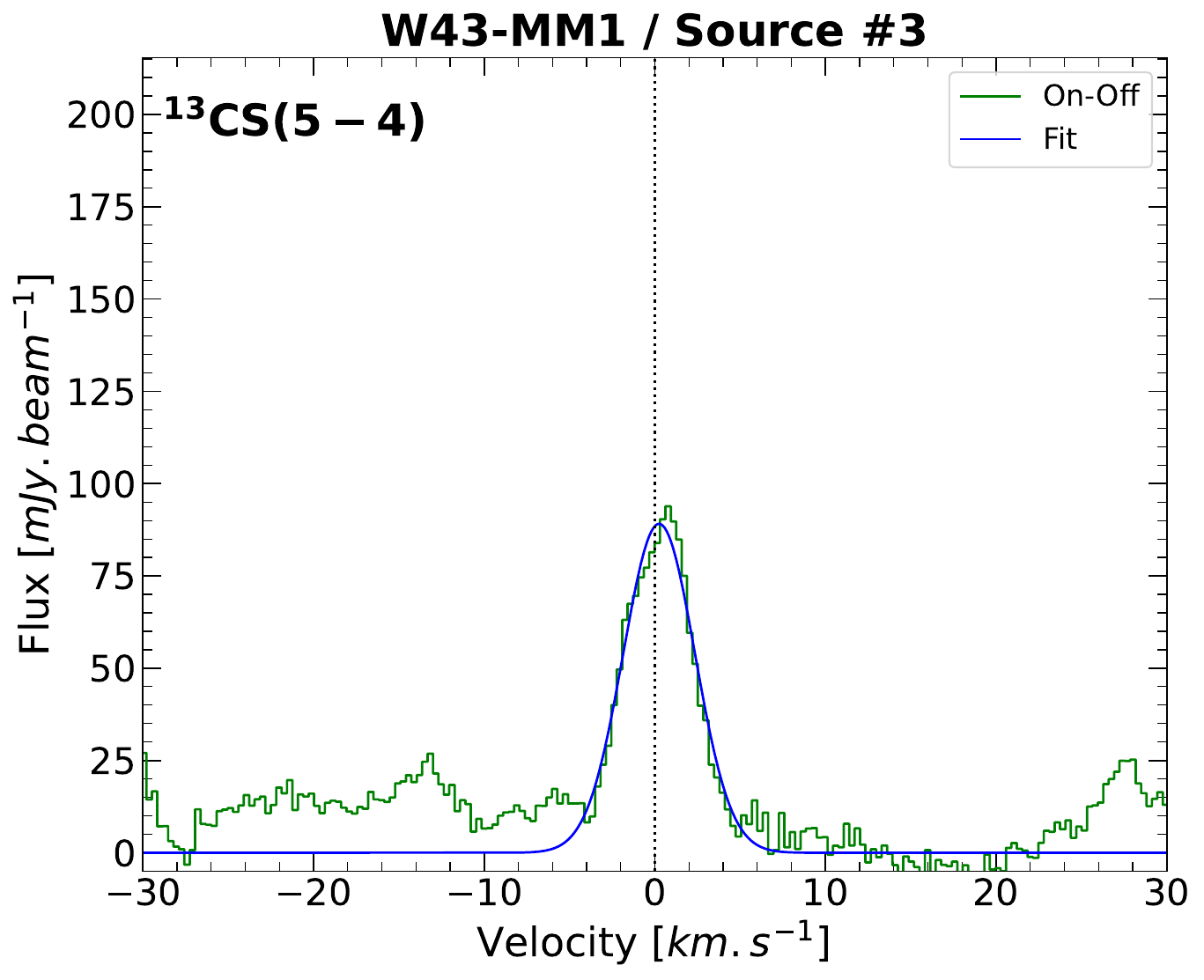}
\includegraphics[width=0.49\textwidth]{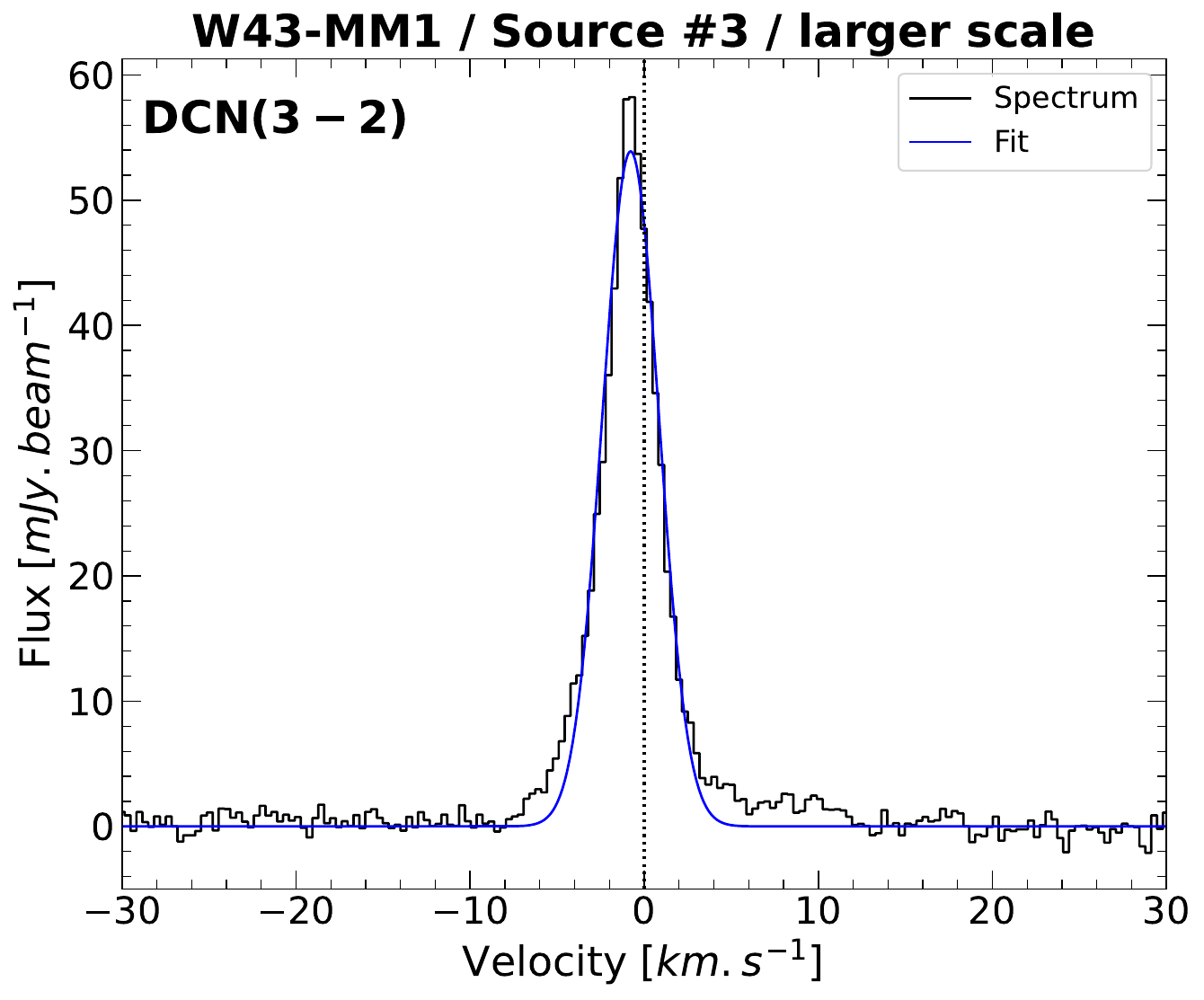}
\includegraphics[width=0.49\textwidth]{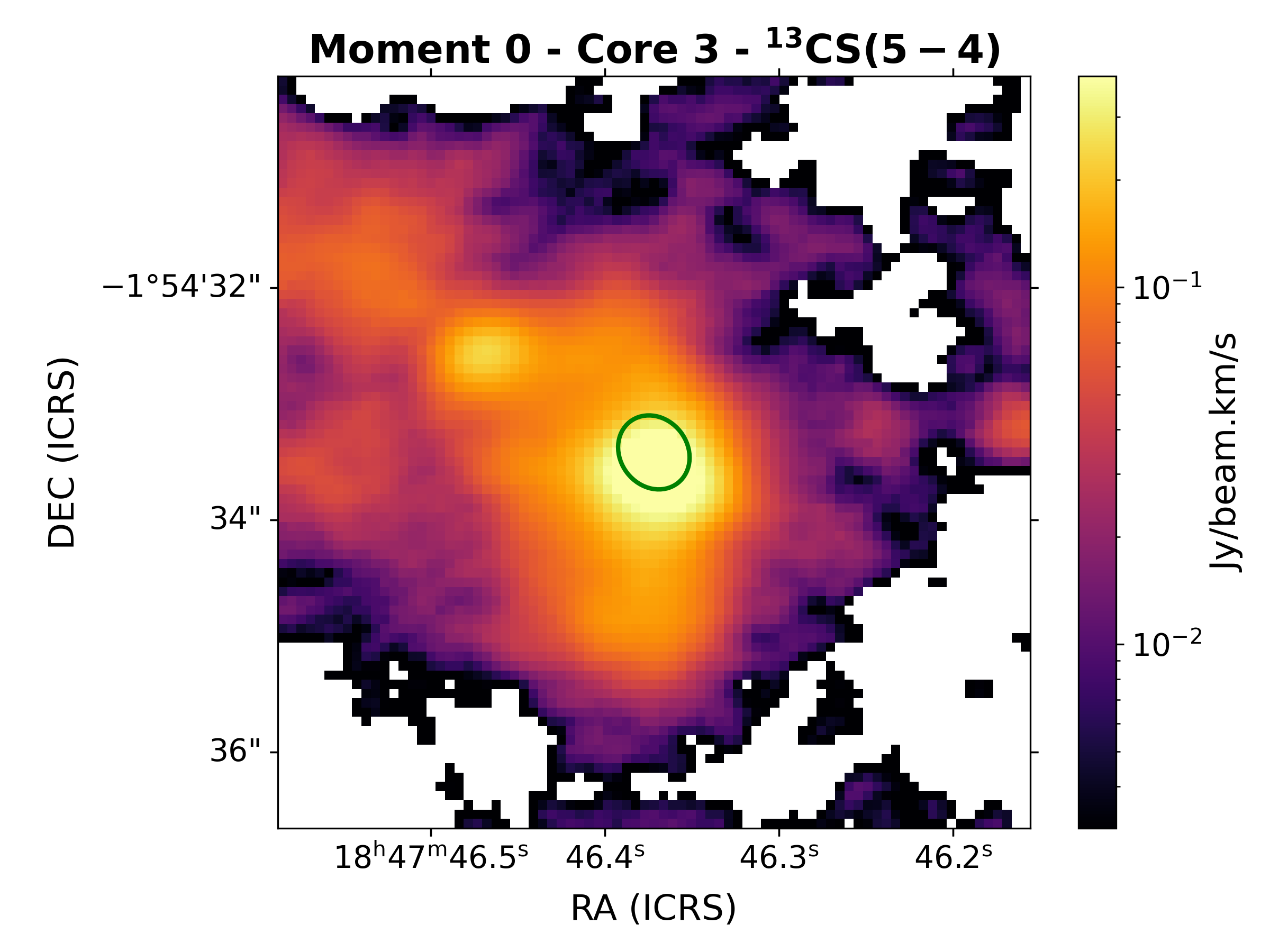}
\includegraphics[width=0.49\textwidth]{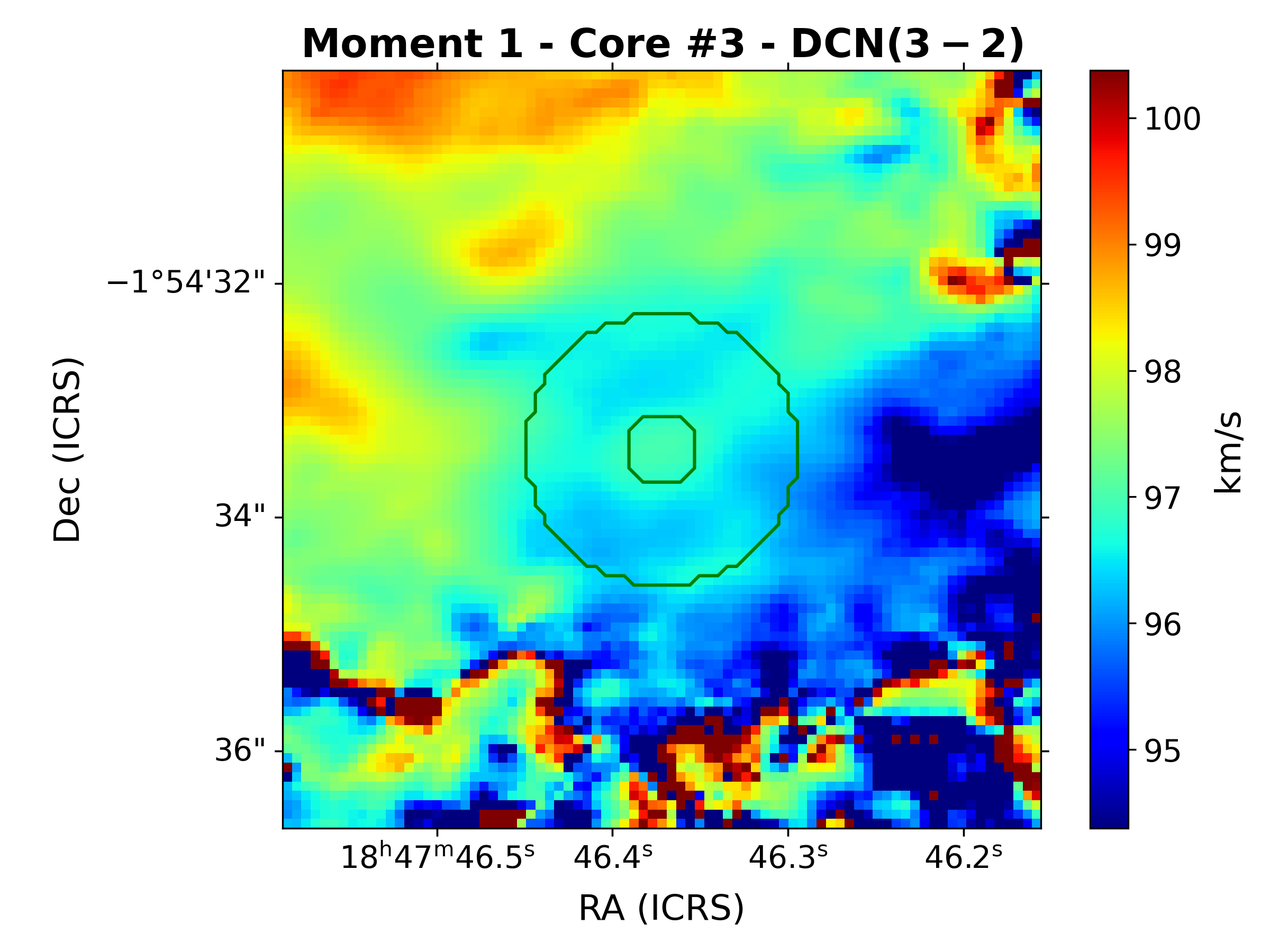}
\caption{Same as Fig.\,\ref{fig:fits_maps_core2} for core \#3.}
\label{fig:fits_maps_core3}  
\end{figure*}

\begin{figure*}[h]
\centering
\includegraphics[width=0.49\textwidth]{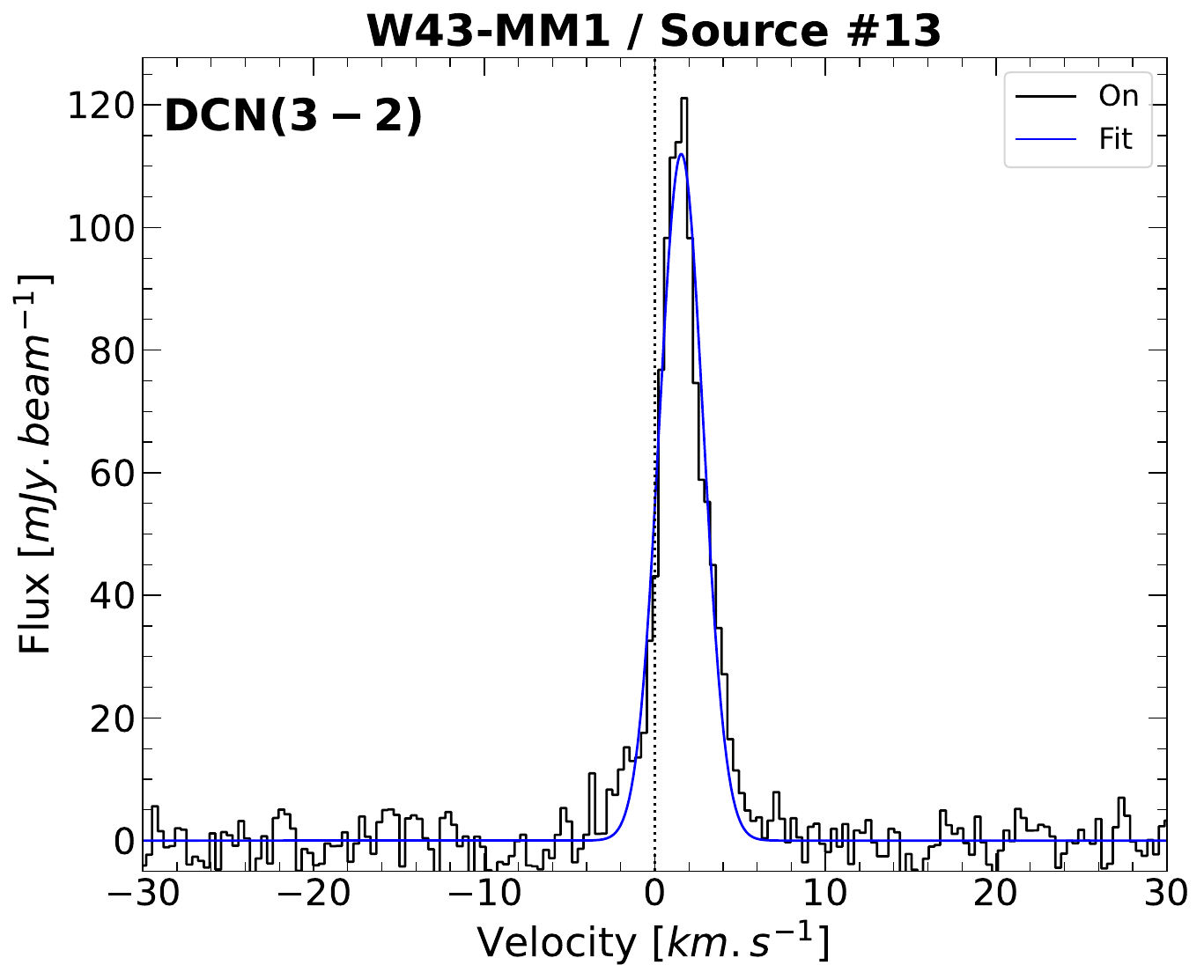}
\includegraphics[width=0.49\textwidth]{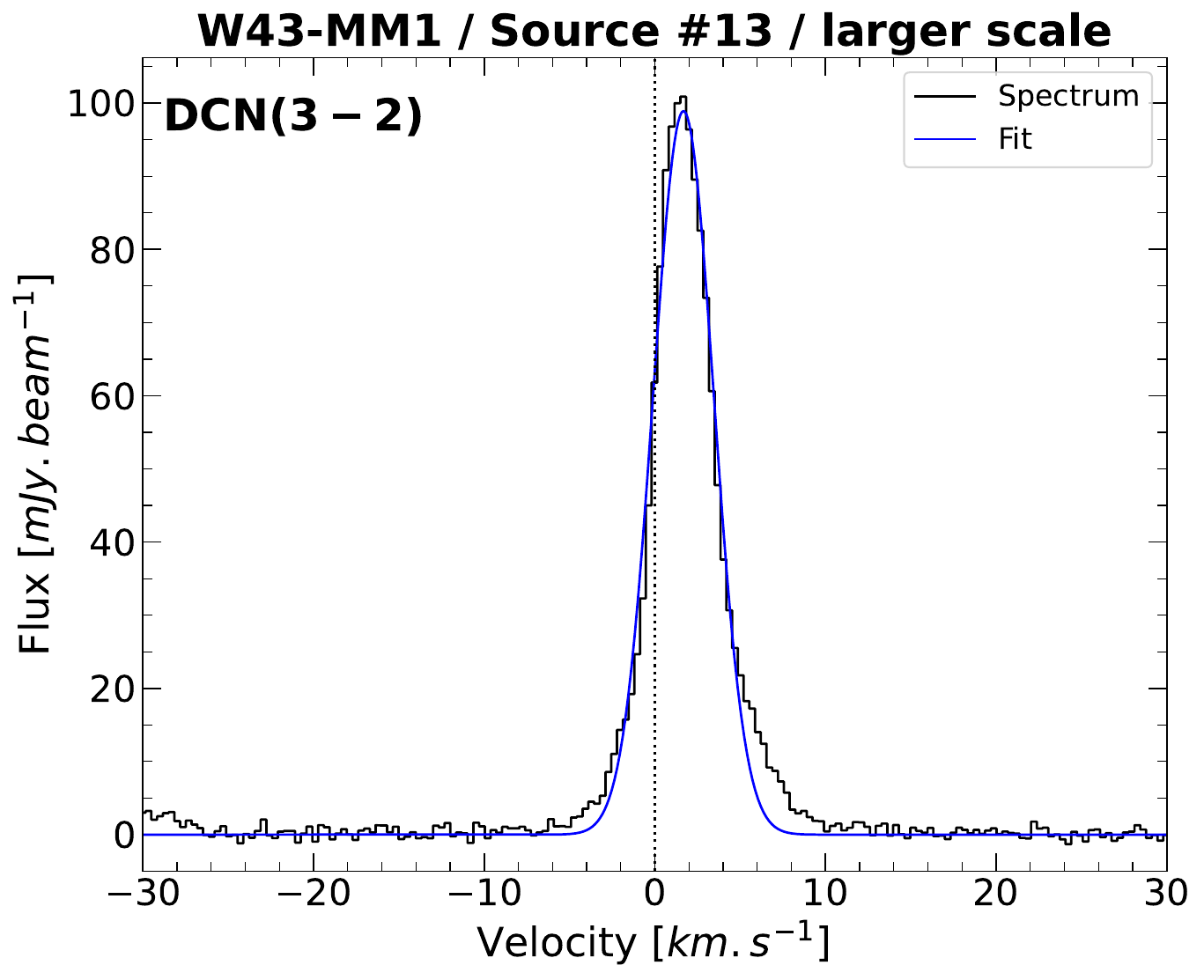}
\includegraphics[width=0.49\textwidth]{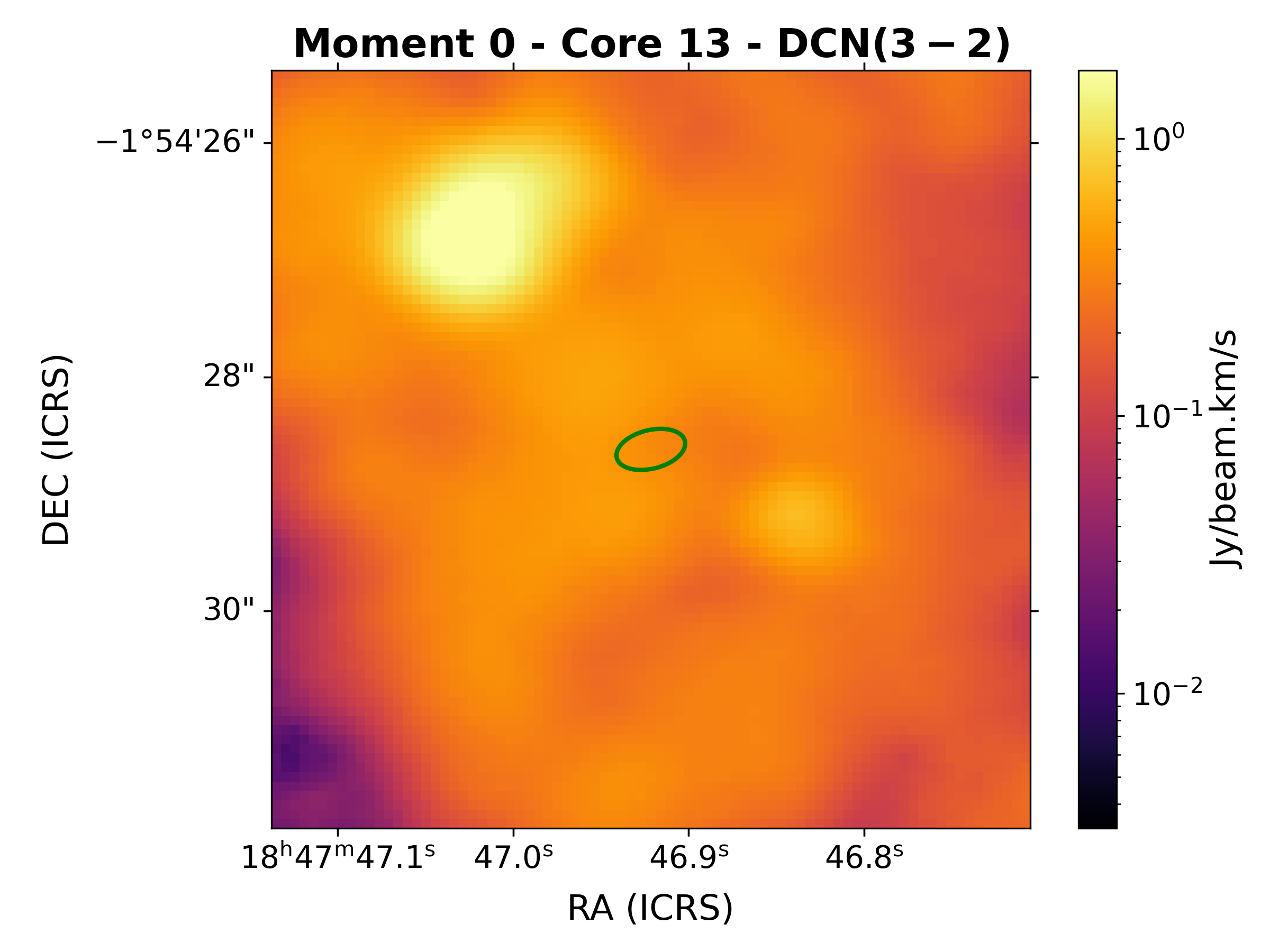}
\includegraphics[width=0.49\textwidth]{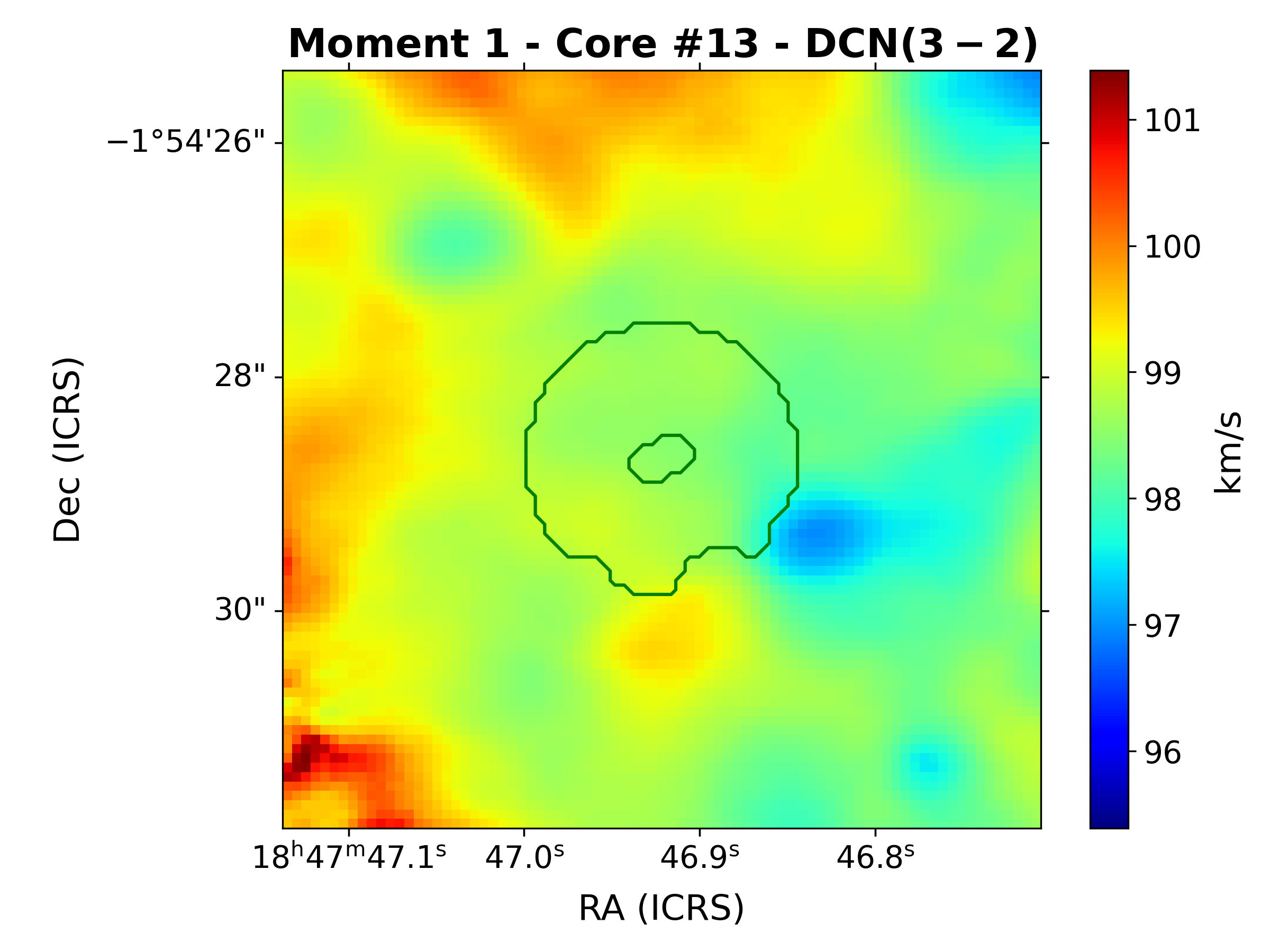}
\caption{Same as Fig.\,\ref{fig:fits_maps_core2} for core \#13.}
\label{fig:fits_maps_core3}  
\end{figure*}

\begin{figure*}[h]
\centering
\includegraphics[width=0.49\textwidth]{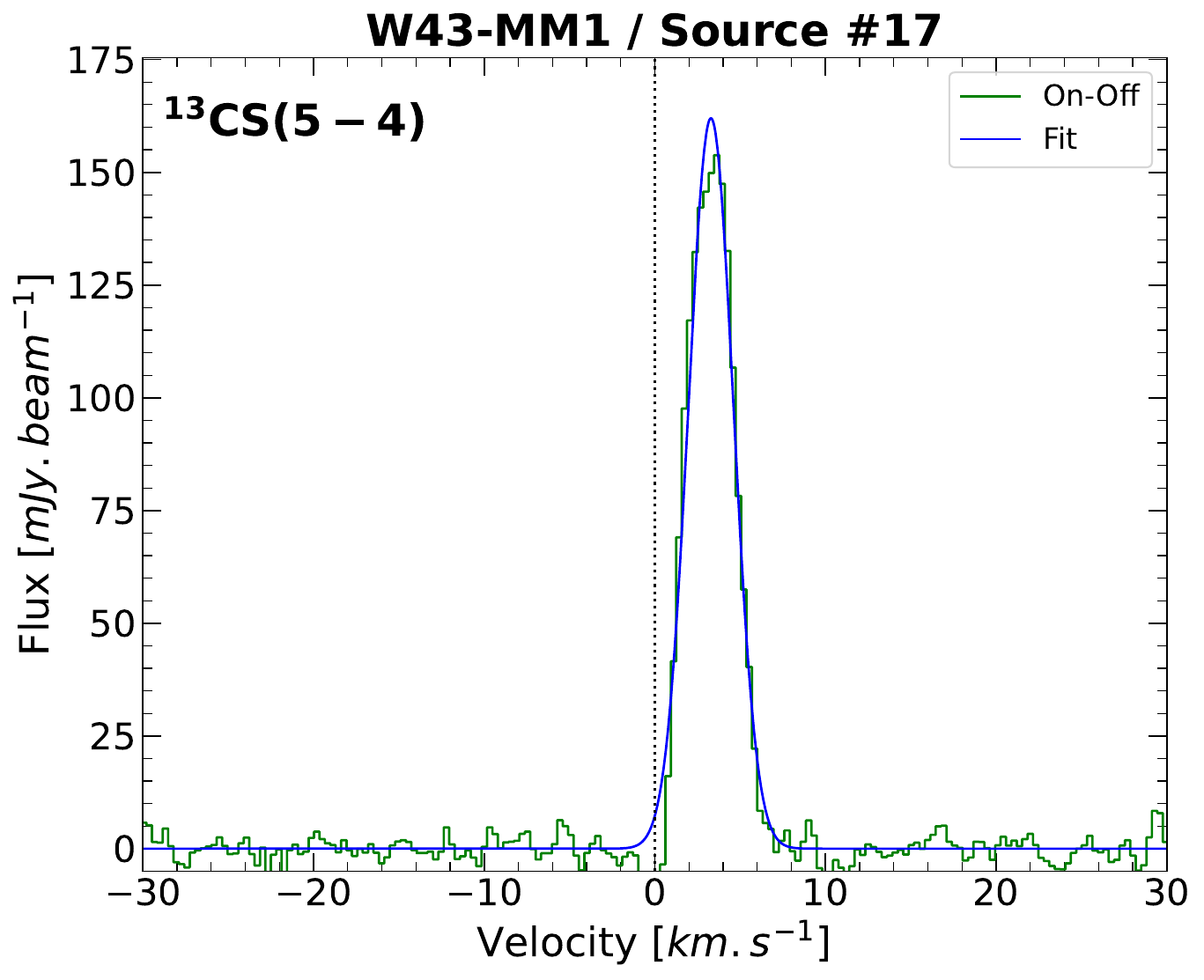}
\includegraphics[width=0.49\textwidth]{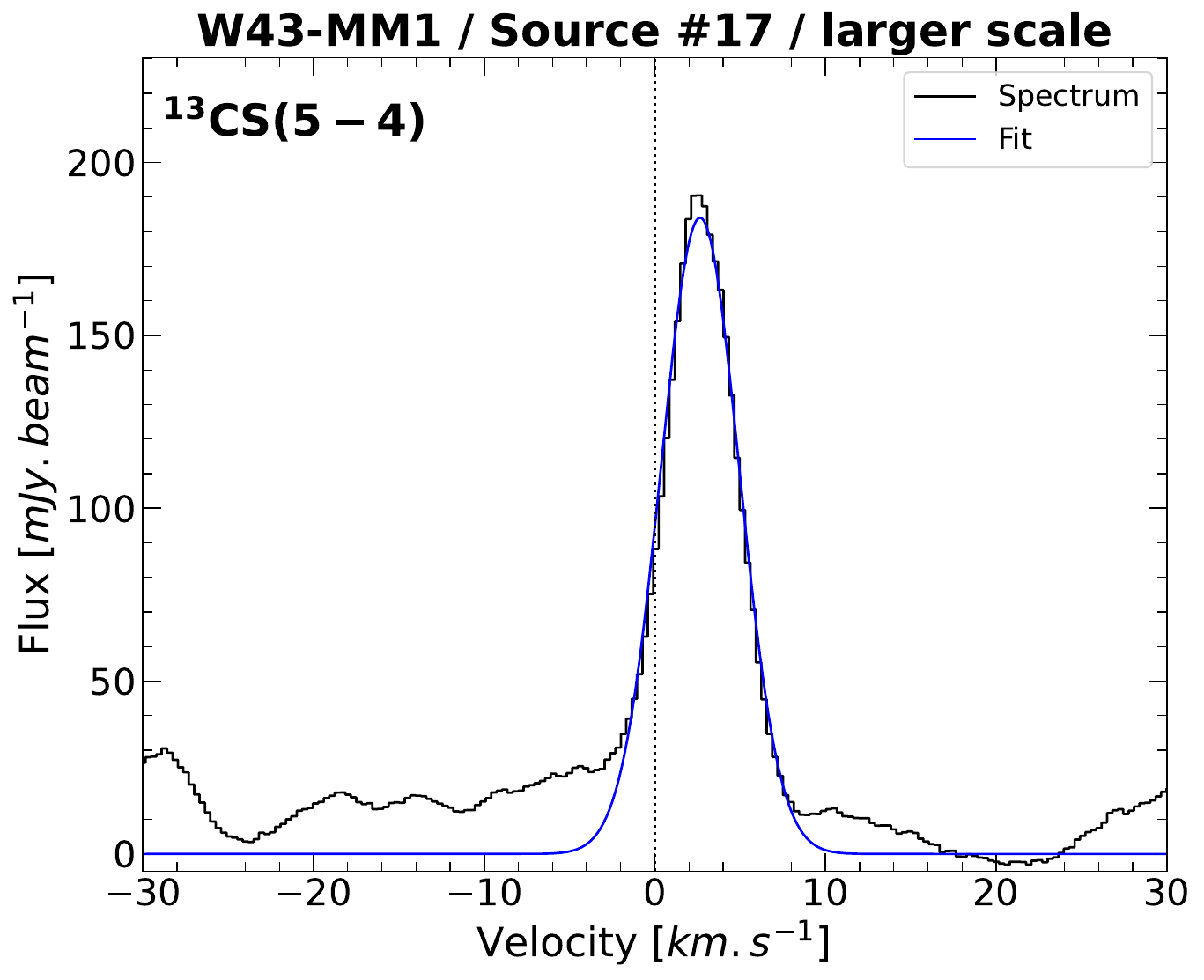}
\includegraphics[width=0.49\textwidth]{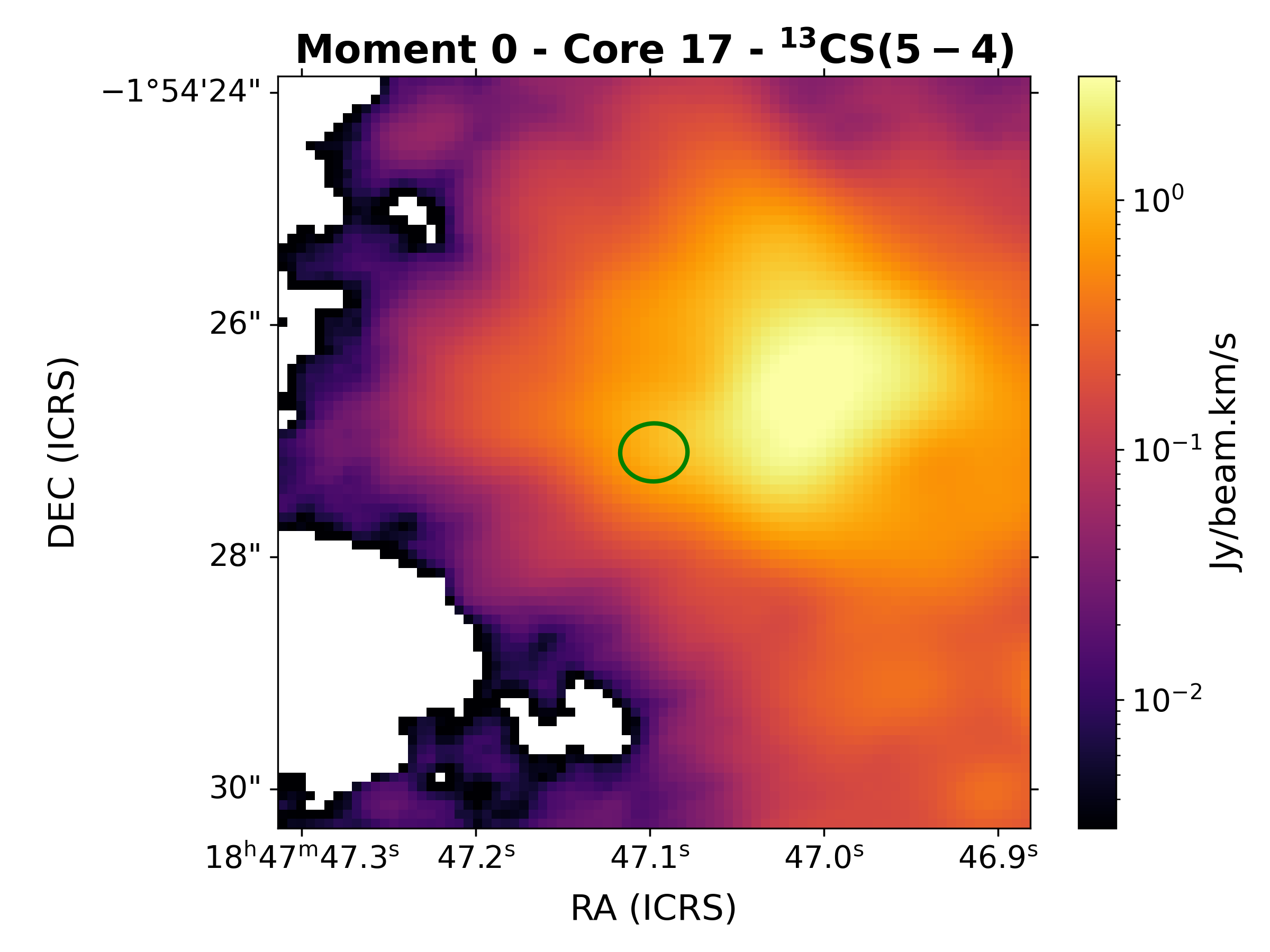}
\includegraphics[width=0.49\textwidth]{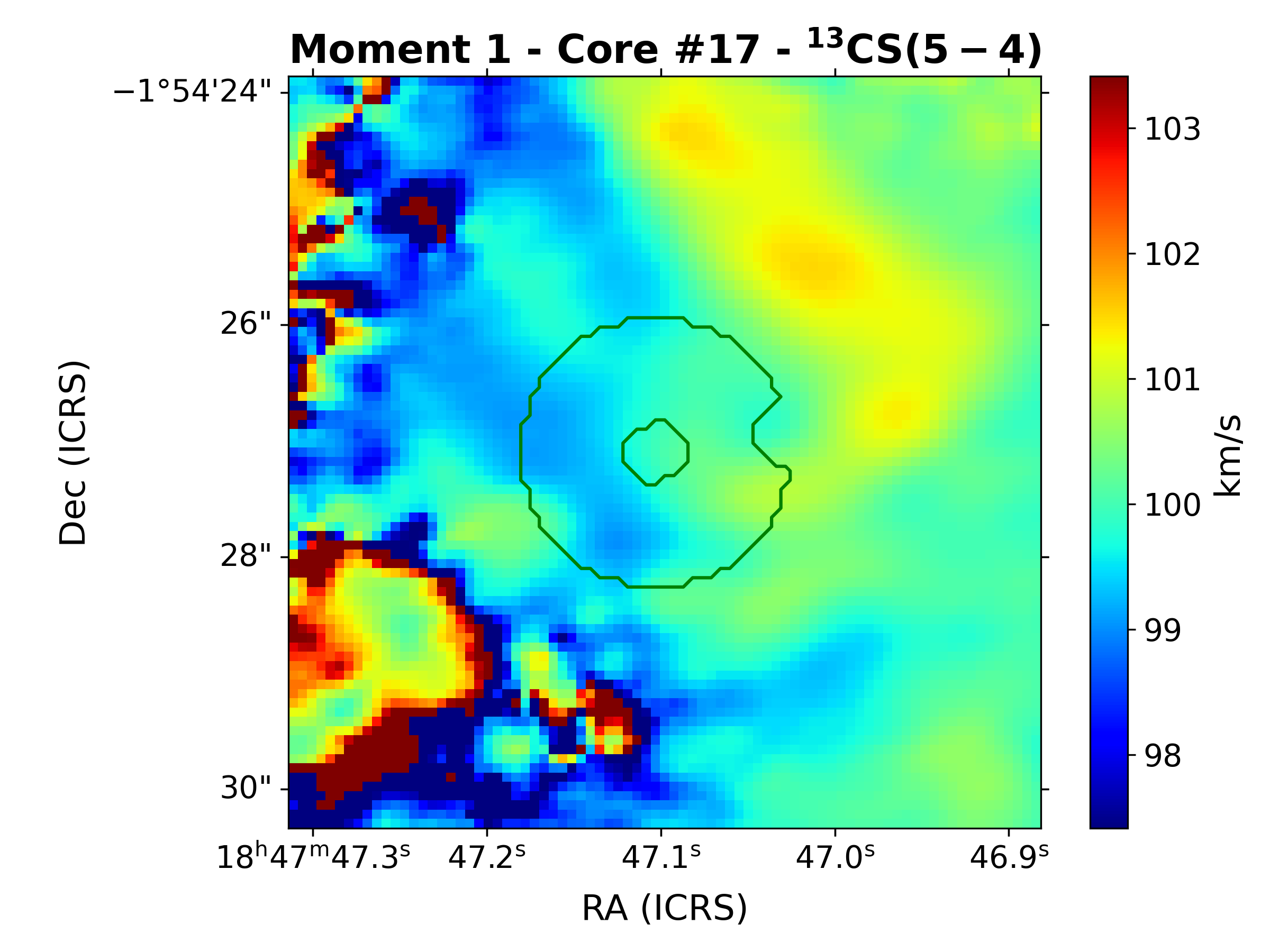}
\caption{Same as Fig.\,\ref{fig:fits_maps_core2} for core \#17.}
\label{fig:fits_maps_core3}  
\end{figure*}

\section{Impact of a common velocity dispersion on core support}\label{app:alpha_tot_appendix}
\begin{figure}[H]
    \centering
    \includegraphics[width=0.49\textwidth]{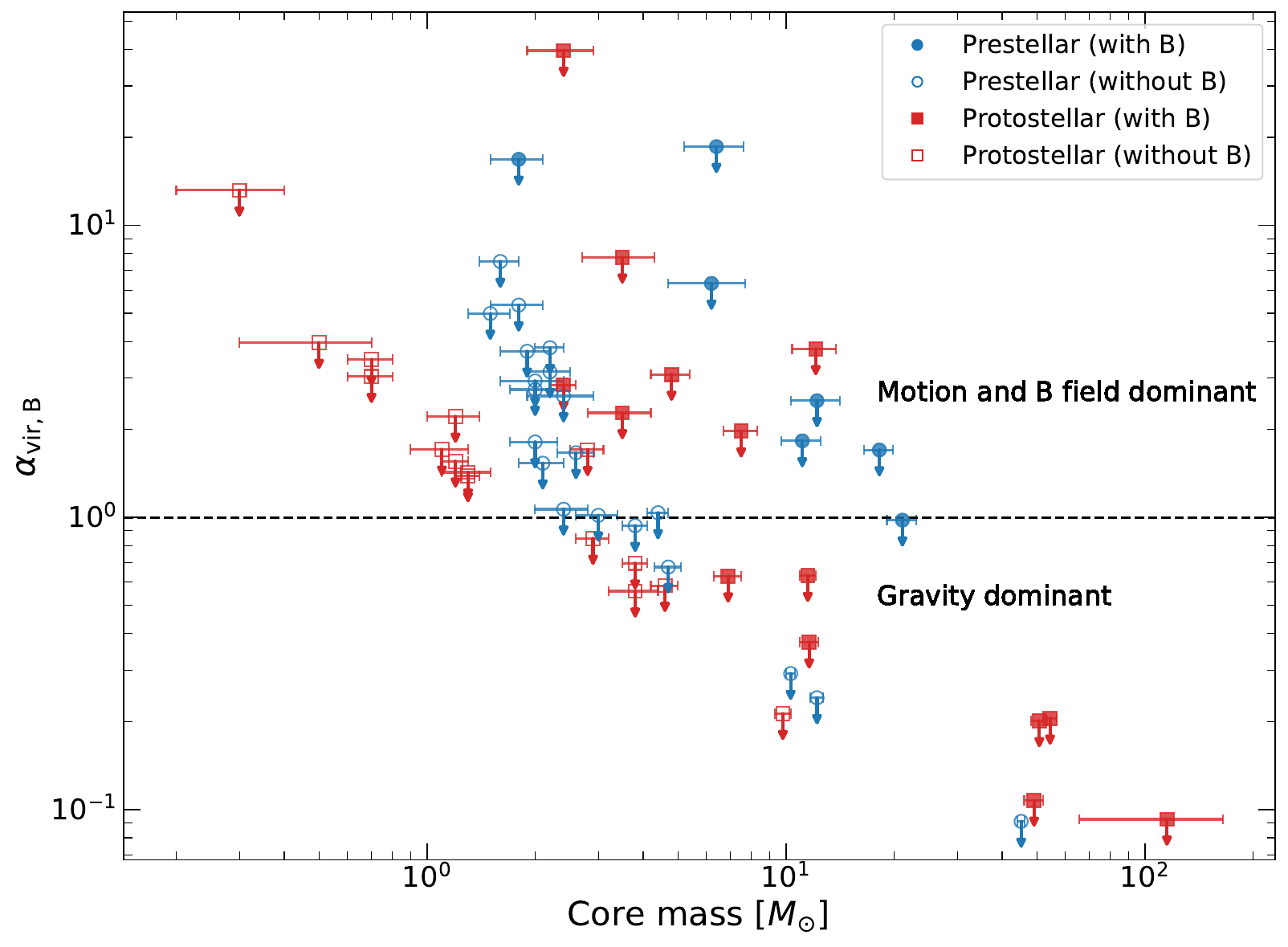}
    \caption{Total virial parameter $\alpha_{\mathrm{vir,B}}$ as a function of core mass, using median velocity dispersions for both kinetic and magnetic support. Prestellar cores are presented as blue dots and protostellar cores with red squares. Filled symbols correspond to cores for which a magnetic field strength could be estimated, while open symbols indicate cores where only kinetic (thermal $+$ non-thermal) motions are considered. The black dashed line displays the critical threshold $\alpha_{\mathrm{vir,B}} = 1$.} 
    \label{fig:alpha_tot_appendix}

\end{figure}

We reproduce in Fig.\,\ref{fig:alpha_tot_appendix} the same figure as Fig.\,\ref{fig:alpha_tot}, but using the median velocity dispersion at both the core-scale and the three-beam scale for all sources. In this configuration, we recover the usual trend reported in studies \citep{Cortes2016, Cortes2019} that rely on a single, global velocity dispersion: the most massive cores appear gravitationally dominated with an anti-correlation between the level of support and the mass of the cores. This demonstrates that adopting core-specific velocity dispersions is a crucial improvement, as it fundamentally alters the resulting virial interpretation.

\end{appendix}
\end{document}